\pdfoutput=1

\documentclass[11pt,twoside,a4paper,cmspaper,final,collab]{cms-tdr}
\def\svnVersion{71598}\def\cmsCernNo{2011-107}\def\cmsCernDate{\today}\def\cmsMessage{Submitted to the Journal of High Energy Physics}

\begin{document}\cmsNoteHeader{EWK-10-005}

\hyphenation{had-ron-i-za-tion}
\hyphenation{cal-or-i-me-ter}
\hyphenation{de-vices}
\RCS$Revision: 71592 $
\RCS$HeadURL: svn+ssh://alverson@svn.cern.ch/reps/tdr2/papers/EWK-10-005/trunk/EWK-10-005.tex $
\RCS$Id: EWK-10-005.tex 71592 2011-07-24 19:12:43Z gdaskal $
\newcommand{\comment}[1]{}
\newcommand{\pb}{\ensuremath{\mathrm{pb}}}%
\newcommand{\pp}{\ensuremath{\mathrm{pp}}}%
\newcommand{\Wo}{\ensuremath{\mathrm{W}}}%
\newcommand{\Wp}{\ensuremath{\mathrm{W^+}}}%
\newcommand{\Wm}{\ensuremath{\mathrm{W^-}}}%
\newcommand{\Zo}{\ensuremath{\mathrm{Z}}}%
\newcommand{\rts}{\ensuremath{\sqrt{s}}}%
\newcommand{\ra}{\ensuremath{\rightarrow}}%
\newcommand{\MN}{\ensuremath{\mu\nu}}%
\renewcommand{\EE}{\ensuremath{\mathrm{e}^+\mathrm{e}^-}}%
\newcommand{\EN}{\ensuremath{\mathrm{e}\nu}}%
\newcommand{\LN}{\ensuremath{\ell\nu}}%
\newcommand{\MW}{\ensuremath{m_\Wo}}%
\newcommand{\MZ}{\ensuremath{m_\Zo}}%
\newcommand{\MT}{\ensuremath{M_\mathrm{T}}}%
\newcommand{\MLL}{\ensuremath{m_{\ell\ell}}}%
\newcommand{\met}{\ensuremath{{E\!\!\!/}_{\!\mathrm{T}}}\xspace}
\renewcommand{\ttbar}{\ensuremath{\mathrm{t}\bar{\mathrm{t}}}}%
\newcommand{\Wmn}{\ensuremath{\Wo \ra \MN}}%
\newcommand{\Wpmn}{\ensuremath{\Wp \ra \mu^+\nu}}%
\newcommand{\Wmmn}{\ensuremath{\Wm \ra \mu^-\overline{\nu}}}%
\newcommand{\Zmm}{\ensuremath{\Zo \ra \MM}}%
\newcommand{\Wtn}{\ensuremath{\Wo \ra \tau\nu}}%
\newcommand{\Ztt}{\ensuremath{\Zo \ra \tau^+\tau^-}}%
\newcommand{\ppZmm}{\pp \ra \Zo(\gamma^*) + X \ra \MM + X}%
\newcommand{\ppWmn}{\pp \ra \Wo + X \ra \MN + X}%
\newcommand{\ppWpmn}{\pp \ra \Wp + X \ra \mu^+\nu + X}%
\newcommand{\ppWmmn}{\pp \ra \Wm + X \ra \mu^-\overline{\nu} + X}%
\newcommand{\Wen}{\ensuremath{\Wo \ra \EN}}%
\newcommand{\Wpen}{\ensuremath{\Wp \ra \mathrm{e}^+\nu}}%
\newcommand{\Wmen}{\ensuremath{\Wm \ra \mathrm{e}^-\overline{\nu}}}%
\newcommand{\Zee}{\ensuremath{\Zo \ra \EE}}%
\newcommand{\ppZee}{\pp \ra \Zo(\gamma^*) + X \ra \EE + X}%
\newcommand{\ppWen}{\pp \ra \Wo + X \ra \EN + X}%
\newcommand{\ppWpen}{\pp \ra \Wp + X \ra \mathrm{e}^+\nu  + X}%
\newcommand{\ppWmen}{\pp \ra \Wm + X \ra \mathrm{e}^-\overline{\nu} + X}%

\newcommand{\Wln}{\ensuremath{\Wo \ra \LN}}%
\newcommand{\Wpln}{\ensuremath{\Wp \ra \ell^+\nu}}%
\newcommand{\Wmln}{\ensuremath{\Wm \ra \ell^-\overline{\nu}}}%
\newcommand{\Zll}{\ensuremath{\Zo \ra \ell^+ \ell^-}}%
\newcommand{\ppZll}{\pp \ra \Zo(\gamma^*) + X \ra \ell^+ \ell^- + X}%
\newcommand{\ppWln}{\pp \ra \Wo + X \ra \LN + X}%
\newcommand{\ppWpln}{\pp \ra \Wp + X \ra \ell^+\nu  + X}%
\newcommand{\ppWmln}{\pp \ra \Wm + X \ra \ell^-\overline{\nu} + X}%

\newcommand{\Wptn}{\ensuremath{\Wp \ra \tau^+\nu}}%
\newcommand{\Wmtn}{\ensuremath{\Wm \ra \tau^-\overline{\nu}}}%

\newcommand{\Wev}{\Wen}
\newcommand{\Wmv}{\Wmn}
\newcommand{\Ztautau}{\Ztt}
\renewcommand{\MET}{\met}
\newcommand{\gammaZ}{\ensuremath{\gamma^{*}\Zo}}
\newcommand{\gammaZmm}{\mbox{$ \gamma^{*}/\Zo\rightarrow \MM$}}
\newcommand{\gammaZee}{\mbox{$ \gamma^{*}\Zo\rightarrow e^{+}  e^{-}$}}
\newcommand{\gammaZtt}{\mbox{$ \gamma^{*}\Zo\rightarrow \tau^{+}  \tau^{-}$}}
\newcommand{\gammaZll}{\mbox{$ \gamma^{*}\Zo\rightarrow \ell^{+}  \ell^{-}$}}
\newcommand{\invpb}{\mbox{$\textrm{pb}^{-1}$}}
\newcommand{\invnb}{\mbox{$\textrm{nb}^{-1}$}}
\newcommand{\mz}{\mbox{$ m_{Z}$}}
\newcommand{\hta}{\mbox{$ \eta$}}
\newcommand{\fh}{\mbox{$ \phi$}}
\newcommand{\etot}{\mbox{$ \epsilon_{tot}$}}
\newcommand{\eclustering}{\mbox{$ \epsilon_{clustering}$}}
\newcommand{\etracking}{\mbox{$ \epsilon_{tracking}$}}
\newcommand{\egsfele}{\mbox{$ \epsilon_{gsfele}$}}
\newcommand{\epreselection}{\mbox{$ \epsilon_{preselection}$}}
\newcommand{\eisolation}{\mbox{$ \epsilon_{isolation}$}}
\newcommand{\eclassification}{\mbox{$ \epsilon_{classification}$}}
\newcommand{\eelID}{\mbox{$ \epsilon_{elID}$}}
\newcommand{\etrigger}{\mbox{$ \epsilon_{trigger}$}}

\newcommand{\DE}{$\Delta\eta_{in}$}
\newcommand{\DP}{$\Delta\phi_{in}$}
\newcommand{\SEE}{$\sigma_{\eta\eta}$~}
\newcommand{\SEP}{$\sigma_{\eta\phi}$}
\newcommand{\SPP}{$\sigma_{\phi\phi}$}
\newcommand{\SXY}{$\sigma_{XY}$}
\newcommand{\pth}{\hat{p}_{\perp}}
\newcommand{\Pt}{p_{\mathrm{T}}}
\renewcommand{\pt}{\Pt}
\newcommand{\Et}{E\,\,_{\!\!\!\mathrm{T}}}
\newcommand{\Lint}{\ensuremath{{\cal L}_{\mathrm{int}}}}
\newcommand{\IRelComb} {I^{\textrm{rel}}_{\textrm{comb}}}%
\newcommand{\ITRK}     {I_{\textrm{trk}}}%
\newcommand{\IECAL}    {I_{\textrm{ECAL}}}%
\newcommand{\IHCAL}    {I_{\textrm{HCAL}}}%
\newcommand{\Nbg}{N_{\mathrm{bkg}}}%

\newcommand{\mumu}{\mu\mu}
\newcommand{\Zmumu}{\Zo_{\mumu}}
\newcommand{\Zmus}{\Zo_{\mu s}}
\newcommand{\Zmut}{\Zo_{\mu t}}
\newcommand{\nonIso}{\mathrm{noniso}}
\newcommand{\ZmumuNonIso}{\Zmumu^\nonIso}
\newcommand{\ZmumuTwoHlt}{\Zmumu^{2\mathrm{HLT}}}
\newcommand{\ZmumuOneHlt}{\Zmumu^{1\mathrm{HLT}}}

\newcommand{\NZtomumu}{N_{\Zmm}}

\newcommand{\Nmumu}{N_{\mumu}}
\newcommand{\Nmus}{N_{\mu s}}
\newcommand{\Nmut}{N_{\mu t}}
\newcommand{\NmumuNonIso}{\Nmumu^\nonIso}
\newcommand{\NmumuTwoHlt}{\Nmumu^{2\mathrm{HLT}}}
\newcommand{\NmumuOneHlt}{\Nmumu^{1\mathrm{HLT}}}

\newcommand{\effHlt}{\epsilon_\mathrm{HLT}}
\newcommand{\effIso}{\epsilon_\mathrm{iso}}
\newcommand{\effTrk}{\epsilon_\mathrm{trk}}
\newcommand{\effSa}{\epsilon_\mathrm{sa}}

\newcommand{\rhoeff} {\rho}
\newcommand{\etaSC}  {\eta_{\mathrm{SC}}}
\newcommand{\effmc}  {\epsilon_{\mathrm{sim}}}
\newcommand{\effdata}  {\epsilon_{\mathrm{data}}}
\newcommand{\effTPmc}  {\epsilon_{\mathrm{\tnp}}(\mathrm{sim)}}
\newcommand{\effTPdata}  {\epsilon_{\mathrm{\tnp}}(\mathrm{data})}
\newcommand{\TNP} {T\&P\xspace}
\newcommand{\tnp} {t\&p}
\def\ERROR#1#2{ \ensuremath{ \pm #1\, (\textrm{#2}) } \xspace }
\def\RESA#1#2#3{ \ensuremath{ #1 \ERROR{#2}{#3} } \xspace}
\def\RESB#1#2#3#4#5{ \ensuremath{ \RESA{#1}{#2}{#3} \ERROR{#4}{#5} } \xspace}
\def\RESC#1#2#3#4#5#6#7{ \ensuremath{ \RESB{#1}{#2}{#3}{#4}{#5} \ERROR{#6}{#7} } \xspace}
\def\RESD#1#2{ \ensuremath{ #1 \pm #2 } }
\def\RESE#1#2#3#4#5#6#7#8#9{ \ensuremath{ \RESC{#1}{#2}{#3}{#4}{#5}{#6}{#7} \ERROR{#8}{#9} } \xspace}
\def\RESGD#1#2#3#4{ \ensuremath{ \RESC{#1}{#2}{stat.}{#3}{syst.}{#4}{lumi.} }}
\def\EFF#1#2{ \ensuremath{ (\RESA{#1}{#2}{stat.})\% } \xspace}
\def\EFFA#1#2{ \ensuremath{ ({#1} \pm {#2})\% } \xspace}
\def\EFFB#1#2#3{ \ensuremath{ (\RESA{#1}{#2}{stat.} \ERROR{#3}{syst.})\% } \xspace}
\def\SIGBR#1#2{  \ensuremath{ \sigma \left( \pp \to #1 X \right) \times {\cal B} \left( #1 \to #2 \right) } \xspace}
\def\SIGBRSHORT#1{\ensuremath{ \sigma\times{\cal B}(#1) }}
\def\RESSIGBR#1#2#3#4#5#6{  \ensuremath{ \SIGBR{#1}{#2} &=& \RESC{#3}{#4}{stat.}{#5}{syst.}{#6}{lumi.} \, \textrm{nb} } \xspace}
\def\RESSIGBRTH#1#2#3#4#5#6#7{  \ensuremath{ \SIGBR{#1}{#2} &=& \RESE{#3}{#4}{stat.}{#5}{syst.}{#6}{th}{#7}{lumi.} \, \textrm{nb} } \xspace}
\def\RATWZ#1#2{ \ensuremath{ {
 \frac{ \sigma(\pp\rightarrow \Wo X)\times {\cal B}(\Wo\rightarrow #1)  }
      { \sigma(\pp\rightarrow \Zo X)\times {\cal B}(\Zo\rightarrow #2)  }   }  } }
\def\RESRATWZ#1#2#3#4#5#6{ \ensuremath{ \RATWZ{#1}{#2} &=& 
                                   #3 \ERROR{#4}{stat.} \ERROR{#5}{syst.} \ERROR{#6}{th.}} }
\def\RATWW#1#2{ \ensuremath{ {
 \frac{ \sigma(\pp\rightarrow \Wp X)\times {\cal B}(\Wp\rightarrow #1)  }
      { \sigma(\pp\rightarrow \Wm X)\times {\cal B}(\Wm\rightarrow #2)  }   }  } }
\def\RESRATWW#1#2#3#4#5#6{ \ensuremath{ \RATWW{#1}{#2} &=& 
                                   #3 \ERROR{#4}{stat.} \ERROR{#5}{syst.} \ERROR{#6}{th.}} }

\def\THEORYSIGBR#1#2{\ensuremath{ \RESD{#1}{#2}~{\mathrm{nb}} }}
\def\THEORYRATIO#1#2{\ensuremath{ \RESD{#1}{#2} }}

\def\EPS#1{ \ensuremath{ \epsilon_{\textrm{#1}} } \xspace}
\def\EPSTNPALL#1{ \ensuremath{ \EPS{\tnp-WP{#1}-ALL}  } \xspace }
\def\EPSTNPREC{ \ensuremath{ \EPS{\tnp-rec} } \xspace }
\def\EPSTNPTRG{ \ensuremath{ \EPS{\tnp-trg} } \xspace }
\def\EPSTNPWP#1{ \ensuremath{ \EPS{\tnp-WP{#1} } } \xspace }
\def\EPSTNPTRGWP#1{ \ensuremath{ \EPS{\tnp-TRG{#1}} } \xspace }  

\newcommand{\THELUMI} {\ensuremath{{35.9\pm 1.4}~\mathrm{pb}^{-1}}}%

\def\WPWIEFFRECO{   \ensuremath{\EFFA{99.7}{1.0}} \xspace} 
\def\WPWIEFFID{     \ensuremath{\EFFA{76.3}{1.9}} \xspace} 
\def\WPWIEFFHLT{    \ensuremath{\EFFA{98.9}{1.3}} \xspace} 
\def\WPWIEFF{       \ensuremath{\EFFA{75.3}{2.3}} \xspace} 
\def\WPWIEFFMC{     \ensuremath{ 79.98\% } \xspace} 
\def\WPWITNPR{      \ensuremath{\RESD{0.941}{0.028}} \xspace}  

\def\WPWIEBEFFRECO{   \ensuremath{\EFFA{97.0}{1.0}} \xspace} 
\def\WPWIEBMCRECO{   \ensuremath{\EFFA{97.78}{0.02}} \xspace}  
\def\WPWIEBRRECO{   \ensuremath{\RESD{0.992}{0.011}} \xspace} 

\def\WPWIEBEFFID{   \ensuremath{\EFFA{84.0}{0.3}} \xspace} 
\def\WPWIEBMCID{   \ensuremath{\EFFA{87.47}{0.05}} \xspace} 
\def\WPWIEBRID{   \ensuremath{\RESD{0.960}{0.004}} \xspace} 

\def\WPZIEBEFFID{   \ensuremath{\EFFA{93.9}{1.5}} \xspace} 
\def\WPZIEBMCID{   \ensuremath{ 96.4\% } \xspace}  
\def\WPZIEBRID{   \ensuremath{\RESD{0.974}{0.016}} \xspace} 

\def\WPWIEBEFFHLT{  \ensuremath{\EFFA{98.0}{0.1}}    \xspace} 
\def\WPWIEBMCHLT{   \ensuremath{\EFFA{97.10}{0.03}}    \xspace} 
\def\WPWIEBRHLT{    \ensuremath{\RESD{1.009}{0.001}} \xspace} 

\def\WPZIEBEFFHLT{  \ensuremath{\EFFA{98.7}{0.2}}    \xspace} 
\def\WPZIEBMCHLT{   \ensuremath{ 99.4\% }           \xspace} 
\def\WPZIEBRHLT{    \ensuremath{\RESD{0.992}{0.002}} \xspace} 

\def\WPWIEBEFF{    \ensuremath{\EFFA{79.8}{0.9}}     \xspace} 
\def\WPWIEBMC{     \ensuremath{\EFFA{83.05}{0.06}}     \xspace} 
\def\WPWIEBR{      \ensuremath{\RESD{0.961}{0.011} } \xspace} 

\def\WPZIEBEFF{     \ensuremath{\EFFA{91.3}{1.5}}    \xspace} 
\def\WPZIEBMC{     \ensuremath{ 94.4\% }            \xspace} 
\def\WPZIEBR{      \ensuremath{\RESD{0.967}{0.016} } \xspace} 

\def\WPWIEEEFFRECO{   \ensuremath{\EFFA{94.3}{1.1}} \xspace} 
\def\WPWIEEMCRECO{   \ensuremath{\EFFA{94.61}{0.05}} \xspace} 
\def\WPWIEERRECO{   \ensuremath{\RESD{0.997}{0.011}} \xspace}  

\def\WPWIEEEFFID{   \ensuremath{\EFFA{73.1}{0.7}} \xspace}  
\def\WPWIEEMCID{   \ensuremath{\EFFA{75.61}{0.06}} \xspace}  
\def\WPWIEERID{   \ensuremath{\RESD{0.966}{0.009}} \xspace}  

\def\WPWIEEEFFHLT{  \ensuremath{\EFFA{97.3}{0.3}}    \xspace} 
\def\WPWIEEMCHLT{   \ensuremath{\EFFA{97.16}{0.04}}    \xspace} 
\def\WPWIEERHLT{    \ensuremath{\RESD{1.001}{0.003}} \xspace} 

\def\WPWIEEEFF{     \ensuremath{\EFFA{67.0}{1.0}} \xspace}  
\def\WPWIEEMC{     \ensuremath{\EFFA{69.51}{0.07}}    \xspace} 
\def\WPWIEER{      \ensuremath{\RESD{0.965}{0.015} } \xspace} 

\def\WPZIEEEFFID{   \ensuremath{\EFFA{90.3}{1.9}} \xspace} 
\def\WPZIEEMCID{   \ensuremath{ 93.9\% } \xspace} 
\def\WPZIEERID{   \ensuremath{\RESD{0.962}{0.020}} \xspace}  

\def\WPZIEEEFF{     \ensuremath{\EFFA{86.1}{1.9}} \xspace}  
\def\WPZIEEMC{     \ensuremath{ 88.3\% } \xspace}
\def\WPZIEER{      \ensuremath{\RESD{0.975}{0.022} } \xspace} 

\def\WPZIEEEFFHLT{   \ensuremath{\EFFA{99.16}{0.02}} \xspace} 
\def\WPZIEEMCHLT{   \ensuremath{ 97.7\% } \xspace} 
\def\WPZIEERHLT{   \ensuremath{\RESD{1.015}{0.0003}} \xspace} 

\def\WEITNPDAT{ \ensuremath{\EFFA{75.1}{0.9}} \xspace} 
\def\WEPTNPDAT{ \ensuremath{\EFFA{75.3}{1.1}} \xspace} 
\def\WEMTNPDAT{ \ensuremath{\EFFA{74.8}{1.1}} \xspace} 

\def\WEITNPMC{ \ensuremath{ \EFFA{78.06 }{0.03}} \xspace} 
\def\WEPTNPMC{ \ensuremath{ \EFFA{77.68 }{0.03}} \xspace} 
\def\WEMTNPMC{ \ensuremath{ \EFFA{78.57 }{0.03}} \xspace} 

\def\WEITNPR{ \ensuremath{\RESD{0.962}{0.012} } \xspace} 
\def\WEPTNPR{ \ensuremath{\RESD{0.969}{0.014} } \xspace} 
\def\WEMTNPR{ \ensuremath{\RESD{0.952}{0.013} } \xspace}

\def\WEIEFFMC{ \ensuremath{(76.40 \pm 0.02)\%} \xspace} 
\def\WEPEFFMC{ \ensuremath{(76.04 \pm 0.03)\%} \xspace} 
\def\WEMEFFMC{ \ensuremath{(76.94 \pm 0.03)\%} \xspace} 

\def\WEIEBEFFMC{ \ensuremath{\RESD{81.51\%} {0.02\%}} \xspace} 
\def\WEPEBEFFMC{ \ensuremath{\RESD{81.38\%} {0.03\%}} \xspace} 
\def\WEMEBEFFMC{ \ensuremath{\RESD{81.69\%} {0.03\%}} \xspace} 

\def\WEIEEEFFMC{ \ensuremath{\RESD{67.63\%} {0.03\%}} \xspace} 
\def\WEPEEEFFMC{ \ensuremath{\RESD{67.47\%} {0.03\%}} \xspace} 
\def\WEMEEEFFMC{ \ensuremath{\RESD{67.90\%} {0.04\%}} \xspace} 

\def\WEIEFF{ \ensuremath{ \EFFA{73.5}{0.9} } \xspace} 
\def\WEPEFF{ \ensuremath{ \EFFA{73.7}{1.0} } \xspace} 
\def\WEMEFF{ \ensuremath{ \EFFA{73.2}{1.0} } \xspace} 

\def\WEISAMPLE{  \ensuremath{235\,687}   \xspace}  
\def\WEPSAMPLE{  \ensuremath{132\,696}   \xspace}  
\def\WEMSAMPLE{  \ensuremath{102\,991}   \xspace}  

\def\WEIAGEN{   \ensuremath{\RESD{0.5201}{0.0003}} \xspace} 
\def\WEPAGEN{   \ensuremath{\RESD{0.5297}{0.0004}} \xspace} 
\def\WEMAGEN{   \ensuremath{\RESD{0.5059}{0.0004}} \xspace} 

\def\WEIAGENGD{   \ensuremath{\RESD{0.520}{0.003}} } 
\def\WEPAGENGD{   \ensuremath{\RESD{0.530}{0.004}} } 
\def\WEMAGENGD{   \ensuremath{\RESD{0.506}{0.007}} } 

\def\WEIAPRIM{   \ensuremath{\RESD{0.3768}{0.0053}} \xspace} 
\def\WEPAPRIM{   \ensuremath{\RESD{0.3815}{0.0061}} \xspace} 
\def\WEMAPRIM{   \ensuremath{\RESD{0.3699}{0.0071}} \xspace} 

\def\WEIACC{   \ensuremath{ \RESD{0.4933 }{ 0.0003 }} \xspace} 
\def\WEPACC{   \ensuremath{ \RESD{0.5017 }{ 0.0004 }} \xspace} 
\def\WEMACC{   \ensuremath{ \RESD{0.4808 }{ 0.0004 }} \xspace} 

\def\WEIEBACC{ \ensuremath{ \RESD{0.3115 }{ 0.0003 }} \xspace} 
\def\WEPEBACC{ \ensuremath{ \RESD{0.3090 }{ 0.0003 }} \xspace} 
\def\WEMEBACC{ \ensuremath{ \RESD{0.3152 }{ 0.0004 }} \xspace} 

\def\WEIEEACC{ \ensuremath{ \RESD{0.1817 }{ 0.0002 }} \xspace} 
\def\WEPEEACC{ \ensuremath{ \RESD{0.1927 }{ 0.0003 }} \xspace} 
\def\WEMEEACC{ \ensuremath{ \RESD{0.1655 }{ 0.0003 }} \xspace} 

\def\WEIYIELD{ \ensuremath{\RESD{136\,328}{386} } \xspace} 
\def\WEPYIELD{ \ensuremath{\RESD{ 81\,568}{297} } \xspace} 
\def\WEMYIELD{ \ensuremath{\RESD{ 54\,760}{246} } \xspace} 

\def\WEIftYIELD{ \ensuremath{\RESD{135\,982}{388} } \xspace} 
\def\WEPftYIELD{ \ensuremath{\RESD{ 81\,286}{302} } \xspace} 
\def\WEMftYIELD{ \ensuremath{\RESD{ 54\,703}{249} } \xspace} 

\def\WEIabYIELD{ \ensuremath{\RESD{136\,003}{498} } \xspace} 
\def\WEPabYIELD{ \ensuremath{\RESD{ 81\,525}{385} } \xspace} 
\def\WEMabYIELD{ \ensuremath{\RESD{ 54\,356}{315} } \xspace} 

\def\rWEIftYIELD{ \ensuremath{\RESD{0.997}{0.005} } \xspace} 
\def\rWEPftYIELD{ \ensuremath{\RESD{0.997}{0.005} } \xspace} 
\def\rWEMftYIELD{ \ensuremath{\RESD{0.999}{0.005} } \xspace} 

\def\rWEIabYIELD{ \ensuremath{\RESD{0.998}{0.007} } \xspace} 
\def\rWEPabYIELD{ \ensuremath{\RESD{0.999}{0.007} } \xspace} 
\def\rWEMabYIELD{ \ensuremath{\RESD{0.993}{0.007} } \xspace} 

\def\WEIKSP{ \ensuremath{ X.XX } \xspace} 
\def\WEPKSP{ \ensuremath{ X.XX } \xspace} 
\def\WEMKSP{ \ensuremath{ X.XX } \xspace} 
\def\WEIKSPCOR{ \ensuremath{ X.XX } \xspace} 
\def\WEPKSPCOR{ \ensuremath{ 0.31 } \xspace} 
\def\WEMKSPCOR{ \ensuremath{ 0.25 } \xspace} 

\def\WEISIGBR{ \ensuremath{  \RESSIGBRTH{\Wo}{\mathrm{e}\nu}
                             {10.48}{0.03}{0.16}{0.09}{0.42} } \xspace} 
\def\WEPSIGBR{ \ensuremath{  \RESSIGBRTH{\Wp}{\mathrm{e}^+\bar{\nu}}
                             {6.15}{0.02}{0.10}{0.07}{0.25} } \xspace} 
\def\WEMSIGBR{ \ensuremath{  \RESSIGBRTH{\Wm}{e^-\nu}
                             {4.34}{0.02}{0.07}{0.06}{0.17} } \xspace} 
\def\ZEESIGBR{ \ensuremath{  \RESSIGBRTH{\Zo}{\mathrm{e}^+\mathrm{e}^-}
                             {0.992}{0.011}{0.018}{0.016}{0.040} } \xspace} 

\def\WEISIGBRX{\ensuremath{ \RESSIGBR{\Wo}{\mathrm{e}\nu}
                             {5.449}{0.015}{0.086}{0.218} } \xspace}
\def\WEPSIGBRX{\ensuremath{ \RESSIGBR{\Wp}{\mathrm{e}^+\nu}
                             {3.257}{0.012}{0.061}{0.130} } \xspace}
\def\WEMSIGBRX{\ensuremath{ \RESSIGBR{\Wm}{\mathrm{e}^-\bar{\nu}}
                             {2.193}{0.010}{0.039}{0.088} } \xspace}
\def\ZEESIGBRX{\ensuremath{ \RESSIGBR{\Zo}{\mathrm{e}^+\mathrm{e}^-}
                             {0.420}{0.005}{0.010}{0.017} } \xspace}

\def\WEISIGBRXGD{\ensuremath{ \RESGD
                             {5.449}{0.015}{0.086}{0.218} }}
\def\WEPSIGBRXGD{\ensuremath{ \RESGD
                             {3.257}{0.012}{0.061}{0.130} }}
\def\WEMSIGBRXGD{\ensuremath{ \RESGD
                             {2.193}{0.010}{0.039}{0.088} }}
\def\ZEESIGBRXGD{\ensuremath{ \RESGD
                             {0.420}{0.005}{0.010}{0.017} }}

\def\ZEESAMPLE{  \ensuremath{8452}   \xspace} 
\def\ZEESAMPLEN{  \ensuremath{8442}   \xspace} 

\def\ZEEYIELD{ \ensuremath{ \RESD{8406}{92} } \xspace} 
\def\ZEEQCDBKG{ \ensuremath{ \RESD{5}{12}  } \xspace} 
\def\ZEEEWKBKG{   \ensuremath{ \RESD{30.8}{0.4}   } \xspace} 
\def\ZEEBKG{ \ensuremath{ \RESD{36}{12}  } \xspace} 
\def\ZEESIG{   \ensuremath{ \RESD{xxx.x}{xx.x} } \xspace} 

\def\ZEEAGEN{     \ensuremath{ \RESD{0.4230}{0.0005} } \xspace}  
\def\ZEEAPRIM{     \ensuremath{ \RESD{0.2577}{0.0045} } \xspace}  

\def\ZEEAGENGD{     \ensuremath{ \RESD{0.423}{0.004} } }  

\def\ZEEACC{     \ensuremath{ \RESD{0.3876}{0.0005} } \xspace} 
\def\ZEEBBACC{   \ensuremath{ \RESD{0.2119}{0.0004} } \xspace} 
\def\ZEEBEACC{   \ensuremath{ \RESD{0.1296}{0.0003} } \xspace} 
\def\ZEEEEACC{   \ensuremath{ \RESD{0.0462}{0.0002} } \xspace} 

\def\ZEETNPDAT{ \ensuremath{ \EFFA{60.7}{1.1} } \xspace} 
\def\ZEETNPMC{ \ensuremath{ \EFFA{66.48}{0.03} } \xspace}  
\def\ZEETNPR{  \ensuremath{ \RESD{0.913}{0.017} } \xspace}  
\def\ZEEEFFMC{ \ensuremath{ \EFFA{66.74}{0.07} } \xspace}  
\def\ZEEEFF{   \ensuremath{ \EFFA{60.9}{1.1} } \xspace}  

\def\EBESCALE{ \ensuremath{ {X.XXX}\pm{X.XXX} } \xspace} 
\def\EEESCALE{ \ensuremath{ {X.XXX}\pm{X.XXX} } \xspace}  

\def\EBESMEAR{ \ensuremath{ {X.XX}\pm{X.XX}~\GeV } \xspace} 
\def\EEESMEAR{ \ensuremath{ {X.XX}\pm{X.XX}~\GeV } \xspace}  

\def\RESRATWZE{ \ensuremath{ \RESRATWZ{\mathrm{e}\nu}{\mathrm{e}^+\mathrm{e}^-}{10.56}{0.12}{0.12}{0.15} } }
\def\RESRATWWE{ \ensuremath{ \RESRATWW{\mathrm{e}^+\nu}{\mathrm{e}^-\bar{\nu}}{1.418}{0.008}{0.022}{0.029} } }

\def\LUMISYST{ \ensuremath{ 4 } \xspace}

\def\WEITNPSYST{ \ensuremath{ 1.4 } \xspace}
\def\WEIESCALESYST{ \ensuremath{ 0.5 } \xspace}
\def\WEPESCALESYST{ \ensuremath{ 0.5 } \xspace}
\def\WEMESCALESYST{ \ensuremath{ 0.6 } \xspace}
\def\WERESCALESYST{ \ensuremath{ 0.1 } \xspace}
\def\WEIMETSYST{ \ensuremath{ 0.3 } \xspace}
\def\WEPMETSYST{ \ensuremath{ 0.3 } \xspace}
\def\WEMMETSYST{ \ensuremath{ 0.3 } \xspace}
\def\WERMETSYST{ \ensuremath{ 0.1 } \xspace}
\def\WEIBKGSYST{ \ensuremath{ 0.35 } \xspace}
\def\WEPBKGSYST{ \ensuremath{ 0.33 } \xspace}
\def\WEMBKGSYST{ \ensuremath{ 0.48 } \xspace}
\def\WERBKGSYST{ \ensuremath{ 0.39 } \xspace}
\def\ZEEBKGSYST{ \ensuremath{ 0.14 } \xspace}
\def\WPWMISID{ \ensuremath{ X.X^{+X.X}_{-X.X} } \xspace}

\def\ZEEESCALESYST{ \ensuremath{ 0.12 } \xspace}
\def\ZEETNPSYST{ \ensuremath{ 1.8 } \xspace}

\def\WEIPDFACCSYST{\ensuremath{ 0.6 }}
\def\ZEEPDFACCSYST{\ensuremath{ 0.9 }}
\def\WEITHSYST{\ensuremath{ 0.7 }}
\def\ZEETHSYST{\ensuremath{ 1.4 }}

\def\WEITOTSYST{\ensuremath{ 1.8 }}
\def\ZEETOTSYST{\ensuremath{ 2.4 }}

\def\WEEXPSYST{\ensuremath{ 1.6 }}
\def\ZEEEXPSYST{\ensuremath{ 1.8 }}
\def\WEITOTTHSYST{   \ensuremath{ 0.9 }}
\def\ZEETOTTHSYST{   \ensuremath{ 1.6 }}

\def\WMUIEFFSA{      \ensuremath{\EFFA{96.4}{0.5}} \xspace}
\def\WMUIMCEFFSA{    \ensuremath{ 97.2\% }  \xspace} 
\def\WMUIRSA{      \ensuremath{\RESD{0.992}{0.005}}  \xspace}
\def\WMUIEFFTRK{      \ensuremath{\EFFA{99.1}{0.4}} \xspace}
\def\WMUIMCEFFTRK{    \ensuremath{ 99.3\% }  \xspace} 
\def\WMUIRTRK{      \ensuremath{\RESD{0.998}{0.003}}  \xspace}
\def\WMUIEFFSEL{      \ensuremath{\EFFA{99.7}{0.3}} \xspace}
\def\WMUIMCEFFSEL{    \ensuremath{ 99.7\% }  \xspace} 
\def\WMUIRSEL{      \ensuremath{\RESD{1.000}{0.003}}  \xspace}
\def\WMUIEFFISO{      \ensuremath{\EFFA{98.5}{0.4}} \xspace}
\def\WMUIMCEFFISO{    \ensuremath{ 99.1\% }  \xspace} 
\def\WMUIRISO{      \ensuremath{\RESD{0.994}{0.004}}  \xspace}
\def\WMUIEFFTRG{      \ensuremath{\EFFA{88.3}{0.8}} \xspace}
\def\WMUIMCEFFTRG{    \ensuremath{ 93.2\% }  \xspace} 
\def\WMUIRTRG{      \ensuremath{\RESD{0.947}{0.009}}  \xspace}
\def\WMUIEFF{      \ensuremath{\EFFA{82.8}{1.0}} \xspace}
\def\WMUIMCEFF{    \ensuremath{ 88.7\% }  \xspace} 

\def\WMIEFFPLS{\ensuremath{ \RESD{0.935}{0.018} }}
\def\WMIEFFMIN{\ensuremath{ \RESD{0.931}{0.019} }}
\def\WMIEFFBAR{\ensuremath{ \RESD{0.955}{0.024} }}
\def\WMIEFFTRA{\ensuremath{ \RESD{0.89}{0.04} }}
\def\WMIEFFEND{\ensuremath{ \RESD{0.92}{0.03} }}

\def\WMIRHOEFF{ \ensuremath{\RESD{0.933}{0.012}} }
\def\WMUIR{        \ensuremath{\WMIRHOEFF}  \xspace}

\def\WMIAGEN{   \ensuremath{\RESD{0.4543}{0.0003}} } 
\def\WMPAGEN{   \ensuremath{\RESD{0.4594}{0.0004}} } 
\def\WMMAGEN{   \ensuremath{\RESD{0.4471}{0.0004}} } 
\def\ZMMAGEN{   \ensuremath{\RESD{0.3977}{0.0048}} } 

\def\WMIAGENGD{   \ensuremath{\RESD{0.464}{0.003}} }
\def\WMPAGENGD{   \ensuremath{\RESD{0.471}{0.005}} }
\def\WMMAGENGD{   \ensuremath{\RESD{0.457}{0.008}} }
\def\ZMMAGENGD{   \ensuremath{\RESD{0.398}{0.005}} }

\def\WMIAPRIM{  \ensuremath{\RESD{0.4618}{0.0051}} } 
\def\WMPAPRIM{  \ensuremath{\RESD{0.4765}{0.0053}} } 
\def\WMMAPRIM{  \ensuremath{\RESD{0.4413}{0.0052}} } 

\def\ZMMBG{ \ensuremath{\RESD{55}{3}} }

\def\WMISAMPLE{  \ensuremath{18\,571} }
\def\WMPSAMPLE{  \ensuremath{10\,682} }
\def\WMMSAMPLE{  \ensuremath{ 7\,889} }
\def\WMISAMPLEH{ \ensuremath{11\,011} }
\def\WMPSAMPLEH{ \ensuremath{ 6\,495} }
\def\WMMSAMPLEH{ \ensuremath{ 4\,516} }
\def\ZMMSAMPLE{  \ensuremath{913} }

\def\WMIYIELD{ \ensuremath{\RESD{12\,257}{111} }} 
\def\WMPYIELD{ \ensuremath{\RESD{ 7\,445}{ 87} }} 
\def\WMMYIELD{ \ensuremath{\RESD{ 4\,812}{ 68} }} 
\def\ZMMYIELD{ \ensuremath{\RESD{13\,728}{121} }} 

\def\WMIEFFSYST{     \ensuremath{ 0.9 }}
\def\WMIEFFPRET{     \ensuremath{ 0.5 }}
\def\ZMMEFFSYST{     \ensuremath{ 1.2 }}
\def\ZMMEFFPRET{     \ensuremath{ 0.5 }}
\def\WMIEFFSYSTPRET{     \ensuremath{ 1.5 }}
\def\WMISCALESYST{   \ensuremath{ 0.22 }}
\def\ZMMSCALESYST{   \ensuremath{ 0.35 }}
\def\WMIQCDSHAPESYST{\ensuremath{ 0.4 }}
\def\WMIRECOILSYST{\ensuremath{ 0.6 }}

\def\WMIMETSYST{ \ensuremath{ 0.2 } }
\def\WMIBKGSYST{\ensuremath{ ? }}
\def\ZMMBKGSYST{\ensuremath{ 0.2 }}
\def\ZMMFITSYST{\ensuremath{ 0.2 }}
\def\ZMMBKGSYST{\ensuremath{ 0.2 }}
\def\ZMMBKGTOTSYST{\ensuremath{ 0.28 }}
\def\ZMMTRIGABSYST{\ensuremath{ 0.1 }}

\def\WMEXPSYST{\ensuremath{ 1.1 }}
\def\ZMMEXPSYST{\ensuremath{ 0.7 }}

\def\WMIPDFACCSYST{\ensuremath{ 0.8 }}
\def\ZMMPDFACCSYST{\ensuremath{ 1.1 }}
\def\WMITHSYST{    \ensuremath{ 0.8 }}
\def\ZMMTHSYST{    \ensuremath{ 1.6 }}
\def\WMITOTSYST{   \ensuremath{ 1.6 }}
\def\ZMMTOTTHSYST{   \ensuremath{ 1.9 }}
\def\WMITOTTHSYST{   \ensuremath{ 1.1 }}
\def\ZMMTOTSYST{   \ensuremath{ 2.0 }}

\def\WMISIGBR{\ensuremath{ \RESSIGBRTH{\Wo}{\mu\nu}
                             {10.18}{0.03}{0.12}{0.11}{0.41}} }
\def\WMPSIGBR{\ensuremath{ \RESSIGBRTH{\Wp}{\mu^+\nu}
                             {5.98}{0.02}{0.07}{0.08}{0.24} } }
\def\WMMSIGBR{\ensuremath{ \RESSIGBRTH{\Wm}{\mu^-\bar{\nu}}
                             {4.20}{0.02}{0.05}{0.07}{0.17} } }
\def\ZMMSIGBR{\ensuremath{  \RESSIGBRTH{\Zo}{\mu^+\mu^-}
                             {0.968}{0.008}{0.007}{0.018}{0.039} } }
\def\WMISIGBRX{\ensuremath{ \RESSIGBR{\Wo}{\mu\nu}
                             {4.723}{0.012}{0.066}{0.189} } \xspace}
\def\WMPSIGBRX{\ensuremath{ \RESSIGBR{\Wp}{\mu^+\nu}
                             {2.815}{0.009}{0.042}{0.113} } \xspace}
\def\WMMSIGBRX{\ensuremath{ \RESSIGBR{\Wm}{\mu^-\bar{\nu}}
                             {1.920}{0.008}{0.027}{0.077} } \xspace}
\def\ZMMSIGBRX{\ensuremath{ \RESSIGBR{Z}{\mu^+\mu^-}
                             {0.385}{0.003}{0.007}{0.015} } \xspace}

\def\WMISIGBRXGD{\ensuremath{ \RESGD
                             {4.736}{0.012}{0.067}{0.189} }}
\def\WMPSIGBRXGD{\ensuremath{ \RESGD
                             {2.815}{0.009}{0.042}{0.113} }}
\def\WMMSIGBRXGD{\ensuremath{ \RESGD
                             {1.921}{0.008}{0.027}{0.077} }}
\def\ZMMSIGBRXGD{\ensuremath{ \RESGD
                             {0.396}{0.003}{0.007}{0.016} }}

\def\RESRATWZM{\ensuremath{ \RESRATWZ{\mu\nu}{\mu^+\mu^-}{10.52}{0.09}{0.10}{0.17} }}
\def\RESRATWWM{\ensuremath{ \RESRATWW{\mu^+\nu}{\mu^-\bar{\nu}}{1.423}{0.008}{0.019}{0.030} }}

\def\WLISIGBR{\ensuremath{ \RESSIGBRTH{W}{\ell\nu}
                             {10.30}{0.02}{0.10}{0.10}{0.41} } }
\def\WLPSIGBR{\ensuremath{ \RESSIGBRTH{\Wp}{\ell^+\nu}
                             {6.04}{0.02}{0.06}{0.08}{0.24} } }
\def\WLMSIGBR{\ensuremath{ \RESSIGBRTH{\Wm}{\ell^-\bar{\nu}}
                             {4.26}{0.01}{0.04}{0.07}{0.17} } }
\def\ZLLSIGBR{\ensuremath{ \RESSIGBRTH{Z}{\ell^+\ell^-}
                             {0.974}{0.007}{0.007}{0.018}{0.039} } }

\def\RESRATWZL{ \ensuremath{ \RESRATWZ{\ell\nu}{\ell^+\ell^-}{10.54}{0.07}{0.08}{0.16} }}
\def\RESRATWWL{ \ensuremath{ \RESRATWW{\ell^+\nu}{\ell^-\bar{\nu}}{1.421}{0.006}{0.014}{0.029} }}

\def\THEORYSIGBR#1#2{\ensuremath{ \RESD{#1}{#2}~{\mathrm{nb}} }}
\def\THEORYSIGBRWI{\ensuremath{ \THEORYSIGBR{10.44}{0.27} }}
\def\THEORYSIGBRWP{\ensuremath{ \THEORYSIGBR{6.15}{0.17} }}
\def\THEORYSIGBRWM{\ensuremath{ \THEORYSIGBR{4.29}{0.11} }}
\def\THEORYSIGBRZ{ \ensuremath{ \THEORYSIGBR{0.97}{0.03} }}
\def\THEORYRATIOWZ{\ensuremath{ \THEORYRATIO{10.74}{0.04} }}
\def\THEORYRATIOWW{\ensuremath{ \THEORYRATIO{1.43}{0.01} }}

\def\RATCMSTHY#1#2#3#4{\ensuremath{ #1\pm #2\,{\mathrm{(exp)}}\pm #3\,{\mathrm{(th)}}\,\, [ \pm #4\,{\mathrm{(tot)}} ] }}
\def\RATCMSTHYWI{\ensuremath{\RATCMSTHY{0.986}{0.009}{0.028}{0.029}}}
\def\RATCMSTHYWP{\ensuremath{\RATCMSTHY{0.982}{0.010}{0.030}{0.031}}}
\def\RATCMSTHYWM{\ensuremath{\RATCMSTHY{0.992}{0.010}{0.029}{0.031}}}
\def\RATCMSTHYZ {\ensuremath{\RATCMSTHY{1.002}{0.010}{0.032}{0.034}}}
\def\RATCMSTHYWZ{\ensuremath{\RATCMSTHY{0.981}{0.010}{0.015}{0.018}}}
\def\RATCMSTHYWW{\ensuremath{\RATCMSTHY{0.990}{0.011}{0.023}{0.025}}}

\cmsNoteHeader{EWK-10-005}

\title{\texorpdfstring{Measurement of the Inclusive W and Z Production Cross Sections
in pp Collisions at $\sqrt{s}=7\TeV$}{Measurement of the Inclusive W and Z Production Cross Sections
in pp Collisions at sqrt(s)=7 TeV}}

\date{\today}

\abstract{
A measurement of inclusive W and Z production
cross sections in pp collisions at $\sqrt{s}=7\TeV$ is presented.
The electron and muon decay channels are analyzed
in a data sample collected with the CMS detector at the LHC and
corresponding to an integrated lu\-mi\-no\-si\-ty of $36~\text{pb}^{-1}$.
The measured inclusive cross sections are
$\sigma(\mathrm{pp} \rightarrow \mathrm{W}X) \times {\cal B}(\mathrm{W} \rightarrow \ell\nu ) =
10.30 \pm 0.02 ~\mathrm{(stat.)} \pm 0.10 ~\mathrm{(syst.)} \pm 0.10 ~\mathrm{(th.)} \pm 0.41 ~\mathrm{(lumi.)}$~nb
and
$\sigma(\mathrm{pp} \rightarrow \mathrm{Z}X) \times {\cal B}(\mathrm{Z} \rightarrow \ell^+\ell^-) = 0.974 \pm 0.007
~\mathrm{(stat.)} \pm 0.007 ~\mathrm{(syst.)} \pm 0.018 ~\mathrm{(th.)} \pm 0.039 ~\mathrm{(lumi.)}$~nb,
limited to the dilepton invariant mass range 60 to 120~\text{Ge\hspace{-.08em}V}.
The luminosity-independent cross section ratios are
${(\sigma(\mathrm{pp} \rightarrow \mathrm{W}X) \times {\cal B}(\mathrm{W} \rightarrow \ell\nu ))}$ / ${(\sigma(\mathrm{pp} \rightarrow \mathrm{Z}X) \times {\cal B}(\mathrm{Z} \rightarrow \ell^+\ell^-))}
= 10.54 \pm 0.07 ~\mathrm{(stat.)} \pm 0.08 ~\mathrm{(syst.)} \pm 0.16 ~\mathrm{(th.)}$
and
${(\sigma(\mathrm{pp} \rightarrow \mathrm{W}^{+}X) \times {\cal B}(\mathrm{W}^{+} \rightarrow \ell^{+}\nu ))}$ / ${(\sigma(\mathrm{pp} \rightarrow \mathrm{W}^{-}X) \times {\cal B}(\mathrm{W}^{-} \rightarrow \ell^{-}\nu ))}
= 1.421 \pm 0.006 ~\mathrm{(stat.)} \pm 0.014 ~\mathrm{(syst.)} \pm 0.029 ~\mathrm{(th.)}$.
The measured values agree with next-to-next-to-leading order QCD cross section calculations
based on recent parton distribution functions.
}

\hypersetup{%
pdfauthor={CMS Collaboration},%
pdftitle={Measurement of the Inclusive W and Z Production Cross Sections in pp Collisions at sqrt(s) = 7 TeV },%
pdfsubject={CMS},%
pdfkeywords={CMS, physics}}

\maketitle 

\section{Introduction}
\label{sec:introduction}

This paper describes a measurement carried out by the Compact Muon Solenoid (CMS) Collaboration
of the inclusive production cross sections for W and Z bosons in pp collisions at
$\sqrt{s} = 7\TeV$. The vector bosons are observed  via their decays to electrons and muons.
In addition, selected cross-section ratios are presented. Precise determination
of the production cross sections and their ratios provide an important
test of the standard model (SM) of particle physics.

The production of the electroweak (EWK) gauge bosons in pp collisions 
proceeds mainly via the weak Drell--Yan (DY) process~\cite{DY} consisting of
the annihilation of a quark and an antiquark.
The production process $\pp \ra \Wo + X$ is dominated by  
$\mathrm{u}\bar{\mathrm{d}}\ra\Wp$ and $\mathrm{d}\bar{\mathrm{u}}\ra\Wm$, 
while  $\pp \ra \Zo + X$ is dominated by $\mathrm{u}\bar{\mathrm{u}}$ and
$\mathrm{d}\bar{\mathrm{d}}\ra\Zo$.

Theoretical predictions of the total W and Z production cross sections
are determined from parton-parton cross sections convolved with parton 
distribution functions (PDFs), incorporating higher-order quantum chromodynamics (QCD) effects. 
PDF uncertainties, as well as higher-order QCD and EWK radiative corrections,
limit the precision of current theoretical predictions, which are available at 
next-to-leading order (NLO)~\cite{nlo1, nlo2, nlo3} and next-to-next-to-leading order 
(NNLO)~\cite{nnlo1, nnlo2, nnlo3, nnlo4, nnlo5} in perturbative QCD.

The momentum fractions of the colliding partons $x_1$, $x_2$ are related to the
vector boson masses ($m_{\Wo/\Zo}^{2} = s x_1 x_2$) and
rapidities ($y = \frac{1}{2}\ln(x_1/x_2)$). 
Within the accepted rapidity interval, $ | y | \le 2.5$, the values 
of $x$ are in the range $10^{-3} \le x \le 0.1$.

Vector boson production in proton-proton collisions requires at least one sea quark,
while two valence quarks are typical of $\mathrm{p}\bar{\mathrm{p}}$ collisions. 
Furthermore, given the high scale of the process, 
${\hat s} = m_{\Wo/\Zo}^2\sim 10^4\GeV^2$, the gluon is the dominant parton 
in the proton so that the scattering sea quarks are mainly generated by the 
$\mathrm{g} \rightarrow \mathrm{q} \bar{\mathrm{q}}$ splitting process.
For this reason, the precision of the cross section predictions
at the Large Hadron Collider (LHC) depends crucially on the uncertainty 
in the momentum distribution of the gluon.
Recent measurements from HERA~\cite{HERApdf} and the
Tevatron~\cite{TevatronPdf_1, TevatronPdf_2, 
TevatronPdf_4, TevatronPdf_5, TevatronPdf_6, TevatronPdf_7, TevatronPdf_8, TevatronPdf_9, TevatronPdf_10} 
reduced the PDF uncertainties, leading to more precise
cross-section predictions at the LHC.

The W and Z production cross sections and their ratios were
previously measured by \mbox{ATLAS}~\cite{WZATLAS:2010} with
an integrated luminosity of 320~nb$^{-1}$ and by 
CMS~\cite{WZCMS:2010} with 2.9~pb$^{-1}$. 
This paper presents an update with the full integrated luminosity recorded by CMS at the LHC 
in 2010, corresponding to 36~pb$^{-1}$.
The leptonic branching fraction and the width of the W boson can be extracted from the 
measured W/Z cross section ratio 
using the NNLO predictions for the total W and Z cross sections and the measured 
values of the Z boson total and leptonic partial widths~\cite{LEPZ}, together with the SM
prediction for the leptonic partial width of the W. 

This paper is organized as follows: in Section~\ref{sec:detector} the CMS detector is presented,
with particular attention to the subdetectors 
used to identify charged leptons and to infer the presence of neutrinos. 
Section~\ref{sec:samples} describes the data sample and simulation 
used in the analysis. The selection of the W and Z candidate 
events is discussed in Section~\ref{sec:eventSelection}. Section~\ref{sec:acceptance} describes 
the calculation of the geometrical and kinematic acceptances. 
The methods used to determine the 
reconstruction, selection, and trigger efficiencies of the leptons within 
the experimental acceptance are presented in Section~\ref{sec:efficiencies}. 
The signal extraction methods for the W and Z channels, as well as 
the background contributions to the
candidate samples, are discussed in Sections~\ref{sec:WsignalExtraction} 
and~\ref{sec:ZsignalExtraction}. Systematic uncertainties are
discussed in Section~\ref{sec:systematics}. The calculation of the 
total cross sections, along with the resulting values of the ratios and derived quantities, 
are summarized in Section~\ref{sec:results}. In the same section we also report the cross 
sections as measured within the fiducial and kinematic acceptance (after final-state QED 
radiation corrections), thereby eliminating the PDF uncertainties from the results.

\section{The CMS Detector}
\label{sec:detector}
\par
The central feature of the CMS apparatus
is a superconducting solenoid of 6~m internal diameter, providing
a magnetic field of $3.8$~T. Within the field volume are a silicon pixel
and strip tracker, an electromagnetic calorimeter (ECAL),
and a hadron calorimeter (HCAL). Muons are detected
in gas-ionization detectors embedded in the steel return
yoke. In addition to the barrel and endcap detectors, CMS has
extensive forward calorimetry.
\par
A right-handed coordinate system is used in CMS, with the origin at the
nominal interaction point, the $x$-axis pointing to the center of
the LHC ring, the $y$-axis pointing up (perpendicular to the LHC plane),
and the $z$-axis along the anticlockwise-beam direction. The polar
angle $\theta$ is measured from the positive $z$-axis and the
azimuthal angle $\phi$ is measured (in radians) in the $xy$-plane.
The pseudorapidity is given by $\eta = -\ln\tan(\theta/2)$.
\par
The inner tracker measures charged particle trajectories in the
pseudorapidity range $|\eta| < 2.5$.   It consists of $1440$ silicon
pixel and 15\,148 silicon strip detector modules.  It provides an
impact parameter resolution of ${\approx}15\mum$ and a transverse
momentum ($\pt$) resolution of about 1\% for charged particles
with $\pt \approx 40\GeV$.
\par
The electromagnetic calorimeter consists of nearly $76\,000$ lead tungstate
crystals, which provide coverage in pseudorapidity $|\eta| < 1.479$ in a
cylindrical barrel region (EB) and $1.479 < |\eta| < 3.0$ in two endcap
regions (EE).
A preshower detector
consisting of two planes of silicon sensors interleaved with a total of
three radiation lengths of lead is located in front of the EE.
The ECAL has an energy resolution of better than $0.5\%$ for
unconverted photons with transverse energies ($\Et$) above $100\GeV$.
The energy resolution is $3\%$ or better for the range of
electron energies relevant for this analysis.
The hadronic barrel and endcap calorimeters are sampling devices with brass
as the passive material and scintillator as the active material.
The combined calorimeter cells are grouped in projective towers of granularity
$\Delta \eta \times \Delta \phi = 0.087\times0.087$ at central rapidities
and $0.175\times0.175$ at forward rapidities.
The energy of charged pions and other quasi-stable hadrons can be measured with 
the calorimeters (ECAL and HCAL combined) with a resolution of 
$\Delta E/E \simeq 100 \%/\sqrt{E(\GeV)} \oplus 5\%$. 
For charged hadrons, the calorimeter resolution improves on the tracker momentum 
resolution only for $\PT$ in excess of 500~GeV. 
The energy resolution on jets and missing transverse energy is substantially improved with
respect to calorimetric reconstruction by using
the particle flow (PF) algorithm~\cite{PFT} which consists in reconstructing and
identifying each single particle with an optimised combination of all sub-detector
information. This approach exploits the very good tracker momentum resolution
to improve the energy measurement of charged hadrons.

\par
Muons are detected in the pseudorapidity window $|\eta|< 2.4$, with
detection planes based on three technologies: drift tubes, cathode strip
chambers, and resistive plate chambers.  A high-$\pt$ muon originating
from the interaction point produces track segments typically
in three or four muon stations.  Matching these segments to tracks
measured in the inner tracker results in a $\pt$ resolution
between 1 and 2\% for $\pt$ values up to $100$~GeV.
\par
The first level (L1) of the CMS trigger system~\cite{cmsTrigger}, composed of custom
hardware processors, is designed to select the most interesting events
in less than $1\mus$,
using information from the calorimeters
and muon detectors. The High Level Trigger (HLT) processor farm~\cite{HLT} further
decreases the event rate
to a few hundred Hz before data storage. 
A more detailed description of CMS can be found elsewhere~\cite{JINST}.

\section{Data and Simulated Samples}
\label{sec:samples}

The W and Z analyses are based on data samples collected
during the LHC data operation periods logged from May through November~2011,
corresponding to an integrated luminosity $\Lint=\THELUMI$.

Candidate events are selected from datasets collected with high-$\Et$ lepton trigger requirements.
Events with high-$\Et$ electrons are selected online if they pass a
L1 trigger filter that requires an energy deposit in a coarse-granularity region
of the ECAL with $\Et > $ 5 or 8~GeV, depending on the data taking period.
They subsequently must pass an HLT filter that requires a minimum $\Et$ threshold of the ECAL cluster
which is well below the offline $\Et$ threshold of 25 GeV. The full ECAL granularity
and offline calibration corrections are exploited by the HLT filter~\cite{CMS-PAS-EGM-10-003}.

Events with high-$\pt$ muons are selected online by a single-muon trigger.
The energy threshold at the L1 is 7~GeV. The $\pt$ threshold at the HLT level
depends on the data taking period and was 9~GeV  for the first 7.5~pb$^{-1}$
of collected data and 15~GeV for the remaining 28.4~pb$^{-1}$.

Several large Monte Carlo (MC) simulated samples are used to
evaluate signal and background efficiencies and to validate the
analysis techniques employed.  Samples of EWK processes with Z
and W bosons, both for signal and background events, are generated
using {\sc powheg}~\cite{Alioli:2008gx, Nason:2004rx, Frixione:2007vw}
interfaced with the {\sc pythia}~\cite{Sjostrand:2006za} parton-shower
generator and the Z2 tune (the PYTHIA6 Z2 tune is identical 
to the Z1 tune described in~\cite{Z1} except that Z2 uses the CTEQ6L PDF, 
while Z1 uses the CTEQ5L PDF). 
QCD multijet events with a muon or electron in the final state and $\ttbar$
events are simulated with {\sc pythia}. Generated events are processed
through the full {\sc Geant4}~\cite{Agostinelli:2002hh, Allison:2006ve}
detector simulation, trigger emulation, and event reconstruction chain
of the CMS experiment.

\section{Event Selection}
\label{sec:eventSelection}

The $\Wln$ events are characterized by a prompt, energetic, and
isolated lepton and significant missing transverse energy, $\MET$.  
No requirement on $\MET$ is applied. Rather, the $\MET$ is used as the main 
discriminant variable against backgrounds from QCD events. 

The Z boson decays to leptons (electrons or muons) are selected based on two 
energetic and isolated leptons.
The reconstructed dilepton invariant mass is required to be consistent with
the known Z boson mass. 

\par
The following background processes are considered:
\begin{itemize}
\item  {\sl QCD multijet events.}
Isolation requirements reduce events with leptons produced inside jets.
The remaining background is estimated with a variety of
techniques based on data.  
\item {\sl High-$\Et$ photons.}
For the $\Wen$ channel only, there is a nonnegligible
background contribution coming from  the conversion of a 
photon from the process $\pp\rightarrow\gamma+$jet(s).
\item {\sl Drell--Yan.}
A DY lepton pair 
constitutes a background for the $\Wln$ channels 
when one of the two leptons is not reconstructed or does not enter a fiducial region.
\item {\sl $\Wtn$ and $\Ztt$ production.}
A small background contribution comes from W and Z events with one or both $\tau$ decaying
leptonically.  The minimum lepton $\Pt$ requirement tends to suppress
these backgrounds.
\item {\sl Diboson production.} The production of boson pairs ($\Wo\Wo$, $\Wo\Zo$, $\Zo\Zo$)
is considered a background to the W and Z analysis
because the theoretical predictions for the vector boson production
cross sections used for comparison with data
do not include diboson production.
The background from diboson production
is very small and is estimated using simulations.
\item {\sl Top-quark pairs.}
The background from $\ttbar$ production is quite small and
is estimated from simulations.
\end{itemize}

The backgrounds mentioned in the first two bullets are referred to 
as ``QCD backgrounds'', the Drell--Yan, $\Wtn$, 
and dibosons as "EWK backgrounds", and the last one as "$\ttbar$ background".
For both diboson and $\ttbar$ backgrounds, the NLO cross sections were used.
The complete selection criteria used to reduce the above backgrounds 
are described below.

\subsection{Lepton Isolation}
\label{sec:isolation}

The isolation variables for the tracker and the electromagnetic 
and hadronic ca\-lo\-ri\-me\-ters are defined:
$\ITRK  = \sum_{\mathrm{tracks}} \pt$\,,
$\IECAL = \sum_{\mathrm{ECAL}} \Et$\,, 
$\IHCAL = \sum_{\mathrm{HCAL}} \Et$\,,
where the sums are performed on all objects falling within a cone of aperture
$\Delta R$ = $\sqrt{(\Delta\eta)^2+(\Delta\phi)^2}$ = 0.3 around
the lepton candidate momentum direction.
The energy deposits and the track associated with the lepton candidate 
are excluded from the sums.

\subsection{Electron Channel Selection}
\label{sec:electronId}

Electrons are identified offline as clusters of ECAL energy deposits
matched to tracks reconstructed in the silicon tracker.
The ECAL clustering algorithm is designed to reconstruct clusters containing a
large fraction of the energy of the original electron, including energy
radiated along its trajectory. The ECAL clusters must fall in the ECAL fiducial volume
of $|\eta| < 1.44$ for EB clusters or $1.57 < |\eta| < 2.5$ for EE clusters.
The transition region $1.44 < |\eta| < 1.57$ is excluded as it leads to lower-quality
reconstructed clusters, due mainly to services and cables exiting between the barrel and
endcap calorimeters. Electron tracks are reconstructed using an algorithm~\cite{GSF} 
(Gaussian-sum filter, or GSF tracking) that accounts for possible energy loss due to 
bremsstrahlung in the tracker layers.

\begin{figure}[htbp]
  \begin{center}
   \includegraphics[width=0.68\textwidth]{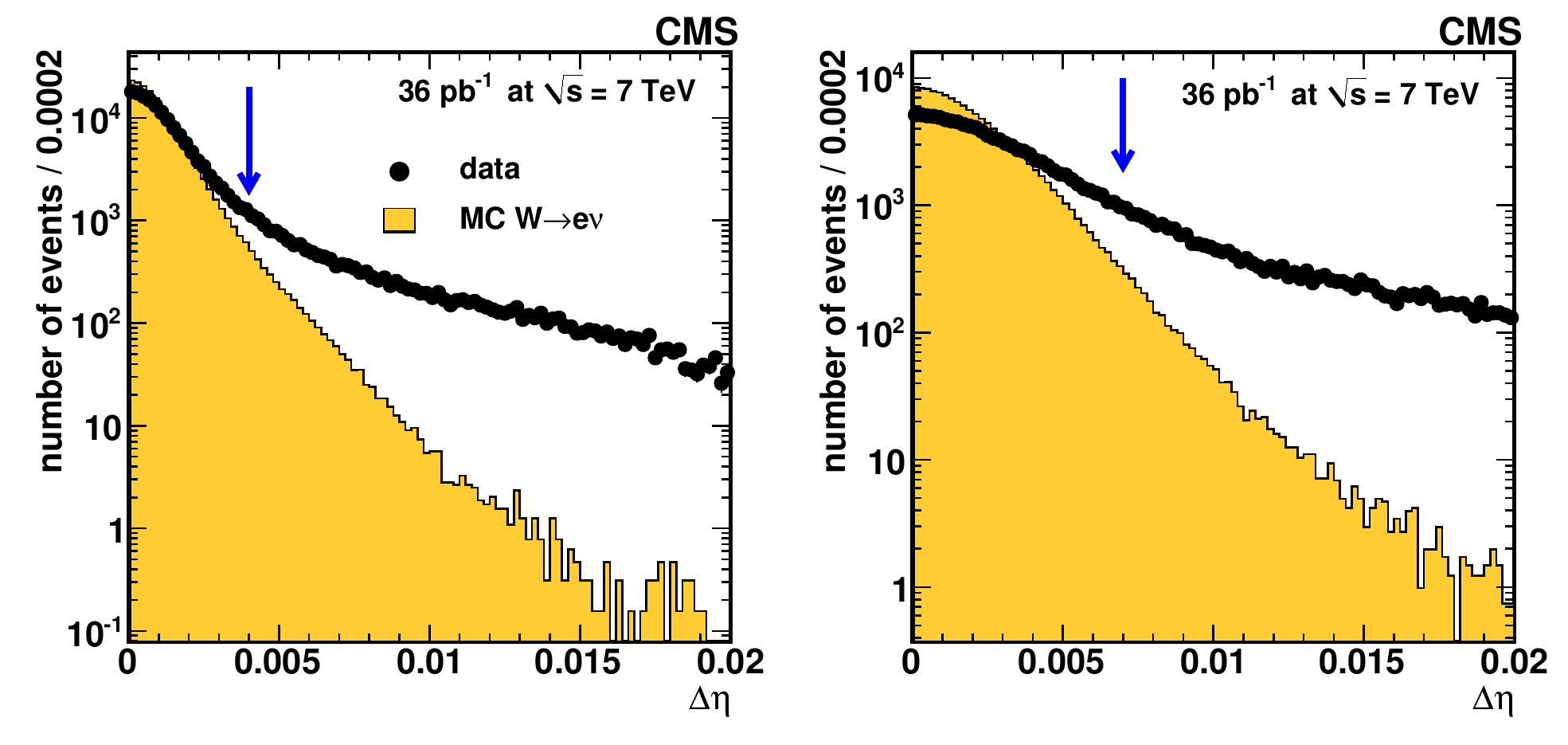}
   \includegraphics[width=0.68\textwidth]{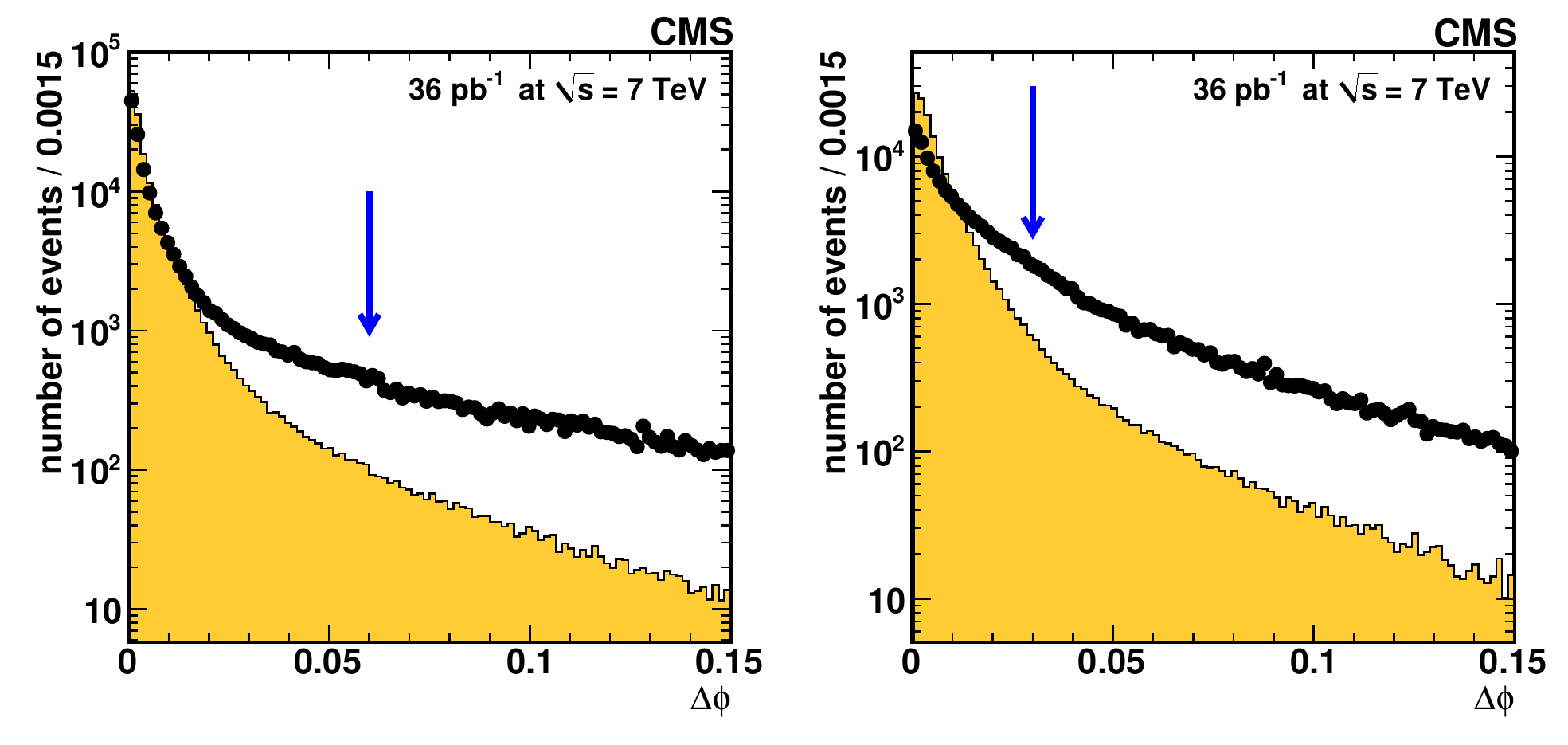}
   \includegraphics[width=0.68\textwidth]{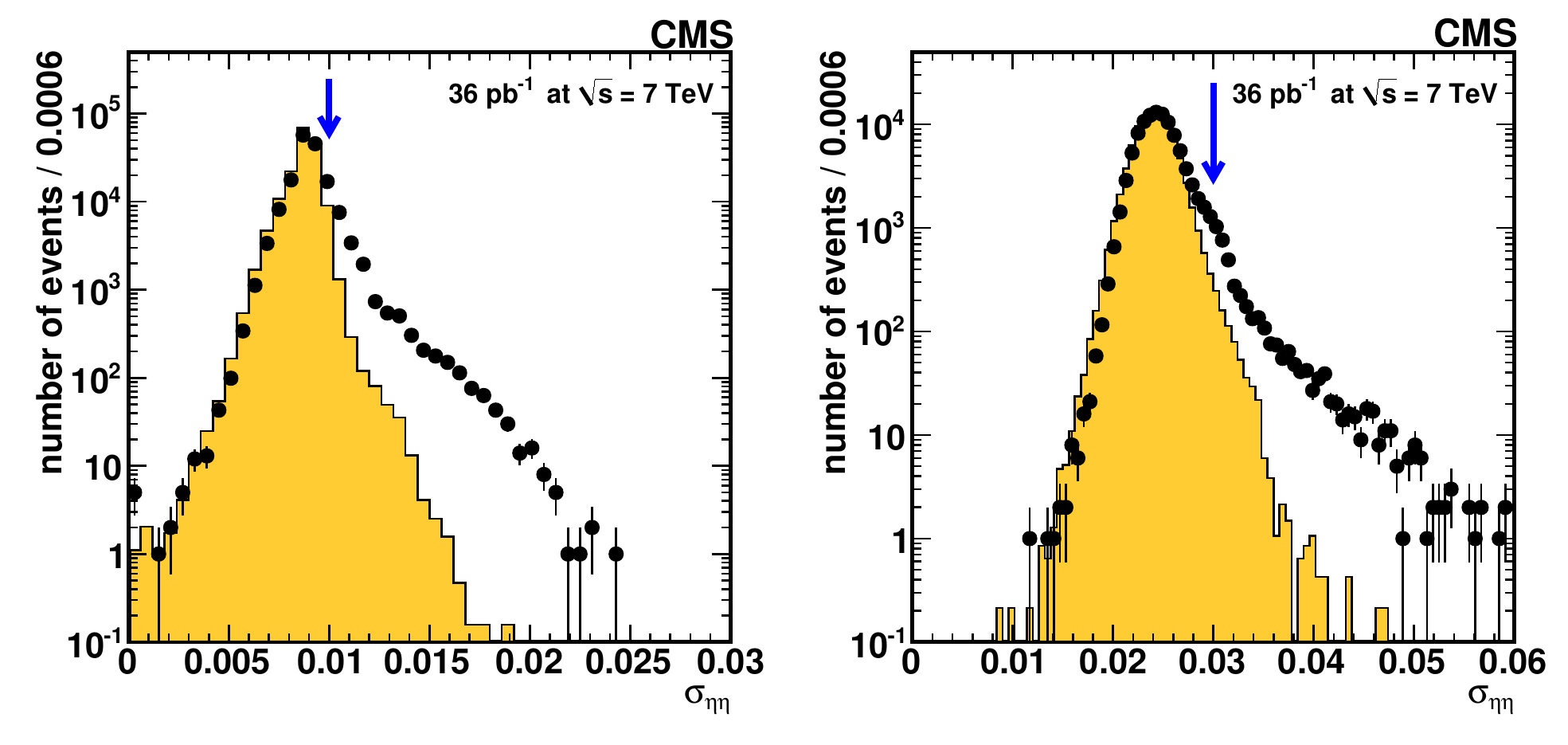}
   \includegraphics[width=0.68\textwidth]{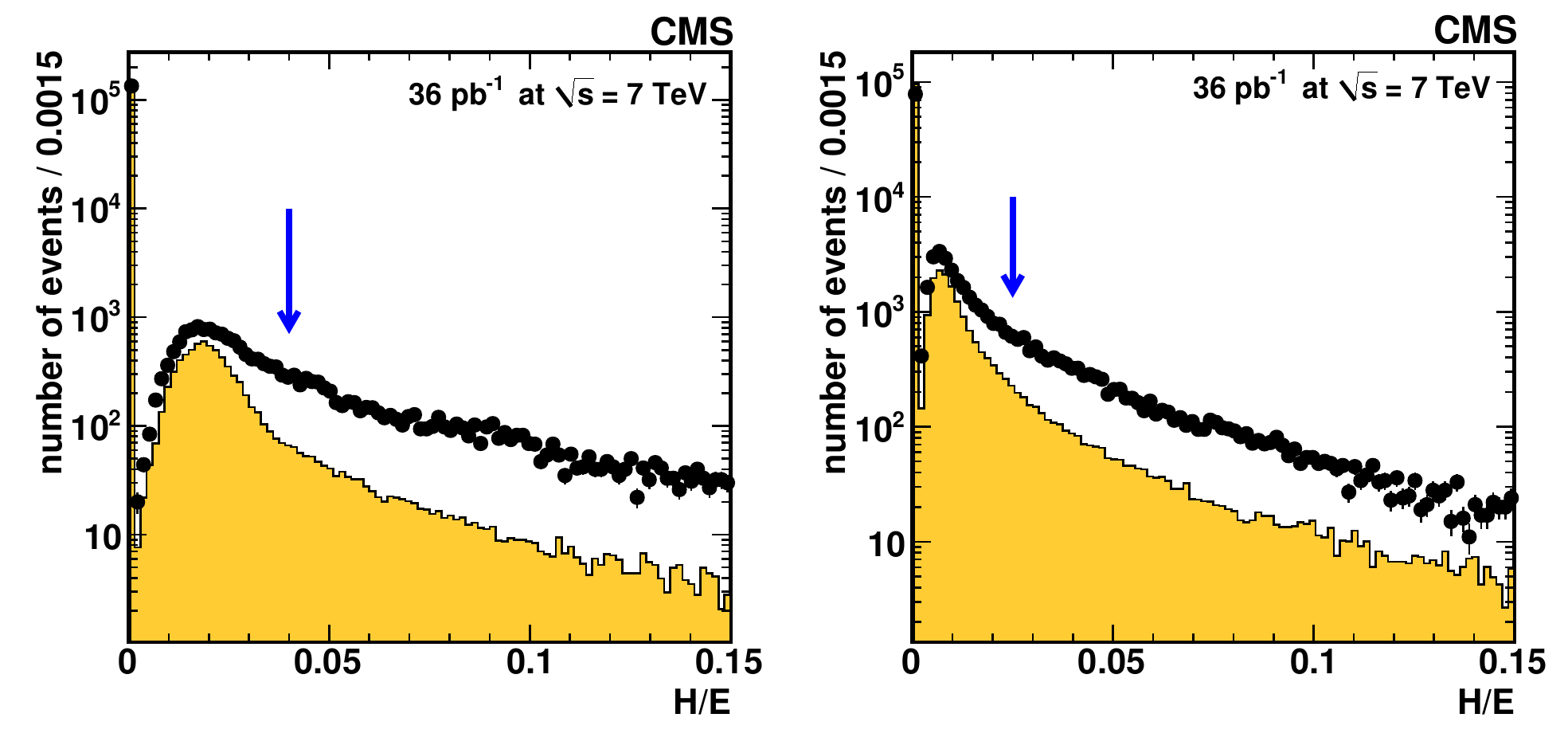}
   \caption{ \label{fig:WenuSelection1}
Distributions of the electron identification variables $\Delta\eta$, $\Delta\phi$, $\sigma_{\eta\eta}$, 
and $H/E$ for data (points with the error bars), for EB (left) and EE (right).
For illustration the simulated $\Wen$ signal (histograms), normalized to the number of events
observed in data, is superimposed.
These distributions are obtained after applying all 
the tight requirements on the selection variables, except that on the presented
variable. The tight requirement on that variable is indicated with an arrow. }
  \end{center}
\end{figure}

\begin{figure}[htbp]
  \begin{center}
   \includegraphics[width=0.68\textwidth]{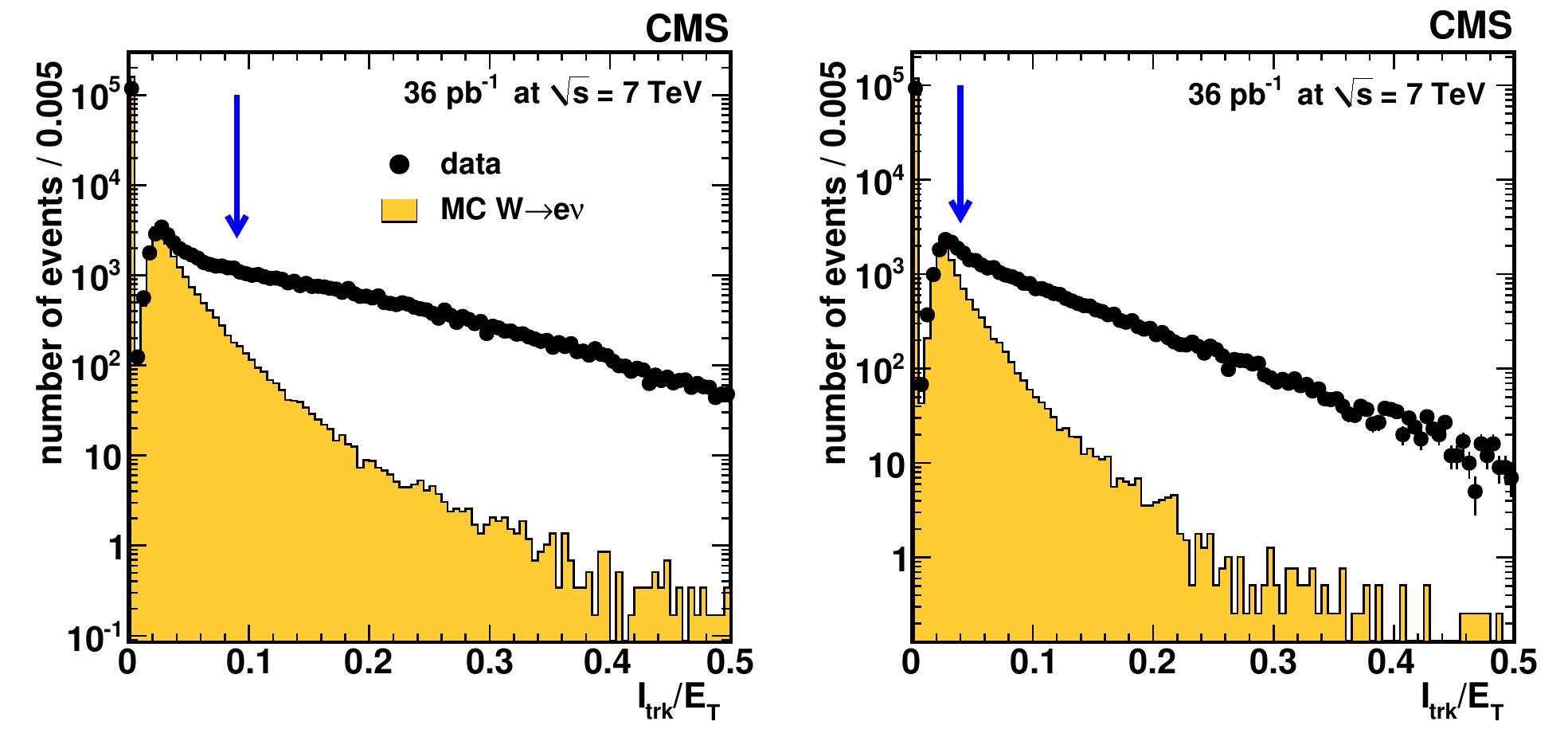}
   \includegraphics[width=0.68\textwidth]{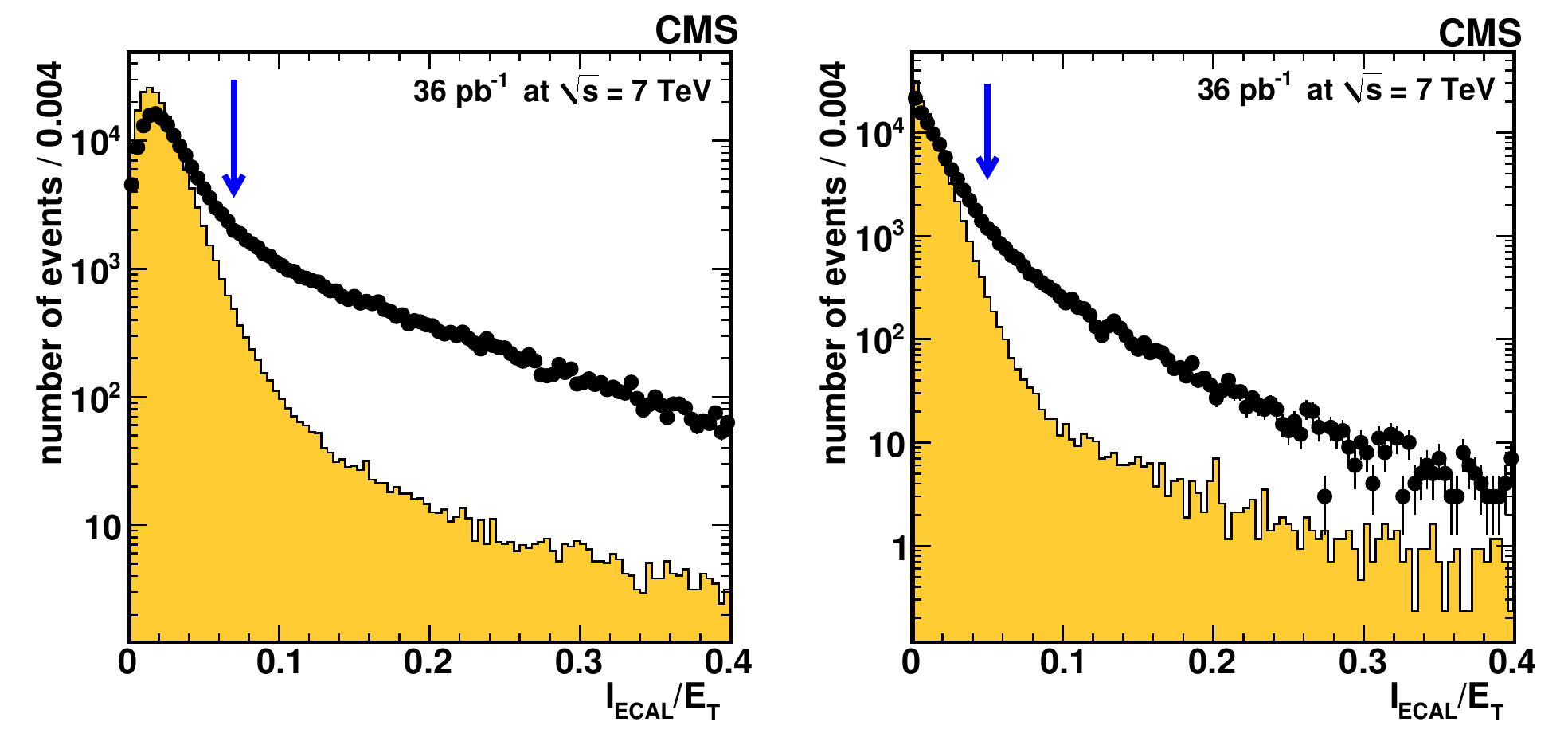}
   \includegraphics[width=0.68\textwidth]{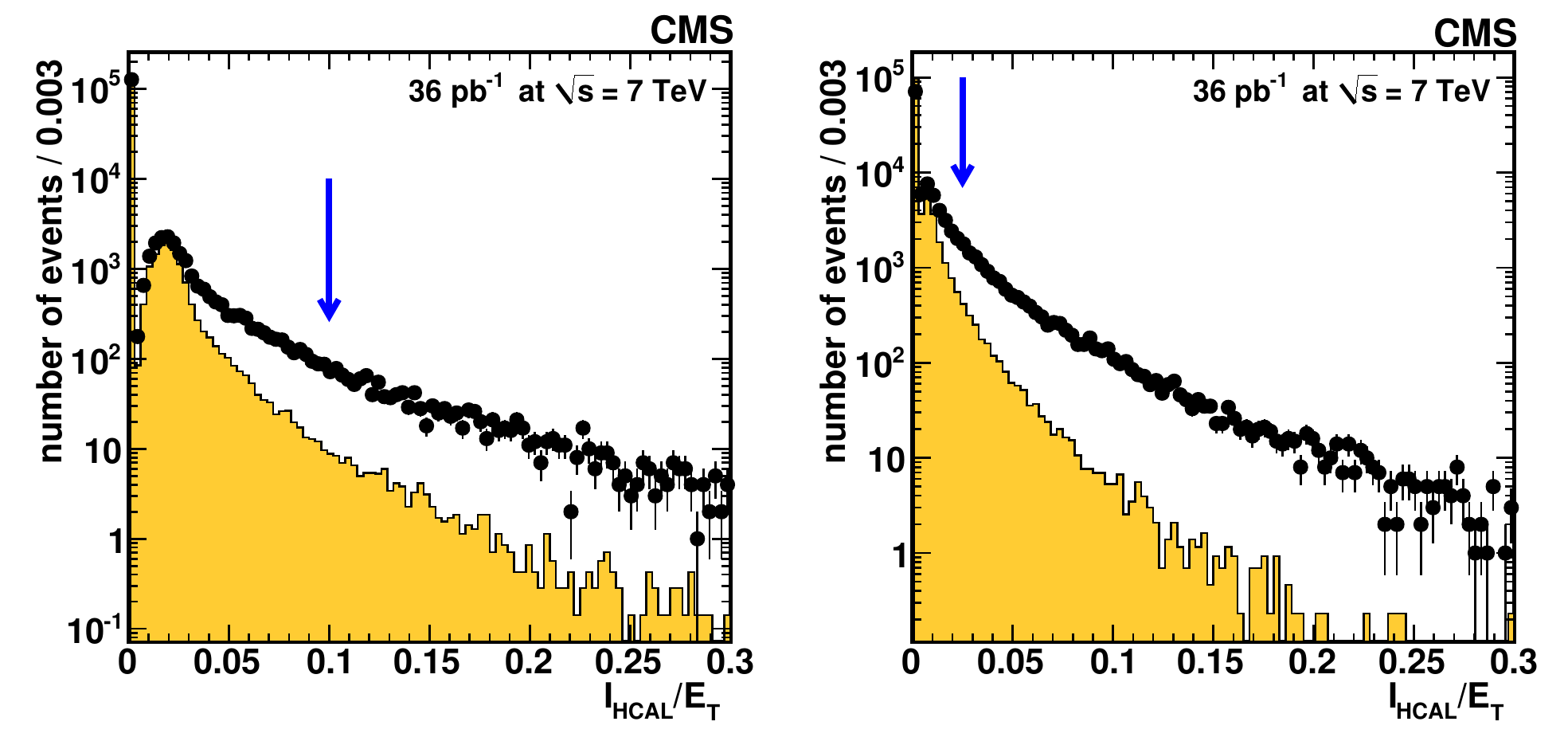}
   \caption{ \label{fig:WenuSelection2}
Distributions of the electron isolation variables $\ITRK/\Et$, $\IECAL/\Et$, and $\IHCAL/\Et$
for data (points with the error bars), for EB (left) and EE (right).
For illustration the simulated $\Wen$ signal (histograms), normalized to the number of events
observed in data, is superimposed.
These distributions are obtained after applying all
the tight requirements on the selection variables, except that on the presented
variable. The tight requirement on that variable is indicated with an arrow. }
  \end{center}
\end{figure}

The radiated photons may convert close to the
original electron trajectory, leading to charge misidentification.
Three different methods are used to determine the electron charge. First, the electron 
charge is determined by the signed curvature of the associated GSF track. Second, the charge 
is determined from the associated trajectory reconstructed in the silicon tracker using a 
Kalman Filter algorithm~\cite{KF}. Third, the electron charge is determined based on the azimuthal 
angle between the vector joining the nominal interaction point and the ECAL cluster 
position and the vector joining the nominal interaction point and innermost hit of the GSF track. 
The electron charge is determined from the two out of three charge estimates that are in agreement.
The electron charge misidentification rate is measured in data using the $\Zee$ data
sample to be within 0.1$\%$--1.3$\%$ in EB and 1.4$\%$--2.1$\%$ in EE, increasing with
electron pseudorapidity.

Events are selected if they contain one or two electrons having $\Et>25~\gev$ 
for the $\Wen$ or the $\Zee$ analysis, respectively. 
For the  $\Zee$ selection there is no requirement on the charges of the electrons.
The energy of an electron candidate with $\Et>25~\gev$ is 
determined by the ECAL cluster energy, while its momentum direction is determined by 
that of the associated track. 

Particles misidentified as electrons are suppressed by requiring that the $\eta$ and $\phi$ coordinates
of the track trajectory extrapolated to the ECAL match those
of the ECAL cluster permitting only small differences ($\Delta\eta$, $\Delta\phi$) 
between the coordinates, by requiring a narrow ECAL cluster width in $\eta$ ($\sigma_{\eta\eta}$), 
and by limiting the ratio of the hadronic energy $H$ to the electromagnetic 
energy $E$ measured in a cone of $\Delta R = 0.15$ around the ECAL cluster direction.
More details on the electron identification variables can be found in Refs.~\cite{EGMid,PhotonQCD}. 
Electron isolation is based on requirements on the three isolation 
variables $\IHCAL/\Et$, $\IECAL/\Et$, and $\ITRK/\Et$.

\par
Electrons from photon conversions are suppressed by requiring the 
reconstructed electron track to have at least one hit in the innermost pixel layer.
Furthermore, electrons are
rejected when a partner track is found that is consistent with a
photon conversion, based on the opening angle and the separation in
the transverse plane at the point where the electron and partner
tracks are parallel.

The electron selection criteria were obtained
by optimizing signal and background levels according to
simulation-based studies. The optimization was done for EB
and EE separately.  

Two sets of electron selection criteria are considered: 
a tight one and a loose one.
Their efficiencies, from simulation studies based on $\Wen$ events, are
approximately 80$\%$ and 95$\%$, respectively. These efficiencies correspond 
to reconstructed electrons within the geometrical and kinematic 
acceptance, which is defined in Section~\ref{sec:acceptance}.  
The tight selection criteria give a purer sample of prompt 
electrons and are used for both the $\Wen$ and $\Zee$ analyses.
The virtue of this choice is to have consistent electron definitions 
for both analyses, simplifying the treatment of systematic 
uncertainties in the $\mathrm{W}/\mathrm{Z}$ ratio measurement. 
In addition, the tight working point, applied to both electrons
in the $\Zee$ analysis, reduces the QCD backgrounds to a negligible level.
Distributions of the selection variables are shown in Figs.~\ref{fig:WenuSelection1}
and~\ref{fig:WenuSelection2}.
The plots show the distribution of data together with the simulated signal 
normalized to the same number of events as the data, after applying all 
the tight requirements on the selection variables except the requirement on the displayed
variable.

For the W analysis, an event is also rejected if there is a second electron 
that passes the loose selection with $\Et > 20~\gev$. This requirement reduces
the contamination from DY events.  
The number of $\Wen$ candidate events selected in the data sample is 
$\WEISAMPLE$, with $\WEPSAMPLE$ positrons and $\WEMSAMPLE$ electrons.

For the Z analysis, two electrons are required within the ECAL acceptance, 
both with $\Et > 25~\gev$ and both satisfying the tight electron selection. 
Events in the dielectron mass region of $60 < m_{\mathrm{ee}} < 120$~GeV are counted.
These requirements select $\ZEESAMPLE$ events.

\subsection{Muon Channel Selection}
\label{sec:muonid}

Muons candidates are first reconstructed separately in the central tracker (referred to simply as ``tracks'' or ``tracker tracks'')
and in the muon detector (``stand-alone muons''). Stand-alone muons are then matched and combined with tracker
tracks to form ``global muons''. Another independent algorithm proceeds from the central tracker outwards, matching
muon chambers hits and producing ``tracker muons''.

The following quality selection are applied to muon candidates.
Global and stand-alone muon candidates must have at least one good hit in the muon chambers.
Tracker muons must match to hits in at least two muon stations.
Tracks, global muons, and tracker muons must have more than 10 hits in the inner tracker, 
of which at least one must be in the pixel detector, and the impact parameter in the 
transverse plane, $d_{xy}$, calculated with respect to the beam axis, 
must be smaller than 2~mm.
More details and studies on muon identification can be found in Refs.~\cite{MUONPAS,MuonPerf}.
\begin{figure}
  \begin{center}
   \includegraphics[width=0.39\textwidth]{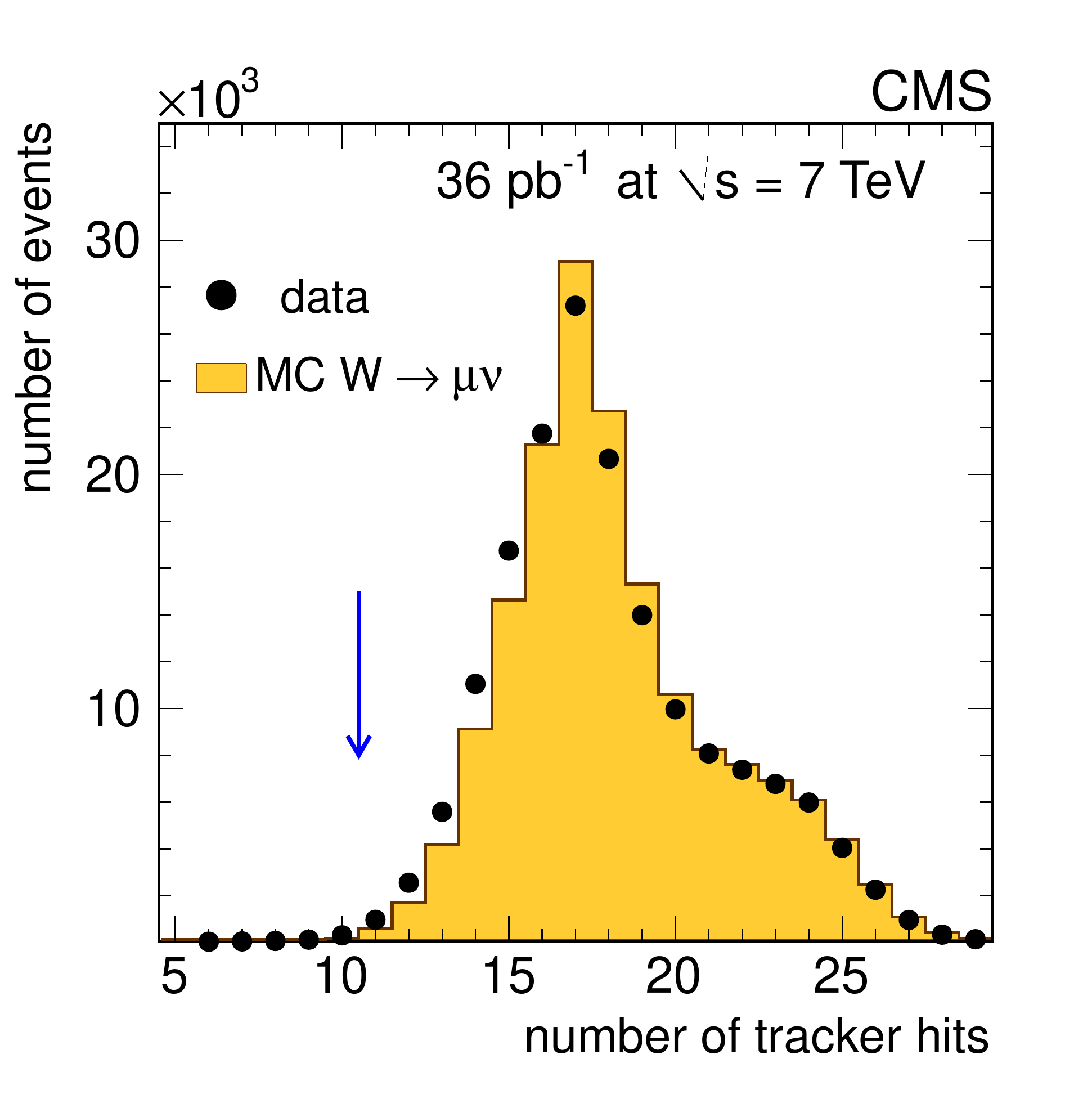}
   \includegraphics[width=0.39\textwidth]{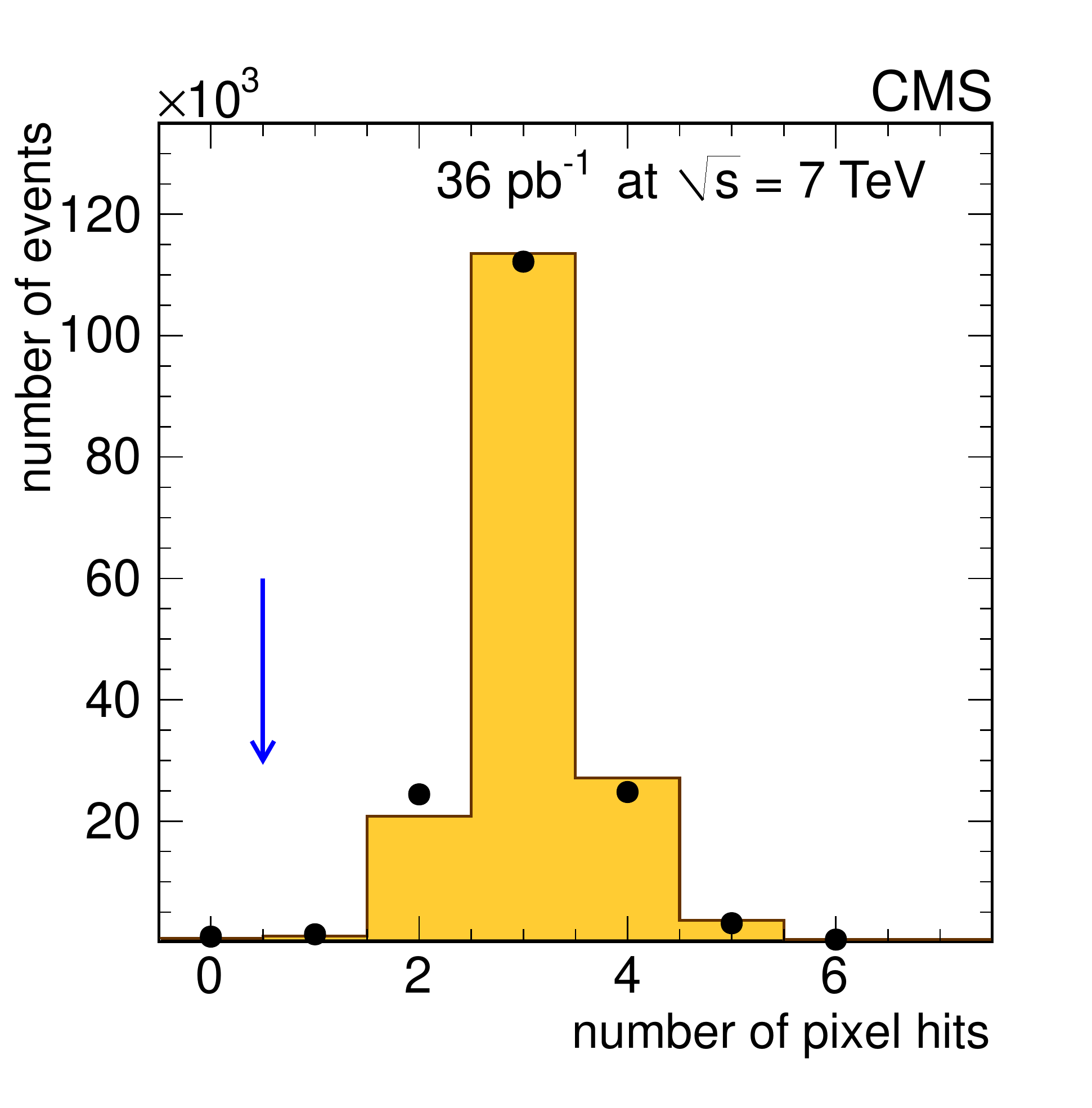}
   \includegraphics[width=0.39\textwidth]{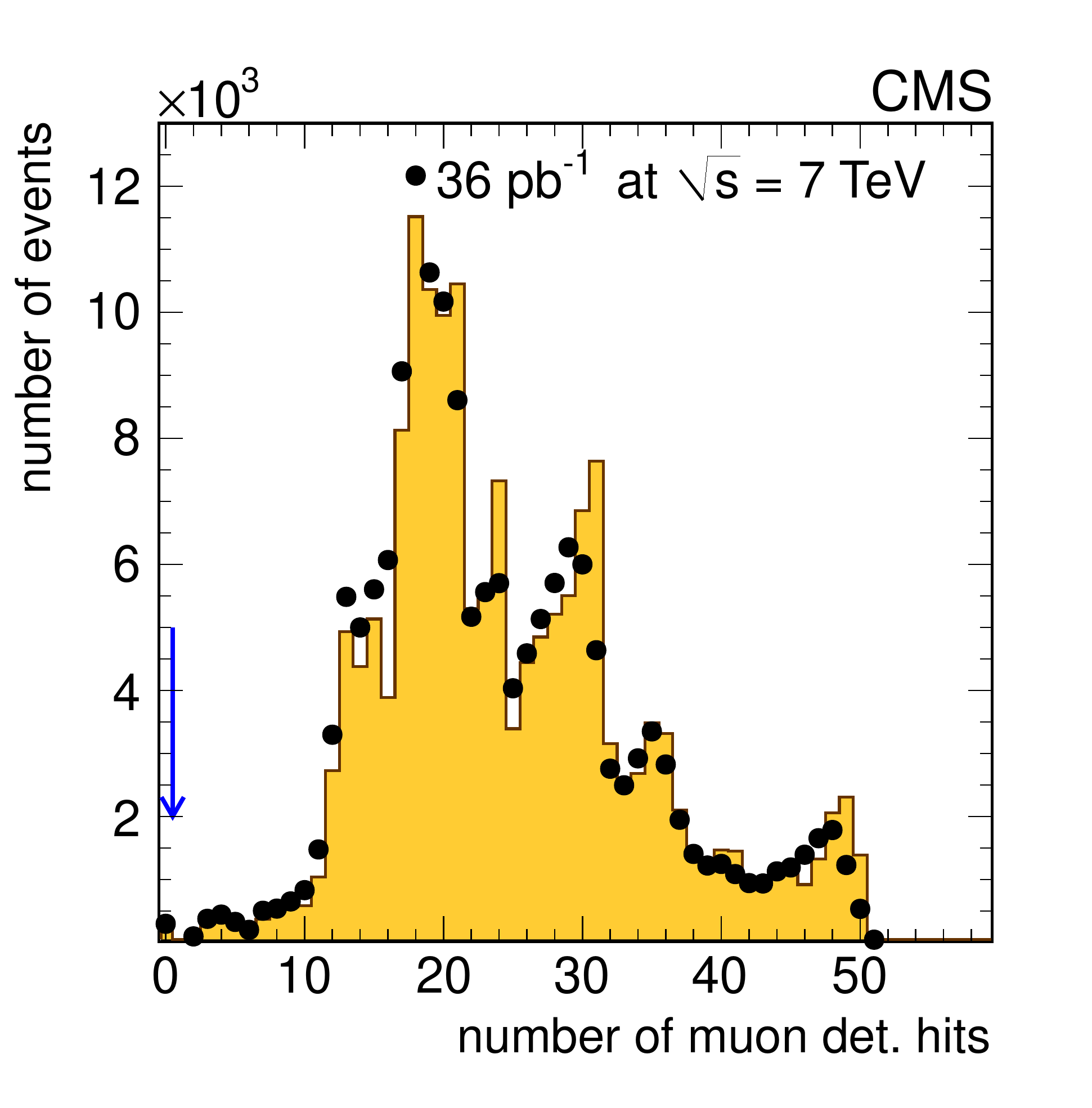}
   \includegraphics[width=0.39\textwidth]{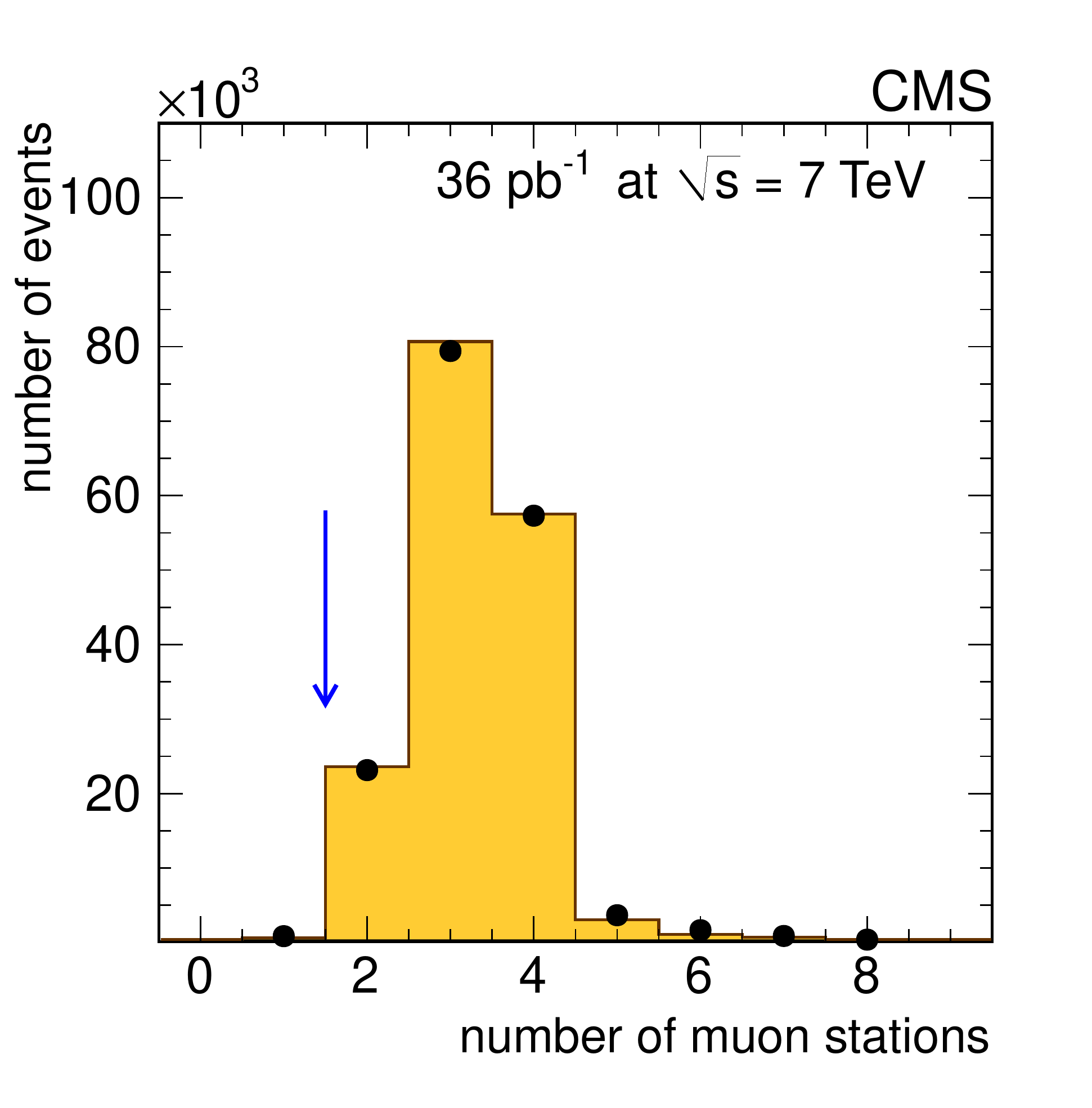}
   \includegraphics[width=0.39\textwidth]{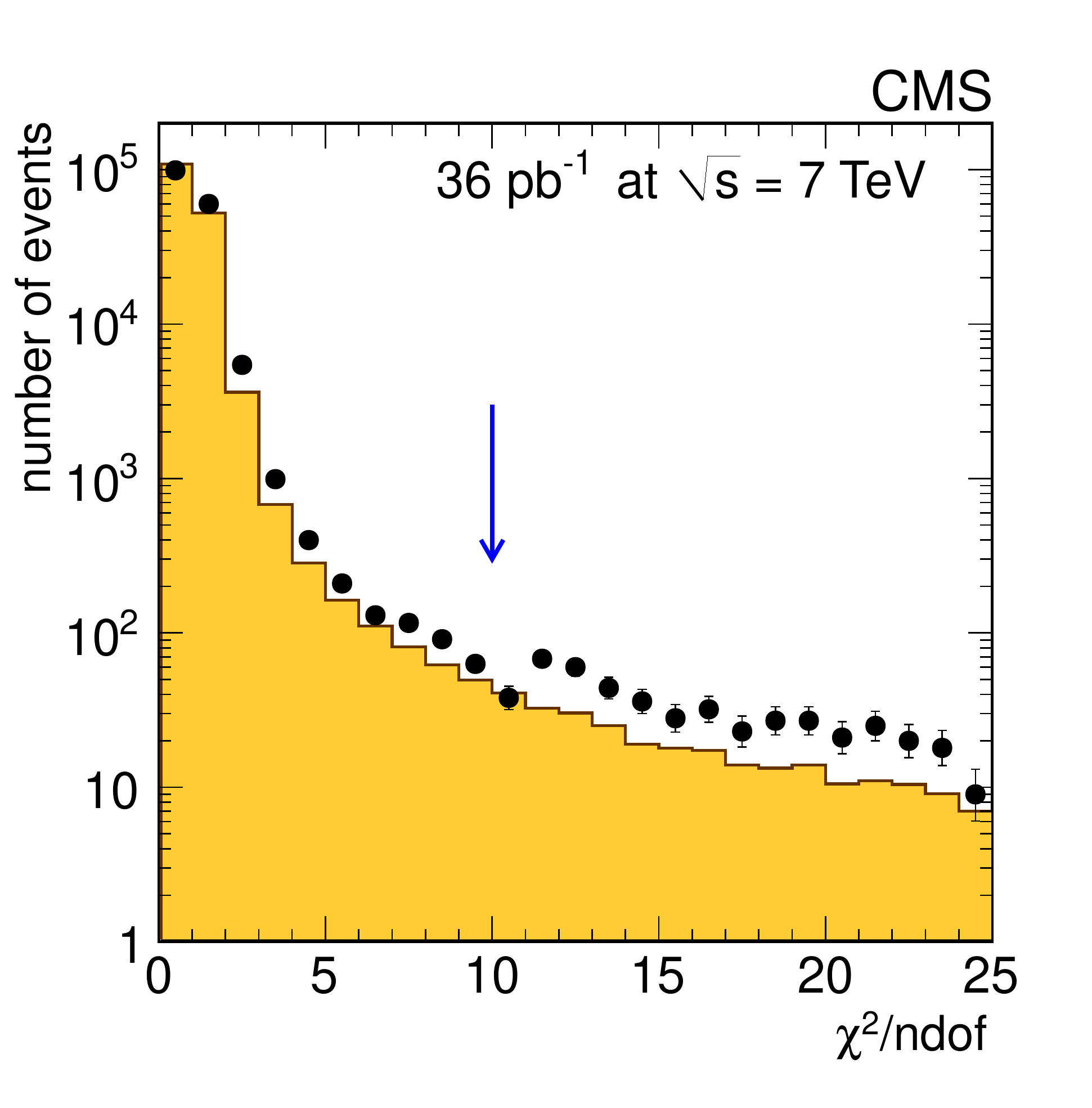}
   \includegraphics[width=0.39\textwidth]{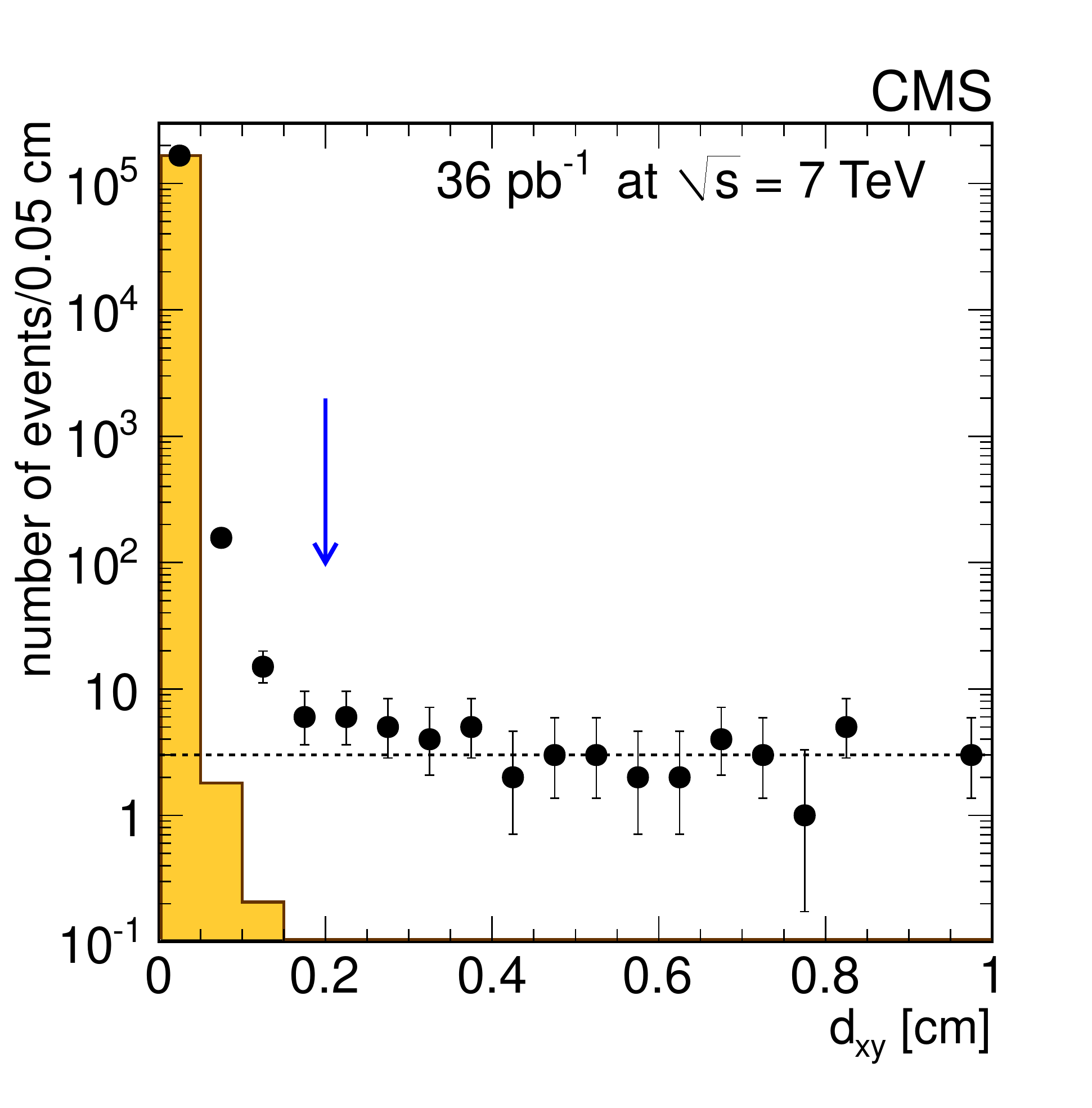}
   \caption{ \label{fig:muonIDvars}
Distribution of number of hits in the inner tracker and in the pixel detector,
number of hits in muon chambers, number of muon segments stations,
$\chi^2$ per degree of freedom, and transverse impact parameter $d_{xy}$ for data
(points with the error bars).
For illustration the simulated $\Wmn$ signal (histogram), normalized to the number of events 
observed in data, is superimposed.
These distributions are for events
satisfying all selection requirements, except that on
the presented variable.
The applied requirement on that variable is indicated as a blue arrow.
In the $d_{xy}$ distribution, the horizontal line shows the average of the 
bins with  $d_{xy}>0.2~\mathrm{cm}$
used to estimate the cosmic-ray muon contamination in the signal region.
The excess of events in data in the region with $d_{xy}<0.2~\mathrm{cm}$ with respect to $\Wmn$ 
signal simulation is due to muons from long-lived
particle decays in the QCD background.
}
  \end{center}
\end{figure}

Muon candidates selected in the $\Wmn$ analysis must be identified both as global and tracker muons.
Moreover, as additional quality selection, the global muon fit must have a $\chi^2$ per degree of freedom less than 10
in order to reject misidentified muons and misreconstructed particles.
The $\Wmn$ candidate events must have a muon candidate in the fiducial volume $|\eta|<2.1$ with 
$\Pt>25~\GeV$. 
The muon must be isolated, satisfying 
$\IRelComb = \left( \ITRK +\IECAL+\IHCAL\right)/\pt < 0.1$.
Events containing a second muon with $\Pt > 10~\GeV$ in the full muon acceptance region
($|\eta|<2.4$) are rejected to minimize the contamination from DY events.
The distributions of the variables used for muon quality selection are shown 
in Fig.~\ref{fig:muonIDvars} after applying all selection requirements, except that on
the presented  variable.

Background due to a cosmic-ray muon crossing the detector in coincidence with a pp collision is very
much reduced by the impact parameter requirement.
The remaining cosmic-ray background is evaluated by extrapolating to the signal region the rate of events with
large impact parameter.
Figure~\ref{fig:muonIDvars} (bottom, right) shows the distribution of the
impact parameter $d_{xy}$ for the $\Wmn$ candidates satisfying all selection requirements, except
that on $d_{xy}$.
Candidates with large $d_{xy}$ are mainly due to cosmic-ray muons and their rate is independent of $d_{xy}$.
A background fraction on the order of $10^{-4}$ in the $d_{xy}<2$~mm region is estimated.

\begin{figure}[htbp] {\centering
    \includegraphics[width=0.5\textwidth]{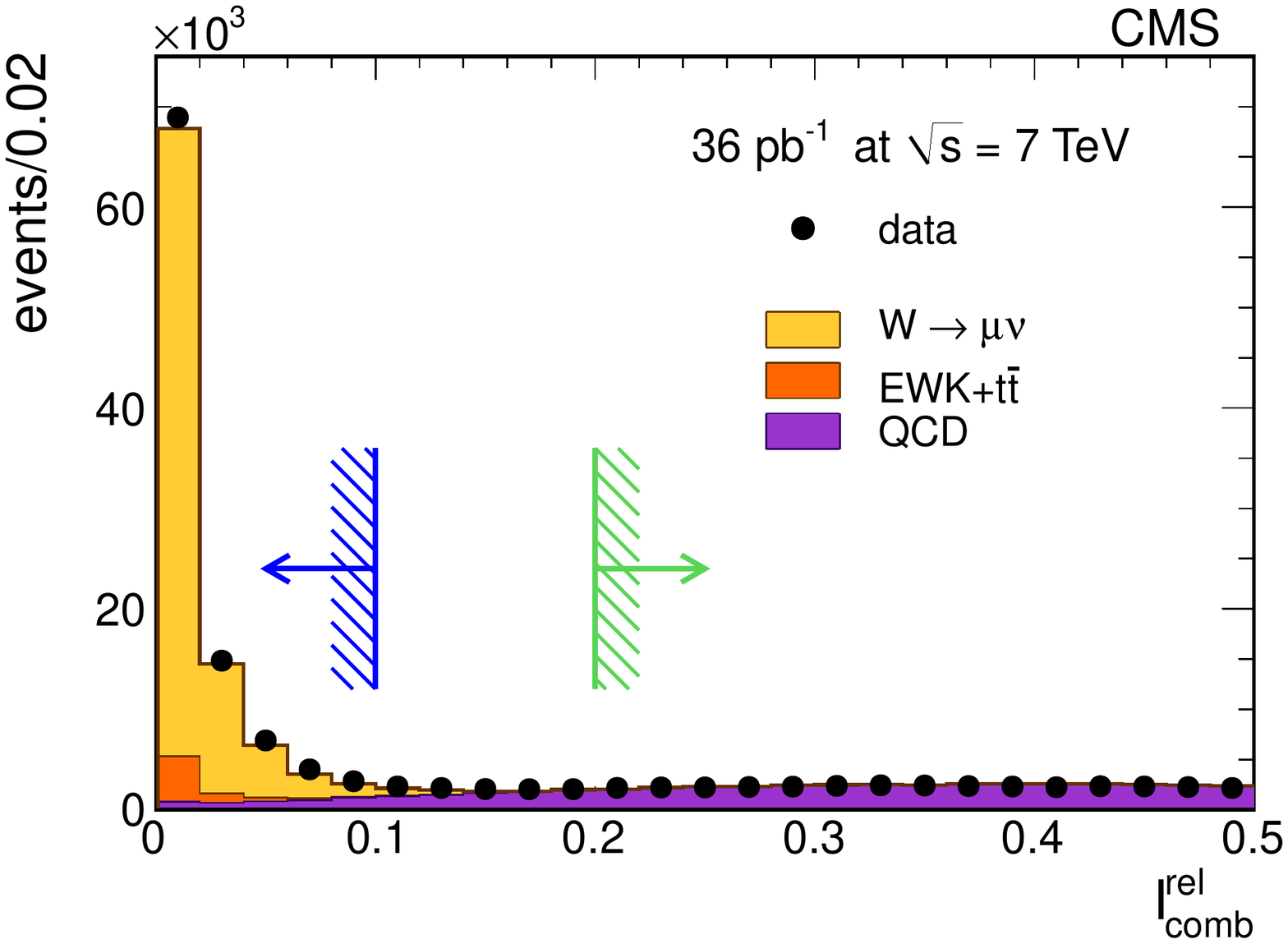}
    \caption{Distribution of $\IRelComb$ for candidates with a good quality muon 
      of $\Pt>25~\GeV$ in the fiducial region $|\eta|<2.1$.
      Points represent the data and the histograms the contribution from the 
      different SM processes.
      The signal selection requirement (dark blue arrow, $\IRelComb<0.1$) and the selection of the QCD-enriched
      control sample (light green arrow, $\IRelComb>0.2$) are shown.}
    \label{figure:Wmunu_iso}}
\end{figure}

The isolation distribution in data, together with the MC expectations, are shown in
Fig.~\ref{figure:Wmunu_iso}. 
Events with $\IRelComb > 0.2$ are mainly from QCD multijet background, and are used as
a control sample (Section~\ref{sec:WQCDbkg}).

After the selection process described, 166\,457 events are selected,
97\,533 of them with a positively charged muon candidate and 68\,924 with a negatively
charged muon candidate.

$\Zmm$ candidate events are selected by pairing a global muon matched to an HLT trigger muon with a second
oppositely charged muon candidate that can be either a global muon, a stand-alone muon, or a track. 
No $\chi^2$ selection or requirement that the muon be reconstructed 
through the tracker-muon algorithm is applied. 
The two muon candidates must both have $\Pt > 20~\mathrm{GeV}$ and $|\eta| < 2.1$,
and their invariant mass must be in the range $60<m_{\mu\mu}<120$~$\mathrm{GeV}$.
Both muon candidates must be isolated according to the tracker isolation requirement $\ITRK<3$~GeV.
The different choice of isolation requirements in $\Wmn$ and $\Zmm$ is
motivated in Section~\ref{sec:Zmumu}.
After the selection process, the number of selected events with two
global muons is 13\,728.

\section{Acceptance}
\label{sec:acceptance}

The acceptance $A_\Wo(\mathrm{e})$ for $\Wen$ is defined
as the fraction of simulated $\Wo$ events having an ECAL cluster within
the ECAL fiducial volume with $\Et>25\GeV$.
The ECAL cluster must match the generated electron after final-state radiation
(FSR) within a cone of $\Delta R=0.2$. No matching in energy is required.

There is an inefficiency in the ECAL cluster reconstruction for
electrons direction within the ECAL fiducial volume
due to a small fraction ($0.5\%$) of noisy or malfunctioning towers
removed from the reconstruction. These are
taken into account in the MC simulation, and no uncertainty is assigned to
this purely geometrical inefficiency. The ECAL cluster selection efficiency is
also affected by a bias in the electron energy scale due to
the $25~\GeV$ energy threshold. The related systematic uncertainty is assigned
to the final $\Wo$ and $\Zo$ selection efficiencies.

The acceptance for the $\Zee$ selection, $A_\Zo(\mathrm{e})$,
is defined as the number of simulated events with
two ECAL clusters  with $\Et>25~\GeV$ within the ECAL fiducial volume and
with invariant mass in the range $60<m_{\mathrm{ee}}<120\,\mathrm{GeV}$, divided by the total number
of signal events in the same mass range, with the invariant mass evaluated using
the momenta at generator level before FSR.
The ECAL clusters must match the two simulated electrons after FSR
within cones of $\Delta R<0.2$. No requirement on energy matching is applied.

For the $\Wmn$ analysis, the acceptance $A_\Wo(\mu)$ is defined as the fraction
of simulated $\Wo$ signal events with muons having transverse momentum $\Pt^{\textrm{gen}}$
and pseudorapidity $\eta^{\textrm{gen}}$, evaluated at the generator level
after FSR, within the kinematic selection: $\Pt^{\textrm{gen}}>25~\GeV$ and $|\eta^{\textrm{gen}}|<2.1$.

The acceptance $A_\Zo(\mu)$ for the $\Zmm$ analysis is defined as the number of
simulated $\Zo$ signal events with both muons passing the kinematic selection with momenta evaluated
after FSR, $\Pt^{\textrm{gen}}>20~\GeV$ and $|\eta^{\textrm{gen}}|<2.1$, and with
invariant mass in the range $60<m_{\mu\mu}<120\,\mathrm{GeV}$, divided by the total number
of signal events in the same mass range, with the invariant mass evaluated using
the momenta at generator level before FSR.

Table~\ref{tab:WZlaccgen} presents the acceptances
for $\Wp$, $\Wm$, and inclusive $\Wo$ and $\Zo$ events,
computed from samples simulated with {\sc powheg} using the CT10 PDF,
for the muon and the electron channels. The acceptances are affected by several
theoretical uncertainties,
which are discussed in detail in Section~\ref{sec:theory}.

\begin{table}[htbp] %
\begin{center}
\caption[.] {\label{tab:WZlaccgen}
Acceptances from {\sc powheg} (with CT10 PDF) for $\Wln$ and $\Zll$ final states,
with the MC statistics uncertainties.\\}
\begin{tabular}{|l|c|c|}
\hline
 {\multirow{2}{*}{Process}} &  \multicolumn{2}{c|}{$A_{\mathrm{W,Z}}$}  \\ \cline{2-3}
  & $\ell=\mathrm{e}$ &$\ell=\mu$ \\
\hline\hline
     $\Wpln$      & \WEPACC & \WMPAGEN \\
     $\Wmln$      & \WEMACC & \WMMAGEN\\
     $\Wln$  & \WEIACC & \WMIAGEN \\
\hline
$\Zll$ & \ZEEACC & $0.3978 \pm 0.0005$ \\
\hline
\end{tabular}
\end{center}
\end{table}

\section{Efficiencies}
\label{sec:efficiencies}

A key component of this analysis is the estimation of lepton efficiencies.
The efficiency is determined for different selection steps:
\begin{itemize}
\item offline reconstruction of the lepton;
\item lepton selection, with identification and isolation criteria;
\item trigger (L1+HLT).
\end{itemize}
The order of the above selections steps is important. Lepton efficiency for each
selection is determined with respect to the prior step.
\par
A tag-and-probe (\TNP) technique is used, as described below, on pure samples of $\Zll$ events.
The statistical uncertainty on the efficiencies is
ultimately propagated as a systematic uncertainty on the cross-section measurements.
This procedure has the advantage
of extracting the efficiencies from a sample of leptons kinematically very similar
to those used in the $\Wo$ analysis and exploits the relatively pure selection
of $\Zll$ events obtained after a dilepton invariant mass requirement around the Z mass.
\par
The \TNP method is as follows: one lepton candidate,
called the ``tag'', satisfies trigger criteria, tight identification
and isolation requirements. The other lepton candidate, called the ``probe'',
is required to pass specific criteria that depend on the efficiency under study.
\par
For each kind of efficiency, the \TNP method is applied to real data and to
simulated samples, and the ratio of efficiencies in data ($\effdata$) and simulation ($\effmc$) is computed:
\begin{equation}
\label{eq:rho}
 \rhoeff = \frac{\effdata}{\effmc}\,,
\end{equation}
together with the associated statistical and systematic uncertainties.

\subsection{Electrons}
\label{sec:ELEefficiencies}

As mentioned in the previous section, 
the tight electron selection is considered for both the W and Z analyses, so:
\begin{equation}
  \EPS{all} = \EPS{rec}\, \EPS{tight}\, \EPS{trg}.
\label{eq:e-eff}
\end{equation}
The reconstruction efficiency $\EPS{rec}$ is relative to ECAL clusters
within the ECAL acceptance, the selection efficiency $\EPS{tight}$ is relative to GSF electrons
within the acceptance, and the trigger efficiency $\EPS{trg}$ is relative to electrons
satisfying the tight selection criteria.

\par
All the efficiencies are determined by the \TNP technique.
Selections with different criteria have been 
tried on the tag electron. It was found that the estimated efficiencies are 
insensitive to the tag selection definition. 
The invariant mass of the \TNP pair
is required to be within the window $60<m_{\mathrm{ee}}<120~\GeV$,
ensuring high purity of the probe sample. No opposite-charge requirement
is enforced.

The number of probes passing and failing the selection is
determined from fits to the invariant mass distribution,
with signal and background components.
Estimated backgrounds, mostly from QCD multijet processes, are in most cases
at the percent level of the overall sample, but can be larger in subsamples
where the probe fails a selection, hence the importance of background
modeling. The signal shape is a Breit--Wigner with nominal Z mass and width convolved with an
asymmetric resolution function (Crystal Ball~\cite{CrystalBall}) with floating parameters.  The
background is modeled by an exponential. Systematic uncertainties that depend on the efficiency
under study are determined by considering alternative signal and background shape models.
Details can be found in Section~\ref{sec:systematics}.

The \TNP event selection efficiencies in the simulation 
are determined from large samples of signal events with no
background added.

The \TNP efficiencies are measured for the EB and EE electrons separately.
Tag-and-probe efficiencies are also determined separately by charge, 
to be used in the measurements of the W$^+$ and W$^-$ cross sections and their ratio.
Inclusive efficiencies and correction factors are summarized in 
Table~\ref{tab:e-eff-summary}. The \TNP measurements of the efficiencies 
on the right-hand side of Eq.~(\ref{eq:e-eff})
are denoted as $\EPSTNPREC$, $\EPS{\tnp-tight}$, and $\EPSTNPTRG$.

\begin{table}[htbp] %
\begin{center}
\caption[.]{\label{tab:e-eff-summary}
Tag-and-probe efficiencies in data and simulation, and the correction factors
used in the electron channels for the barrel (EB) and endcaps (EE). The combined statistical and systematic 
uncertainties are quoted. }
\begin{tabular}{|l|c|c|c|}
\hline
Efficiency & Data & Simulation & Data/simulation ($\rhoeff$) \\
\hline
\hline
\multicolumn{4}{|c|}{EB} \\
\hline
 \EPS{\tnp-rec}      & \WPWIEBEFFRECO  & \WPWIEBMCRECO & \WPWIEBRRECO \\
 \EPS{\tnp-tight}    & \WPWIEBEFFID    & \WPWIEBMCID   & \WPWIEBRID   \\
 \EPS{\tnp-trg}      & \WPWIEBEFFHLT   & \WPWIEBMCHLT  & \WPWIEBRHLT  \\
\hline
 \EPS{\tnp-all}  & \WPWIEBEFF  & \WPWIEBMC & \WPWIEBR \\
\hline
\hline
\multicolumn{4}{|c|}{EE} \\
\hline
 \EPS{\tnp-rec}       & \WPWIEEEFFRECO  & \WPWIEEMCRECO & \WPWIEERRECO \\
 \EPS{\tnp-tight}    & \WPWIEEEFFID    & \WPWIEEMCID   & \WPWIEERID   \\
 \EPS{\tnp-trg} & \WPWIEEEFFHLT   & \WPWIEEMCHLT  & \WPWIEERHLT  \\
\hline
 \EPS{\tnp-all}  & \WPWIEEEFF  & \WPWIEEMC & \WPWIEER \\
\hline
\end{tabular}
\end{center}
\end{table}

\par
Event selection efficiencies are measured with respect to the W events
within the ECAL acceptance.
Simulation efficiencies estimated from {\sc POWHEG} $\Wo$ samples 
are shown in Table~\ref{tab:el-Weff}.
These are efficiencies at the event level,
e.g.: they include efficiency loss due to the second electron veto.  
Given the acceptances listed in Table~\ref{tab:WZlaccgen} and the \TNP 
efficiencies listed in Table~\ref{tab:e-eff-summary},
the overall efficiency correction factors for electrons from $\Wo$ decays are computed.
The overall $\Wo$ signal efficiencies, obtained as products of simulation efficiencies
with data/simulation correction factors, are listed in Table~\ref{tab:el-Weff}.
\begin{table}[ht] %
  \begin{center}
  \caption{ Simulation efficiencies and the final corrected selection efficiencies for the 
$\Wp$, $\Wm$, and their average, in the $\Wen$ analysis. The quoted uncertainties are 
statistical for $\effmc$ and include both statistical and systematic uncertainties 
for the corrected efficiencies $\effmc \times \rhoeff$.
  \label{tab:el-Weff}}
  \begin{tabular}{|l|c|c|}
    \hline
     & $\effmc$  &  $\effmc \times \rhoeff$ \\
    \hline\hline
 $\Wpen$   & \WEPEFFMC  & \WEPEFF \\
 $\Wmen$   & \WEMEFFMC  & \WEMEFF \\
 $\Wen$  & \WEIEFFMC  & \WEIEFF \\
    \hline
    \end{tabular}
  \end{center}
\end{table}

\begin{figure}[htbp]
\begin{center}
  \includegraphics[width=0.48\textwidth]{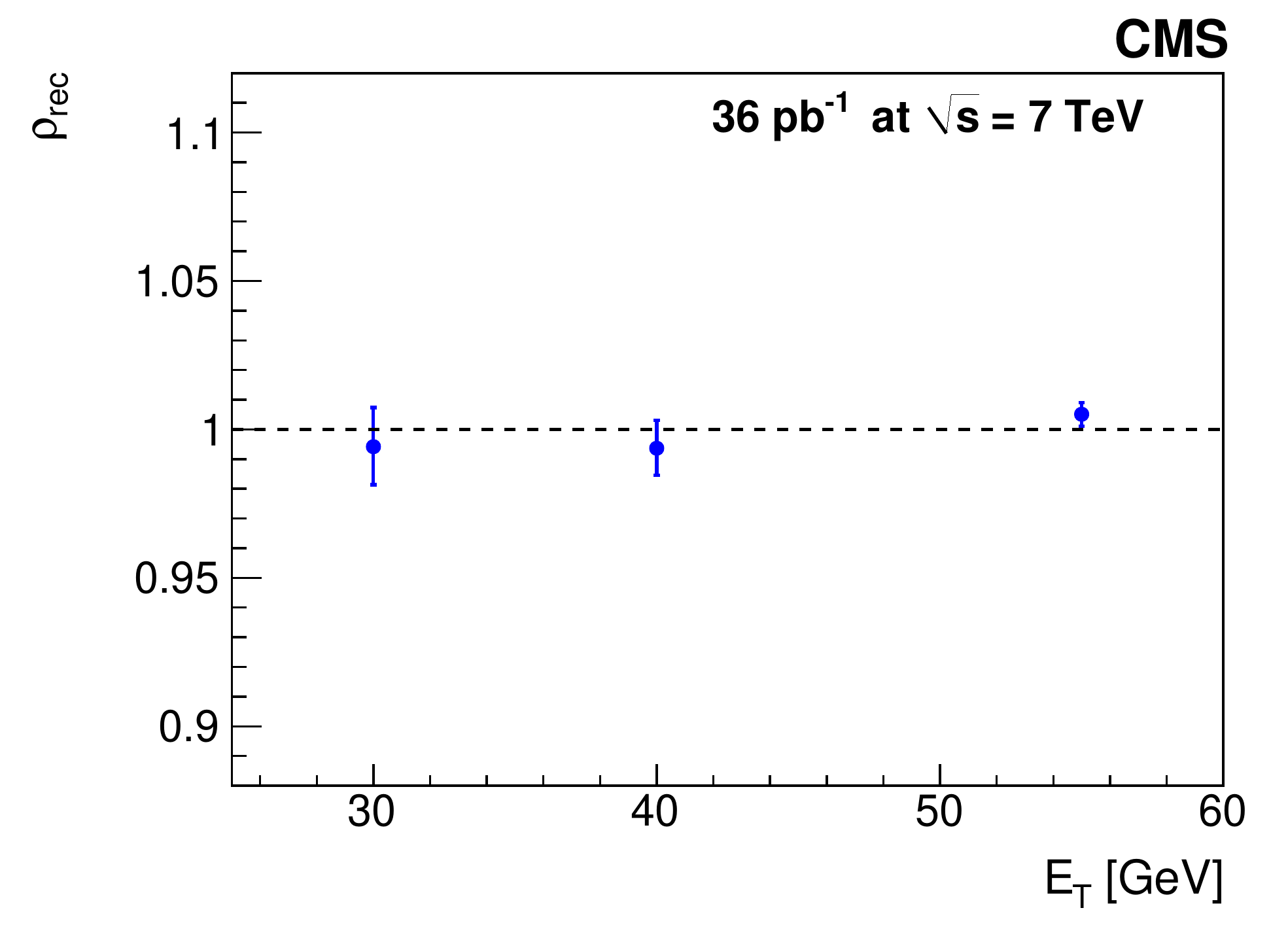}
  \includegraphics[width=0.48\textwidth]{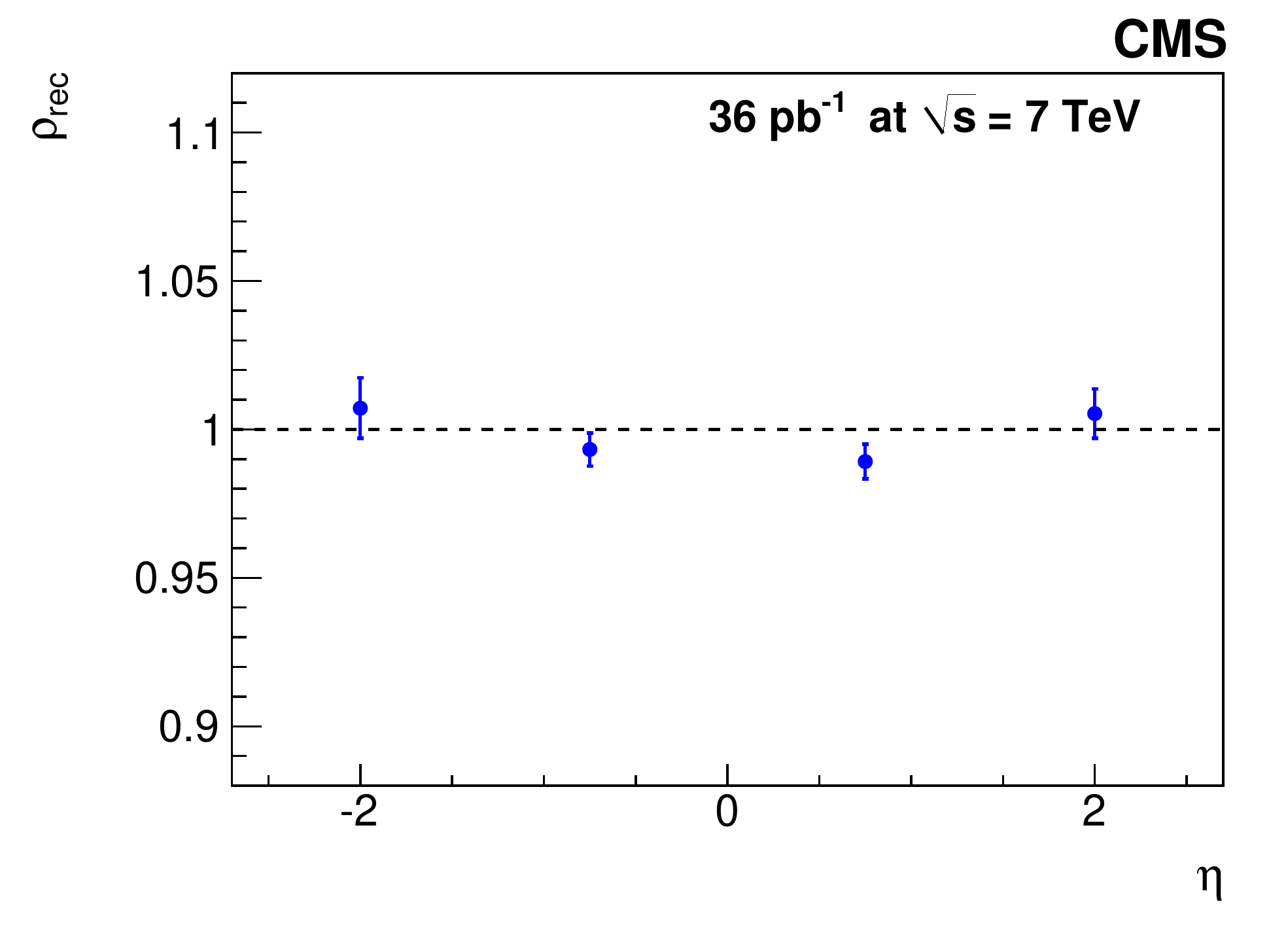}
  \includegraphics[width=0.48\textwidth]{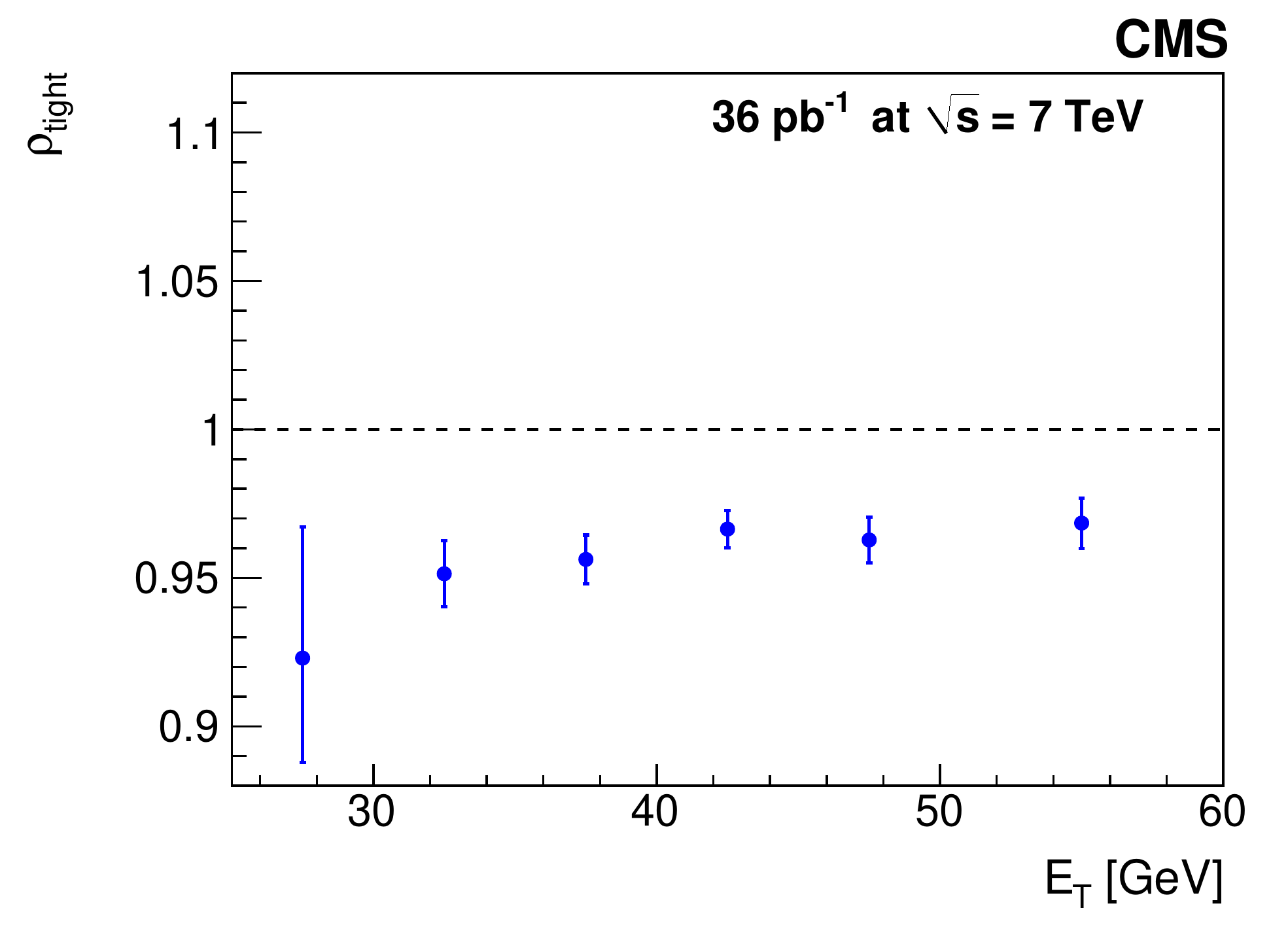}
  \includegraphics[width=0.48\textwidth]{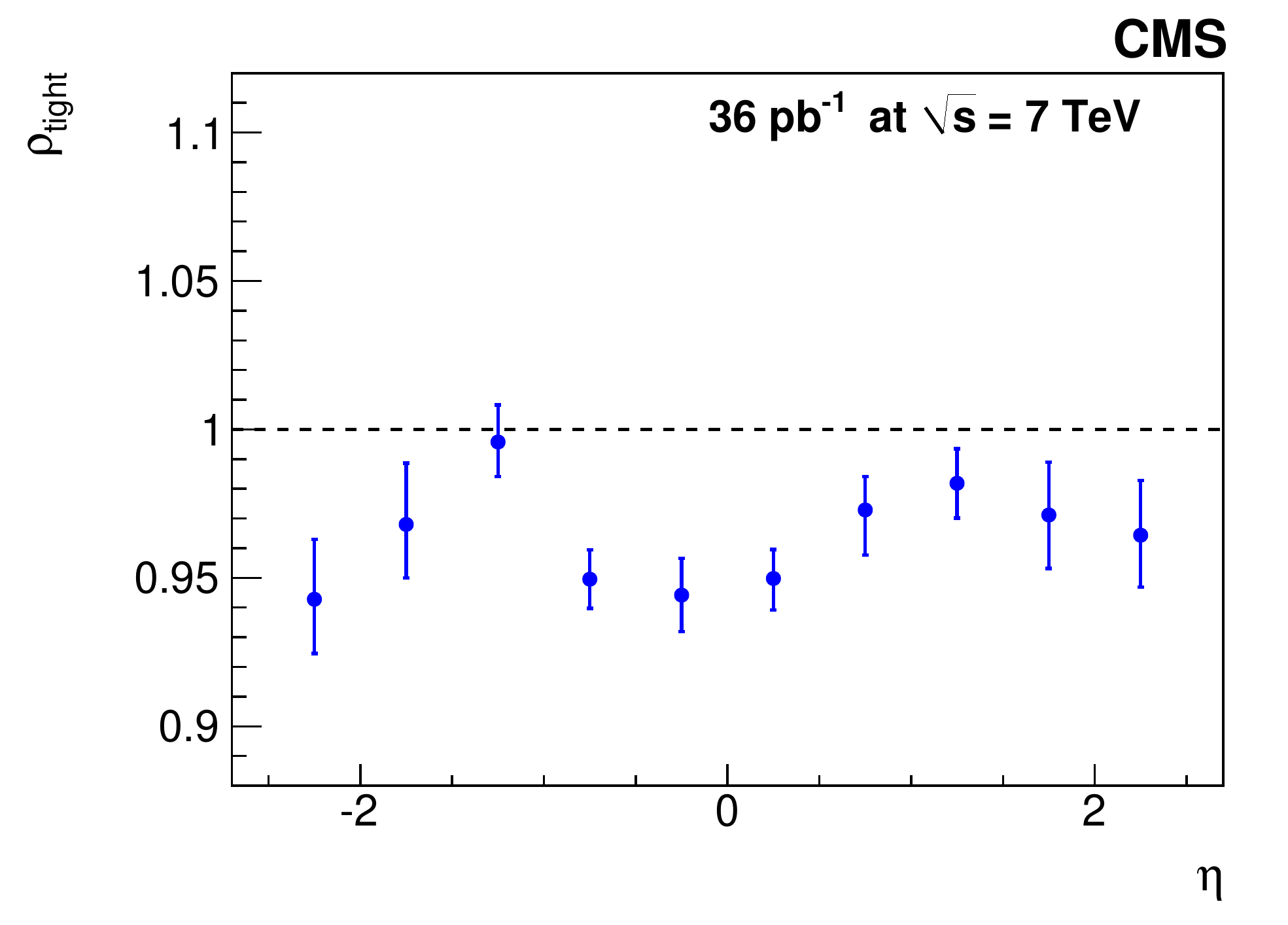}
  \includegraphics[width=0.48\textwidth]{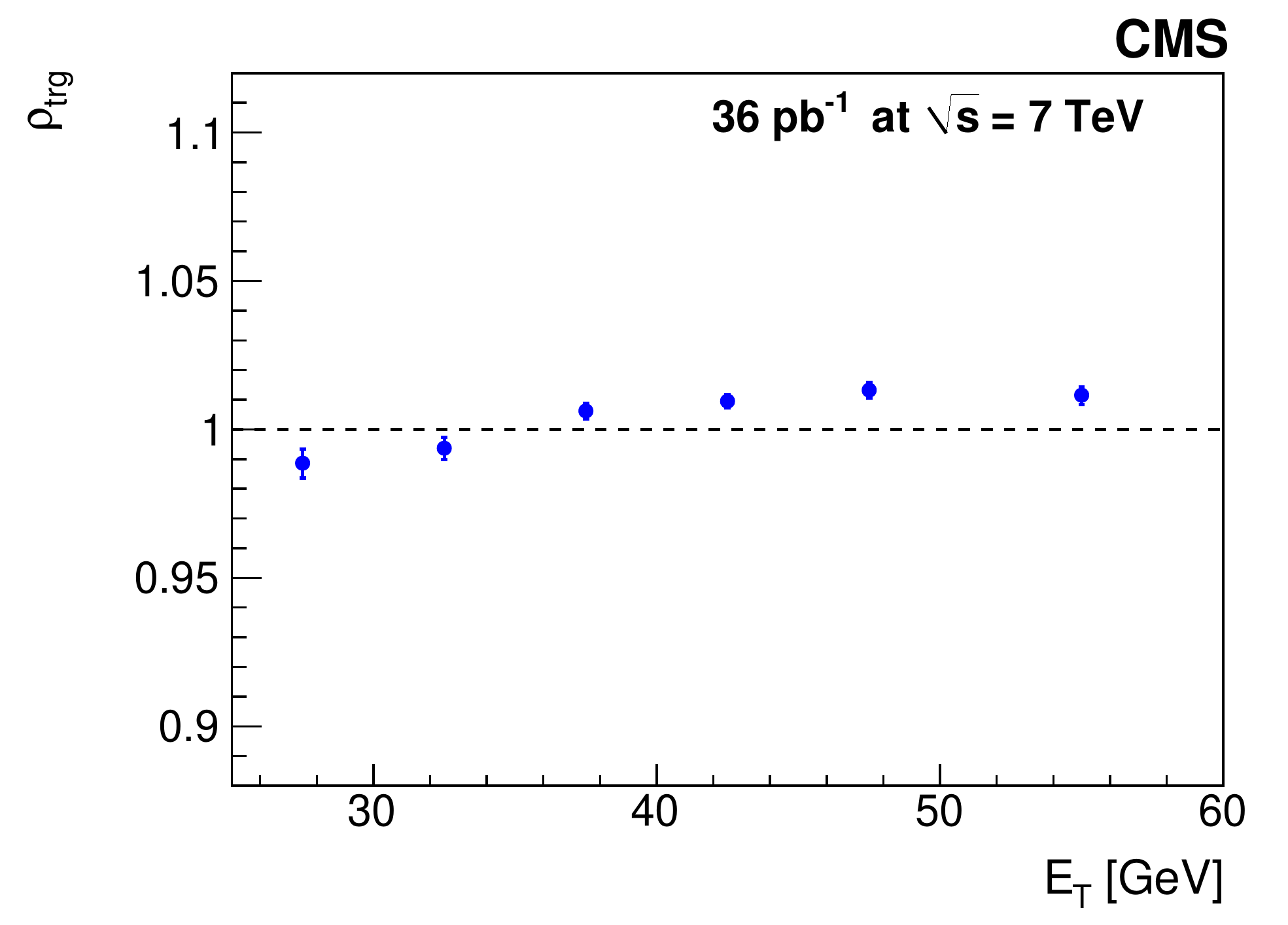}
  \includegraphics[width=0.48\textwidth]{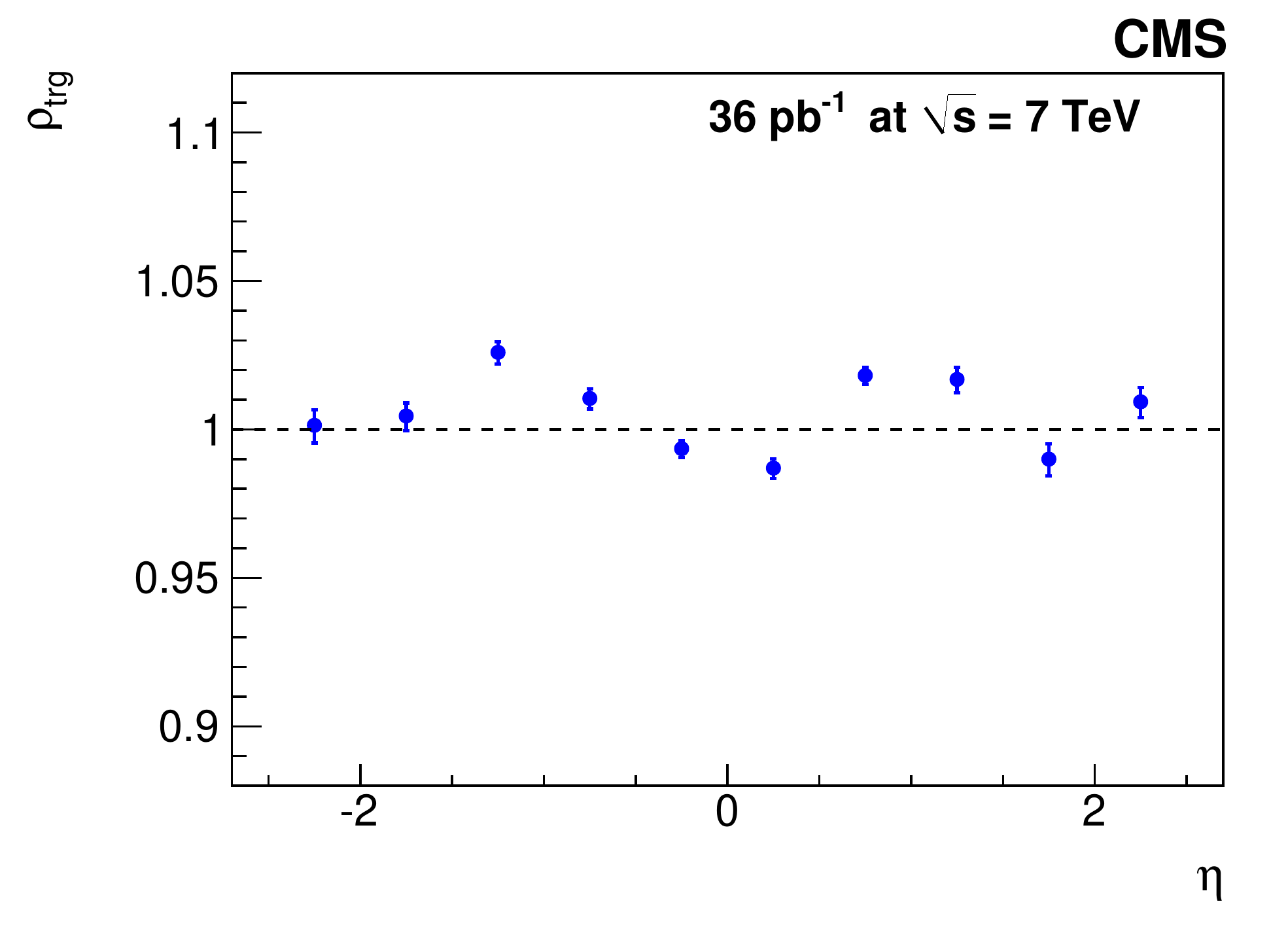}
 \end{center}
\caption{Data/simulation \TNP ratios versus electron $\Et$ (left column) and $\eta$ (right column).
The ratios are presented for the reconstruction ($\rho_\mathrm{rec}$, top row), selection 
($\rho_\mathrm{tight}$, middle row), and trigger ($\rho_\mathrm{trg}$, bottom row) efficiencies.
Points with error bars represent the ratio measured in data; dashed lines correspond to a constant ratio of one.
\label{fig:e-TnPratios}}
\end{figure}

The efficiencies and the data/simulation ratios are also estimated in bins of the electron 
$\Et$ and $\eta$ in order to examine in detail the detector performance and take into 
account the differences in the W and Z kinematic distributions. 
The data/simulation ratios for reconstruction, selection, and trigger are shown 
in Fig.~\ref{fig:e-TnPratios} as functions of the electron $\Et$ and $\eta$.

The reconstruction data/simulation ratios appear to be uniform with respect to $\Et$ and $\eta$, so
a smaller number of bins is sufficient for the determination of their values.
The data/simulation ratios for the selection and trigger efficiencies show a dependence 
that is estimated using ten $\eta$ bins and six $\Et$ bins. Data/simulation ratios are estimated for 
both electron charges as well. 

The binned ratios and simulation efficiencies are transferred into the W analysis by properly weighting 
their product in each ($\Et$, $\eta$) bin by the relative ECAL cluster abundance
estimated from {\sc POWHEG} simulations. The corrected efficiencies are compared with the 
two-bin case in which the efficiencies are estimated in two bins of $\eta$ (EB and EE). 
The multibin corrected efficiencies are found to be consistent with the  two-bin 
corrected efficiencies within the assigned uncertainties. 
In order to be sure that no hidden systematic uncertainty is missed,  half of the 
maximum difference between the multibin and two-bin corrected efficiencies 
is propagated as an additional systematic uncertainty on the two-bin efficiencies used to estimate the cross 
sections. The additional relative uncertainty is at the level of 0.6$\%$.

The $\Zo$ selection efficiencies for data and simulation are obtained based
on the \TNP efficiencies
listed in Table~\ref{tab:e-eff-summary}  and
the event acceptances given in Table~\ref{tab:WZlaccgen}.
The $\Zo$ efficiencies are first determined after
reconstruction and identification (as products of
single-electron efficiencies).
The event trigger efficiency is computed
as the probability that at least one of the two electrons
satisfies the L1+HLT requirement. 
The overall selection efficiency for the $\Zo$ analysis is
the product of the reconstruction, identification, and trigger 
efficiencies. The simulation efficiency obtained from the {\sc POWHEG} $\Zo$ 
samples, together with the final corrected $\Zo$ selection efficiency 
$\effmc \times \rhoeff$, are shown in Table~\ref{tab:el-Zeff}.
These efficiencies are relative to the $\Zo$ events with both electrons within the ECAL acceptance.
\begin{table}[ht] %
  \begin{center}
  \caption{ Simulation efficiency and the final corrected selection efficiency for the ~$\Zee$ analysis.
The quoted uncertainties are statistical for $\effmc$ and include both statistical and 
systematic uncertainties for the corrected efficiency $\effmc \times \rhoeff$.
 \label{tab:el-Zeff}}
  \begin{tabular}{|l|c|c|}
    \hline
     & $\effmc$ & $\effmc \times \rhoeff$ \\
    \hline\hline
    $\Zee$  & \ZEEEFFMC & \ZEEEFF \\
    \hline
    \end{tabular}
  \end{center}
\end{table}

\par

\subsection{Muons}
\label{sec:muonEff}

For the $\Wmn$ cross section determination
the single-muon efficiency combines the efficiencies of all the steps in the muon selection: 
triggering on the muon, reconstructing it in the muon and central detectors, and applying the
quality selection and the isolation requirement. 
In the procedure followed in this analysis, the reconstruction efficiency in the central tracker is factorized 
and  computed independently, while the remaining terms are computed globally, without further 
factorizing them into different terms.

An initial preselection of Z events for the \TNP method is performed by selecting 
events that contain tracks measured in the central tracker having $\Pt>25$~GeV, $|\eta|<2.1$, 
and, when combined with an oppositely charged track, give an 
invariant mass in the range $60<m_{\mu^+\mu^-}<120$~GeV. We 
further require the presence in the event of a ``tag'' muon, defined as a global muon, that is matched to one of
the preselected tracks, passes the selection described in Section~\ref{sec:muonid}, and corresponds
to an HLT muon. The number of tag muons selected in data is about $22\, 000$. All the other preselected 
tracks are considered as probes to evaluate the muon efficiency.
The background present in this sample is subtracted with a fit to the dimuon invariant mass spectrum of the 
sum of a Z component and a linear background contribution. 
The shape of the Z component is taken from simulation.

\begin{figure}
  \begin{center}
  \includegraphics[width=7cm]{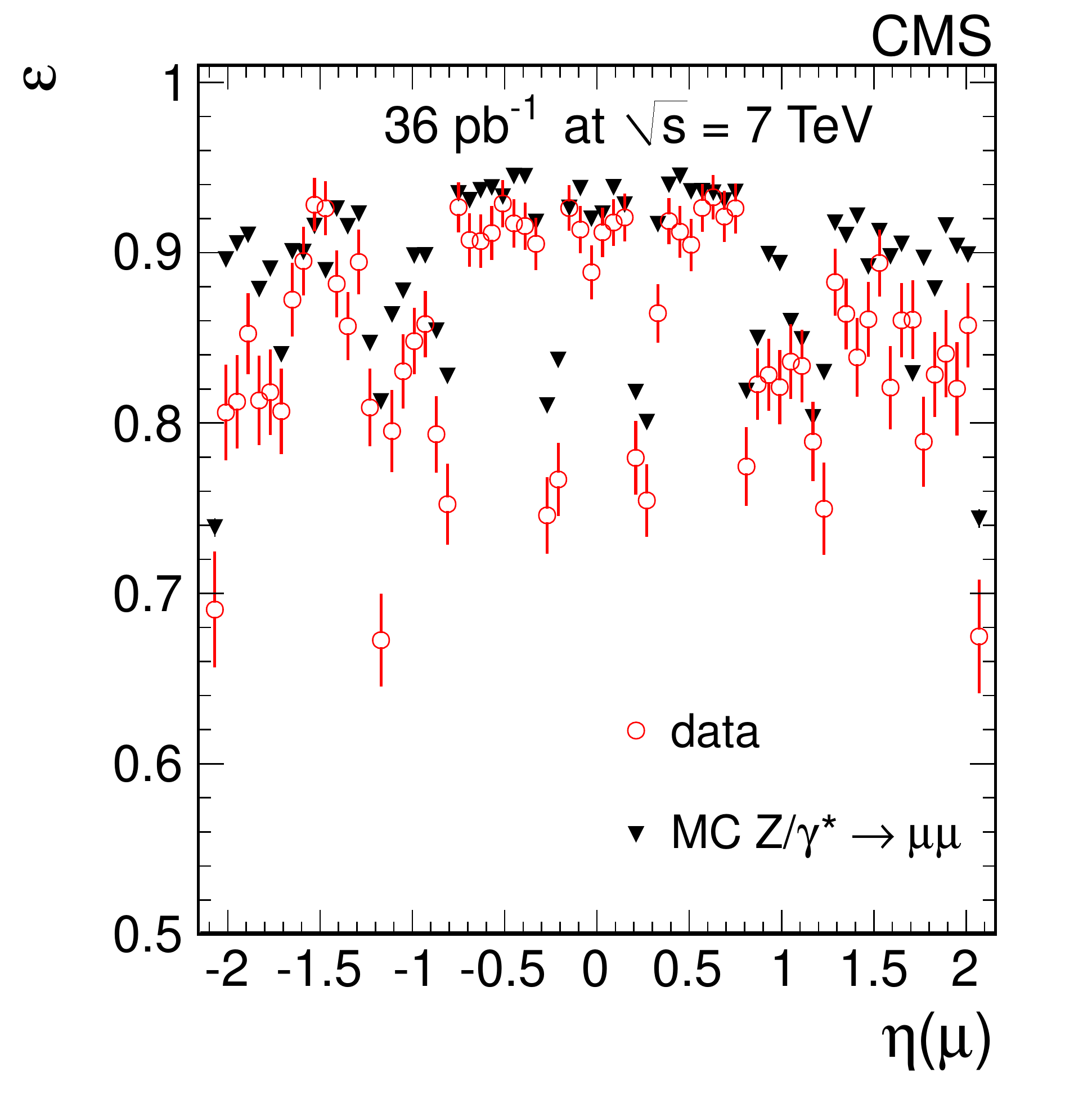}%
  \includegraphics[width=7cm]{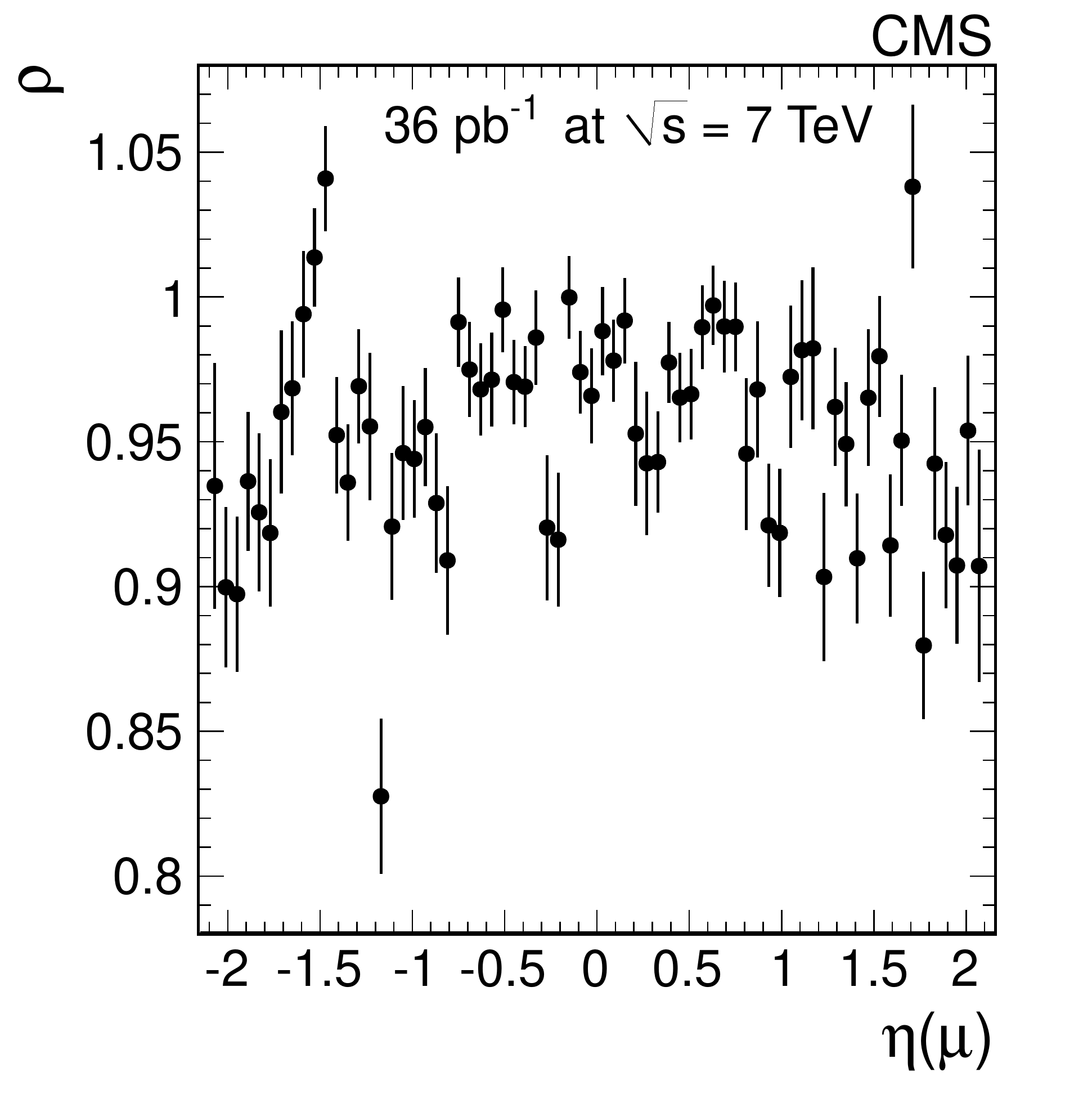}
  \end{center}
\caption{Single-muon efficiencies (left) for data (red circles with error bars)
 and simulation (black triangles), and the ratio between them (right), as a function of the muon $\eta$.}
\label{rho_eta}
\end{figure}

\begin{figure}
  \begin{center}
  \includegraphics[width=7cm]{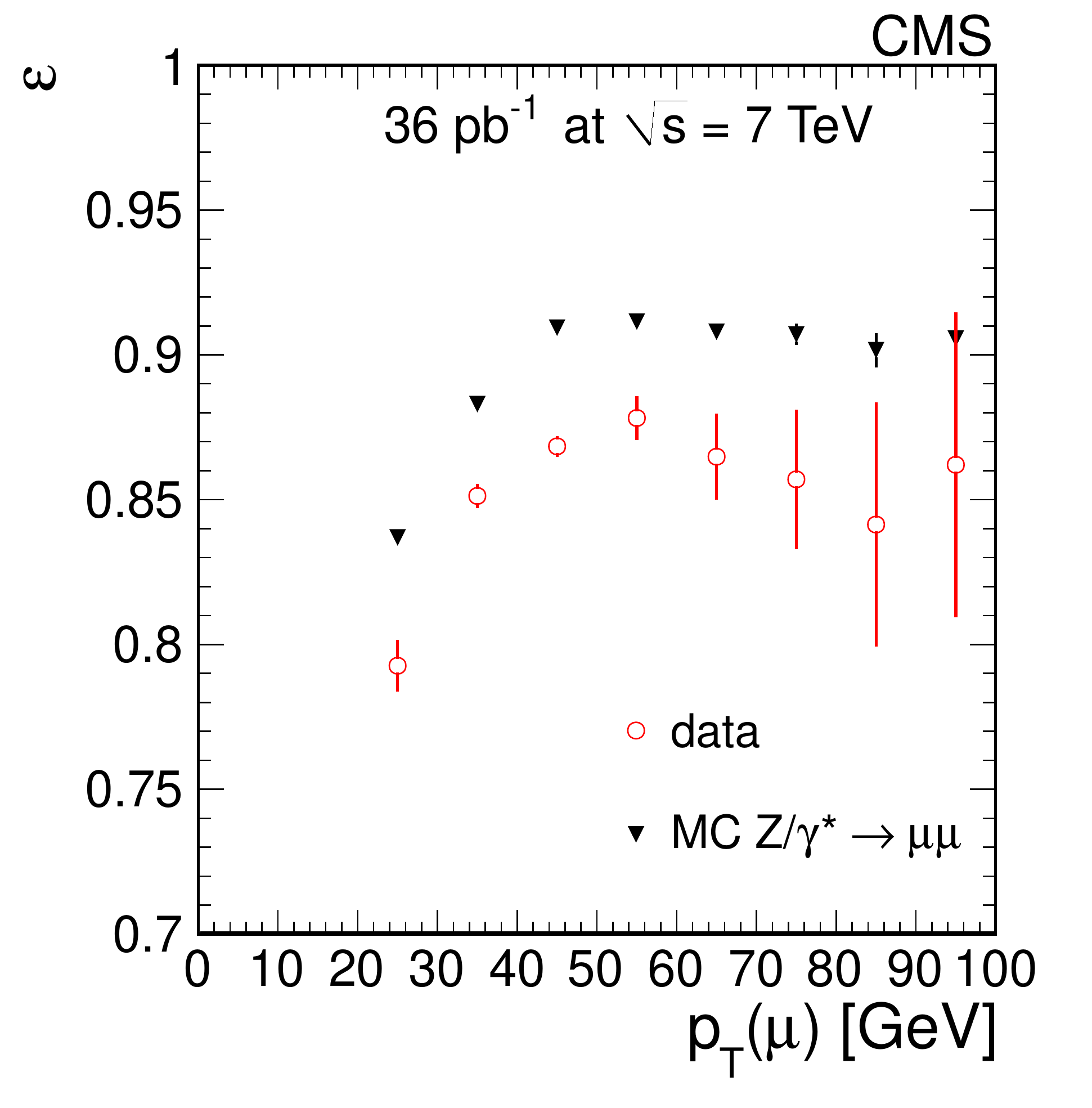}%
  \includegraphics[width=7cm]{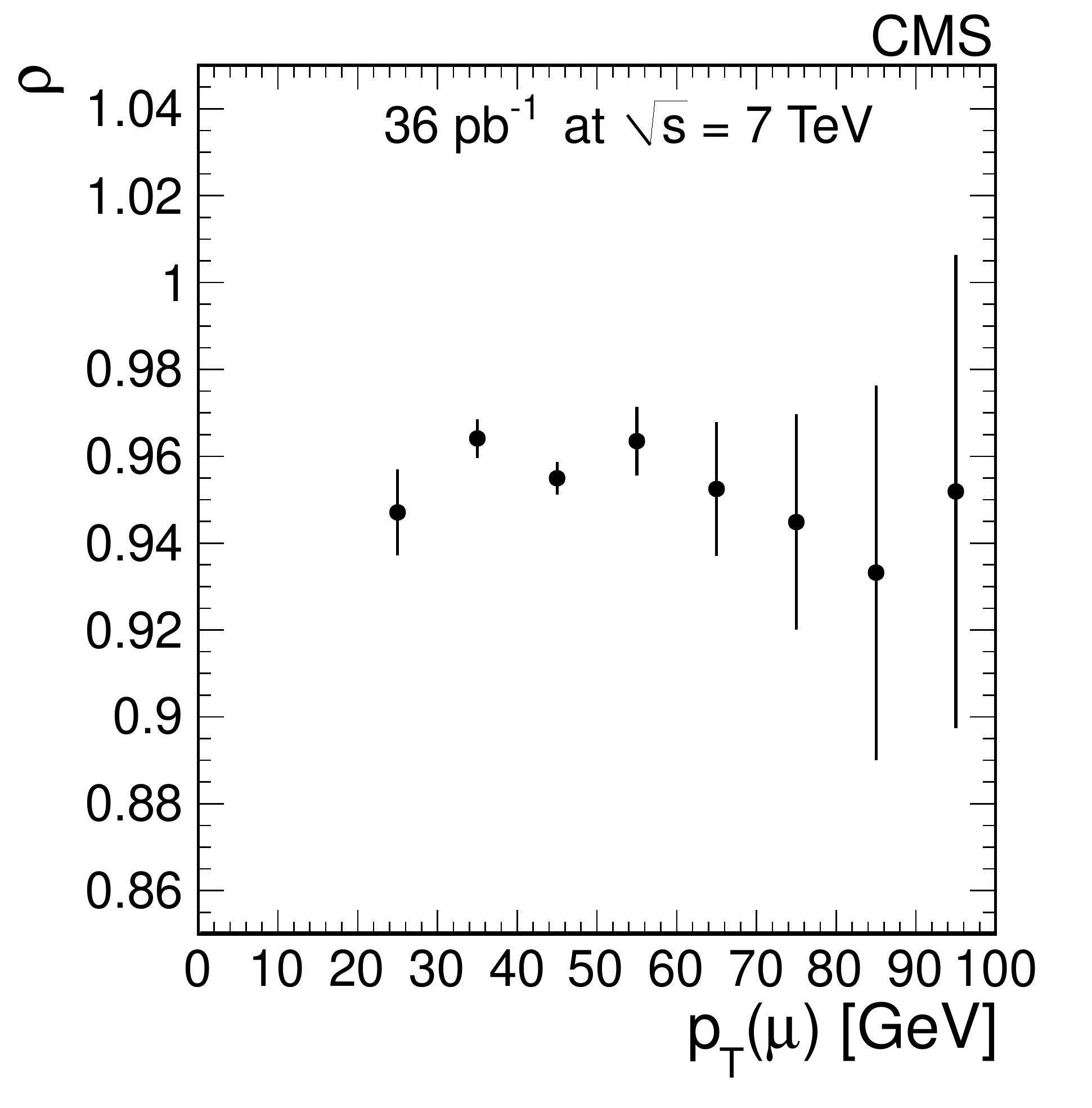}
  \end{center}
\caption{Single-muon efficiencies (left) for data (red circles with error bars) 
and simulation (black triangles) and the ratio between them (right), 
as a function of the muon $\Pt$.}
\label{rho_pt}
\end{figure}

The efficiency is studied as a function of the muon $\eta$ and $\Pt$.
A dependence on $\eta$ is observed (Fig.~\ref{rho_eta}, left) because different 
regions are covered by different muon detectors. 
This behavior is not fully reproduced in 
the simulation, as reflected in the corresponding $\rho$ values (Fig.~\ref{rho_eta}, right). 
The efficiency also exhibits a dependence on $\Pt$ (Fig.~\ref{rho_pt}, left), but this trend is 
similar in data and in simulation, and the correction factors can be taken as approximately constant up to 
$\Pt$ = 100~GeV (Fig.~\ref{rho_pt}, right).
 These binned correction factors are applied to the W analysis during signal modeling (Section 8):
W simulated events are weighted with the $\rho$ factor corresponding to the ($\Pt$, $\eta$) bin of the muon. 
The slightly difference between the kinematic characteristics of the muons and those from W decays is thus taken into account.


The average single-muon efficiencies and correction factors are reported in Table~\ref{tab:effic_charge}
for positively and negatively charged muons separately, and inclusively.
The statistical uncertainties reflect the size of the available Z sample. Systematic uncertainties
on $\effTPdata$ and the correction factors $\rho$ are discussed in Section~\ref{sec:muonSyst}.

\begin{table}[htb] %
    \begin{center}
    \caption{Tag-and-probe efficiencies in data and simulation and correction factors 
for positively and negatively charged muons.
The errors on  $\effTPmc$ are statistical only, while the systematic uncertainty is included for
the other quantities.
}
    \label{tab:effic_charge}
      \begin{tabular}{|l|c|c|c|} \hline
                    &            $\mu^+$            &            $\mu^-$           &                     $\mu^\pm$ \\ \hline\hline
        $\effTPdata$  & $(86.0 \pm 0.8)\%$ & $(85.0 \pm 0.8)\%$ & $(85.58 \pm 0.8)\%$ \\
        $\effTPmc$    & $(89.25 \pm 0.05 )\%$         & $(89.38 \pm 0.05)\%$         & $(89.32 \pm 0.04)\%$        \\
        $\rhoeff$   & $(96.3 \pm 0.9)\%$ & $(95.1 \pm 0.9)\%$& $(95.7 \pm 0.9)\%$\\

      \hline
      \end{tabular}
    \end{center}
  \end{table}

A small fraction of muon events are lost because of L1 muon trigger prefiring, i.e., the assignment
of a muon segment to an incorrect bunch crossing, occurring with a 
probability of a few per mille per segment.
The effect is only sizable in the drift-tube system. The efficiency correction in the barrel region
is estimated for the current data to be ${\sim}1\%$ per muon. This estimate is
obtained from studies of muon pairs selected by online and offline single-muon trigger paths at
the wrong bunch crossing, that have an invariant mass near the Z mass. 
Tracker information is not present in the case of prefiring, precluding the building of a 
trigger muon online or a global muon in the offline reconstruction. 
Since this effect is not accounted for in the efficiency from \TNP,
the measured $\Zmm$ and $\Wmn$ cross sections are increased by 1\% and 0.5\%, 
respectively (including barrel and endcap regions)
to correct for the effect of trigger prefiring. The uncertainty on those corrections
is taken as a systematic uncertainty.

The $\Wmn$ efficiencies from simulation are shown in Table~\ref{tab:mu-MCeff} for the $\Wp$ and $\Wm$ samples separately
and combined after applying the binned corrections estimated with the \TNP method using Z events.

\begin{table}[ht] %
  \begin{center}
  \caption{Simulation efficiencies and final corrected efficiencies for the $\Wmn$ analysis.
The quoted uncertainties are statistical for $\effmc$ and include both statistical and systematic uncertainties
for the corrected efficiencies $\effmc \times \rhoeff$.
  \label{tab:mu-MCeff}}
  \begin{tabular}{|l|c|c|}
    \hline
   & $\effmc$ & $\effmc \times \rhoeff$ \\
    \hline\hline
 $\Wpmn$   & $(89.19\pm 0.03)\%$ & $(85.4\pm 0.8)\%$ \\
 $\Wmmn$   & $(89.19\pm 0.03)\%$ & $(84.1\pm 0.8)\%$ \\
 $\Wmn$  & $(89.19\pm 0.03)\%$ & $(84.8\pm 0.8)\%$ \\
    \hline
    \end{tabular}
  \end{center}
\end{table}

For the $\Zmm$ cross section measurement, the muon efficiencies are determined together with the Z yield using 
a simultaneous fit described in Section~\ref{sec:Zmumu}.

\section{\texorpdfstring{The $\Wln$  Signal Extraction}{The W-> l nu  Signal Extraction}}
\label{sec:WsignalExtraction}

The signal and background yields are obtained by fitting
the $\MET$ distributions for $\Wen$ and $\Wmn$ to different
functional models.
An accurate $\MET$ measurement is essential for distinguishing
a $\PW$ signal from QCD multijet backgrounds.
We profit from the application of the PF
algorithm, which provides superior $\MET$
reconstruction performance~\cite{PFMET1} with respect to alternative
algorithms at the energy scale of the $\PW$ boson.

The $\MET$ is the magnitude of the transverse component of the missing momentum
vector, computed as the negative of the vector sum of all
reconstructed transverse momenta of particles identified with
the PF algorithm. The algorithm combines the information from
the inner tracker, the muon chambers, and the calorimeters
to classify reconstructed objects according to particle type
(electron, muon, photon, or charged or neutral hadron),
thereby allowing precise energy corrections.
The use of the tracker information reduces the sensitivity of $\MET$ to miscalibration of the calorimetry.

\par
The QCD multijet background is one of the most significant backgrounds in W analyses.
At high $\MET$, EWK backgrounds, in particular $\Wtn$ and DY,
also become relevant, leading to contamination levels on the
order of $10\%$.

The $\MET$ model is fitted to the observed distribution as the sum of three contributions:
the W signal, and the QCD and EWK backgrounds.
The EWK contributions are normalized to the W signal yield in the fit
through the ratios of the theoretical cross sections.

Simultaneous fits are performed to the two $\MET$ spectra of W$^+$ and W$^-$ candidates,
fitting either the total W cross section and
the ratio of positive and negative W cross sections, or
the individual positive and negative W cross sections.
In both cases the overall normalization of QCD multijet events is determined from the fit.
The diboson and $\ttbar$ contributions,
taken from simulations, are negligible (Section~\ref{sec:EWKbkgds}).

\par
In the following sections the modeling
of the $\MET$ shape for the signal and the EWK backgrounds are presented,
and the methods used to determine
the $\MET$ shape for the QCD multijet background from data are
 described.
Finally, the extraction of the signal yields is discussed.

\subsection{\texorpdfstring{Signal ${E\!\!/}_{\!\mathrm{T}}$  Modeling}{Signal ET Modeling}}
\label{sec:WsignalMETtemplate}

The $\Wln$ signal is extracted with methods that employ
simulation predictions of the $\MET$ distribution in signal events.
These predictions
rely on the modeling of the vector-boson recoil and detector effects that
can be difficult to simulate accurately. Discrepancies could result
from deficiencies in the modeling of the
calorimeter response and resolution, and from an incomplete description of
the underlying event.
These residual effects are addressed using corrections
determined from the study of Z-boson recoil in data, discussed in the following paragraph.

The recoil to the vector boson is defined as the negative of the vector
sum of transverse energy vectors of all particles reconstructed with the PF algorithm
in W and Z events, after subtracting the contribution from the daughter lepton(s).
The recoil is determined for each event in $\Zll$ data and simulated $\Zll$
and $\Wln$ samples.
We fit the distributions of the recoil components (parallel and perpendicular to the
boson p$_T$ direction) with a double Gaussian, whose mean and width vary with the boson
transverse momentum.
For each sample, we fit polynomials to the extracted mean and width of the recoil
distributions as functions of the boson transverse momentum.
The ratios of data to simulation fit-parameters from the $\Zo$ samples are used as scale
factors to correct the polynomials parameters of the W simulated recoil curves.
For each $\Wo$ simulated event, the recoil is replaced with a value drawn from the
distribution obtained with the corrected parameters corresponding to the $\Wo$ p$_T$.
The $\MET$ value is calculated by adding back the energy of the $\Wo$ lepton.
The energy of the lepton used in the
calculation is corrected for the energy-scale and resolution effects.  Statistical
uncertainties from the fits are propagated into the $\MET$ distribution as systematic
uncertainties.  An additional systematic uncertainty is included to account for possible
differences in the recoil behavior of the W and Z bosons.

\begin{figure}
\begin{center}
\includegraphics[width=0.48\textwidth]{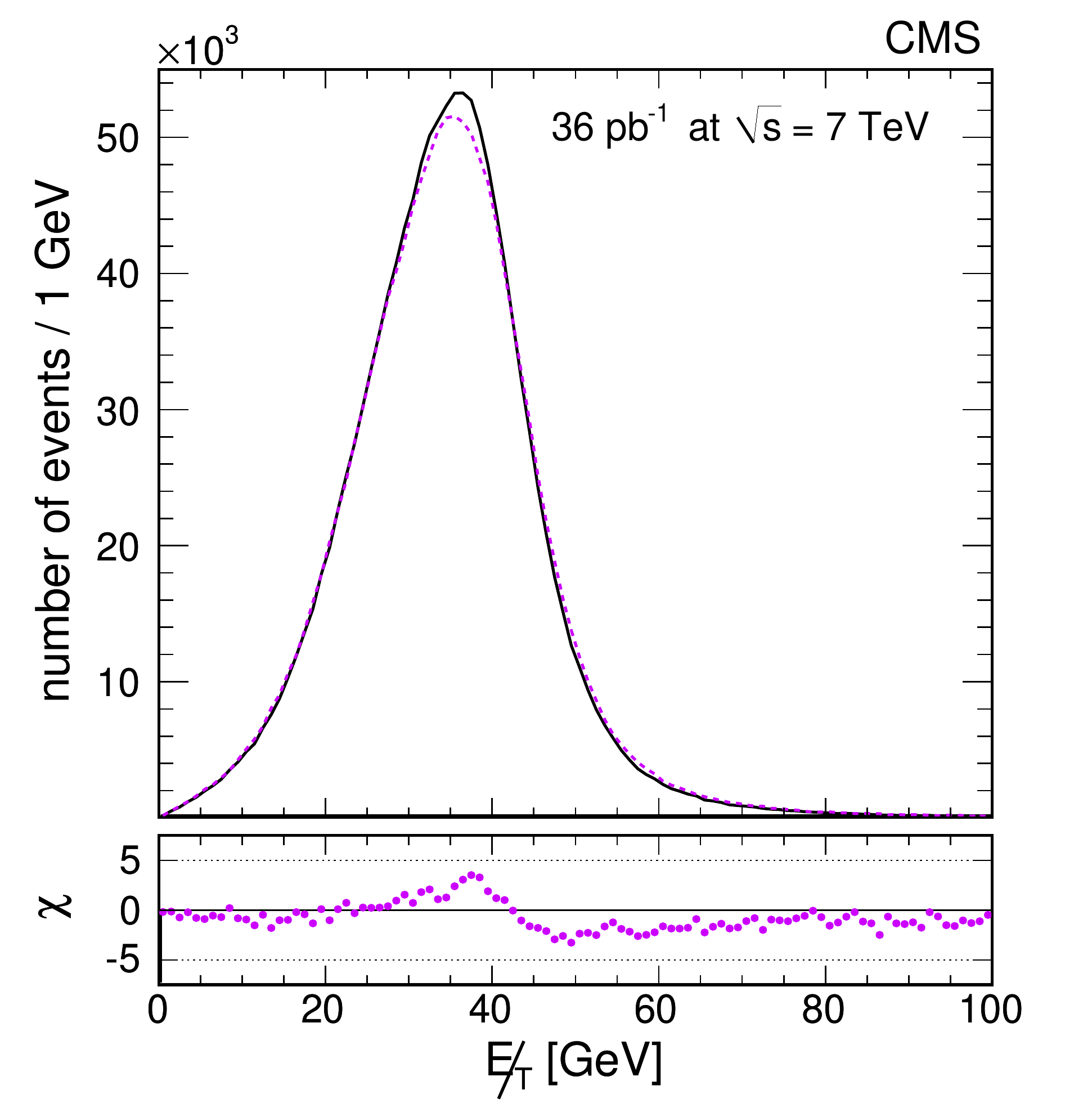}
\includegraphics[width=0.48\textwidth]{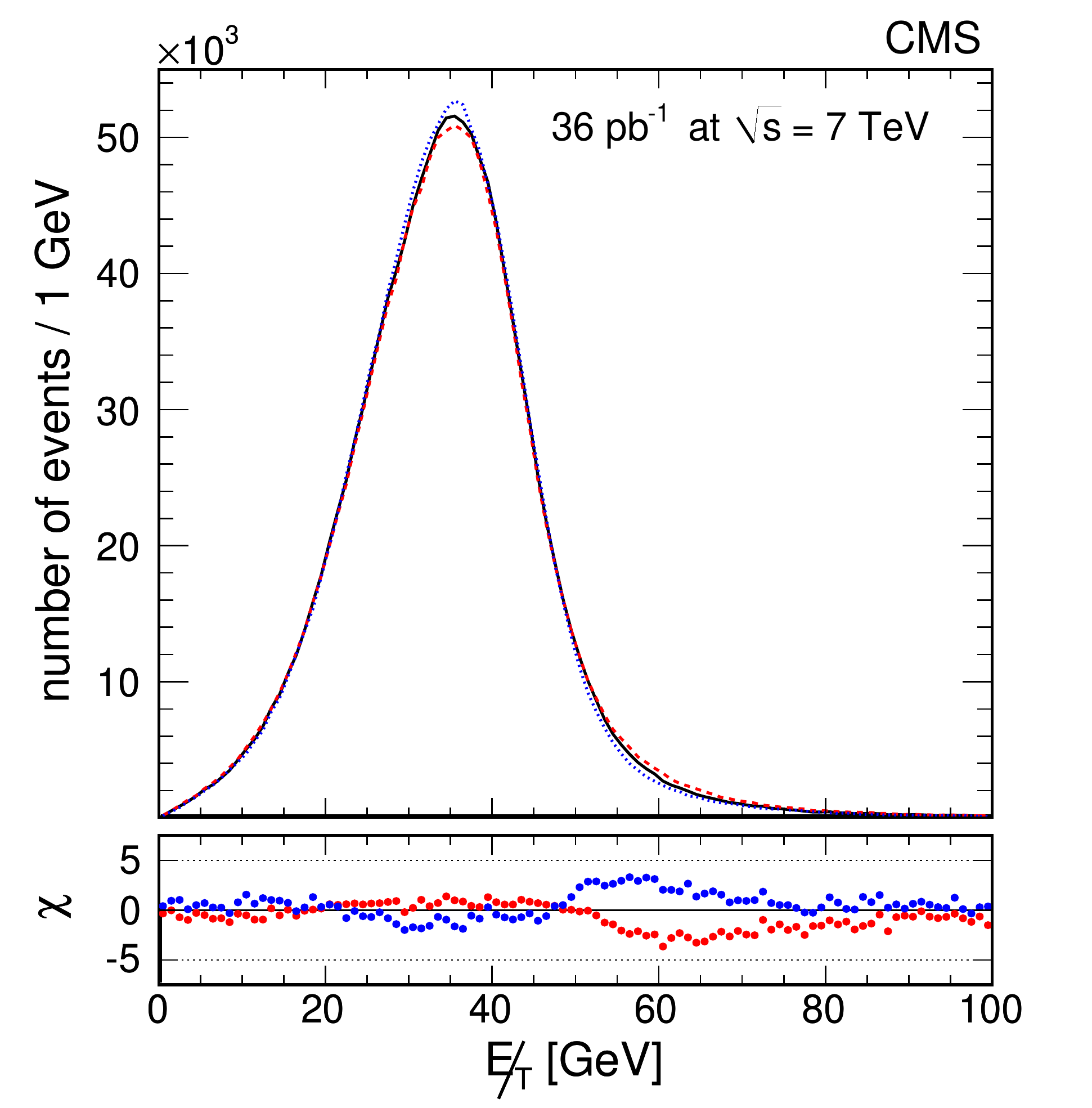}
\caption{ \label{fig:Recoil}
Left: simulated $\MET$ distribution in $\Wen$ events before (continuous black line)
and after (dashed red line) recoil corrections.
Right: the uncertainties from the recoil method propagated
to the corrected $\MET$ shape of $\Wen$ events (continuous black line, identical to the dashed
red line on the left-hand side plot) are presented with the red-dashed and blue-dotted lines.
These two shapes are obtained when the recoil systematic uncertainties are varied
by one standard deviation.
At the bottom of each plot is shown the distribution of the residuals, $\chi$, defined
as the per-bin difference of the two distributions, divided by the
corresponding statistical uncertainty.
}
\end{center}
\end{figure}

The same strategy is followed for the recoil corrections in the electron and muon analyses.
As an example, Fig.~\ref{fig:Recoil} (left) shows the effect of the recoil
corrections on the $\MET$ shape for simulated events in the electron channel, while Fig.~\ref{fig:Recoil} (right)
shows the uncertainty from the recoil method propagated to the corrected $\MET$ shape
of $\Wen$ events. The distribution of the residuals, $\chi$, is shown at the bottom of each plot,
where $\chi$ is defined as the per-bin difference of the two distributions, divided by the
corresponding statistical uncertainty. The same definition is used throughout this paper.

The systematic uncertainties on the signal $\MET$ shape are propagated as systematic
uncertainties on the extracted signal yield through the fitting procedure.
Signal shapes are determined for the W$^+$ and W$^-$ separately.

\subsection{Electroweak Backgrounds}
\label{sec:EWKbkgds}

A certain fraction of the events passing the selection criteria for $\Wln$
are due to other EWK processes. Several sources of
contamination have been identified. The events with $\Zll$ 
(DY background), where one of the two leptons lies
beyond the detector acceptance and escapes
detection, mimic the signature of $\Wln$ events. Events from $\Ztt$ and $\Wtn$,
with the tau decaying leptonically, have in general a lower-momentum lepton than signal
events and are strongly suppressed by the minimum $\Pt$ requirements.

The $\MET$ shape for the EWK vector boson and ${\mathrm t}\bar{\mathrm t}$ contributions are
evaluated from simulations. For the main EWK backgrounds ($\Zll$ and $\Wtn $), the $\MET$ shape is 
corrected by means of the procedure described in Section~\ref{sec:WsignalMETtemplate}.
The $\MET$ shapes are evaluated separately for $\Wptn$ and $\Wmtn$.

A summary of the background fractions in the $\Wen$ and $\Wmn$ analyses can be found in Table~\ref{tab:WlnBG}.
The fractions are similar for the $\Wen$ and $\Wmn$ channels, 
except for the DY background which is higher in the $\Wen$ channel. 
The difference is mainly due to the tighter definition of the DY veto in the $\Wmn$ channel, 
which is not compensated by the larger geometrical acceptance of electrons 
($|\eta|<2.5$) with respect to muons ($|\eta|<2.1$).

\begin{table} %
\begin{center}
\caption{\label{tab:WlnBG}
Estimated background-to-signal ratios in the $\Wen$ and $\Wmn$ channels.}

\begin{tabular}{|l|c|c|}
\hline
{\multirow{2}{*}{Processes}} & \multicolumn{2}{c|}{Bkg. to sig. ratio}  \\ \cline{2-3}
                           & $\Wen$ & $\Wmn$ \\ 
\hline \hline
$\Zee,\, \mu^+\mu^-,\, \tau^+\tau^-$ (DY)               & 7.6\%  &  4.6\% \\
$\Wtn $                    & 3.0\%  & $3.0$\%    \\
$\Wo\Wo$+$\Wo\Zo$+$\Zo\Zo$ & 0.1\% & $0.1$\%   \\
$\ttbar$                   & 0.4\% & $0.4$\%   \\
\hline
Total EWK                  & 11.2\%& $8.1$\% \\
\hline
\end{tabular}
\end{center}
\end{table}

\subsection{\texorpdfstring{Modeling of the QCD Background and $\Wen$ Signal Yield}{Modeling of the QCD Background and W-> e nu Signal Yield}}
\label{sec:WQCDbkg}

Three signal extraction methods are used, which give consistent
signal yields. The method described in Section~\ref{sec:AnalyticalFunction}
is used to extract the final result.

\subsubsection{Modeling the QCD Background Shape with an Analytical Function}
\label{sec:AnalyticalFunction}

The $\Wen$ signal is extracted using an unbinned maximum likelihood
(UML) fit to the $\MET$ distribution.

\begin{figure}[htbp]
\begin{center}
\includegraphics[width=0.50\textwidth]{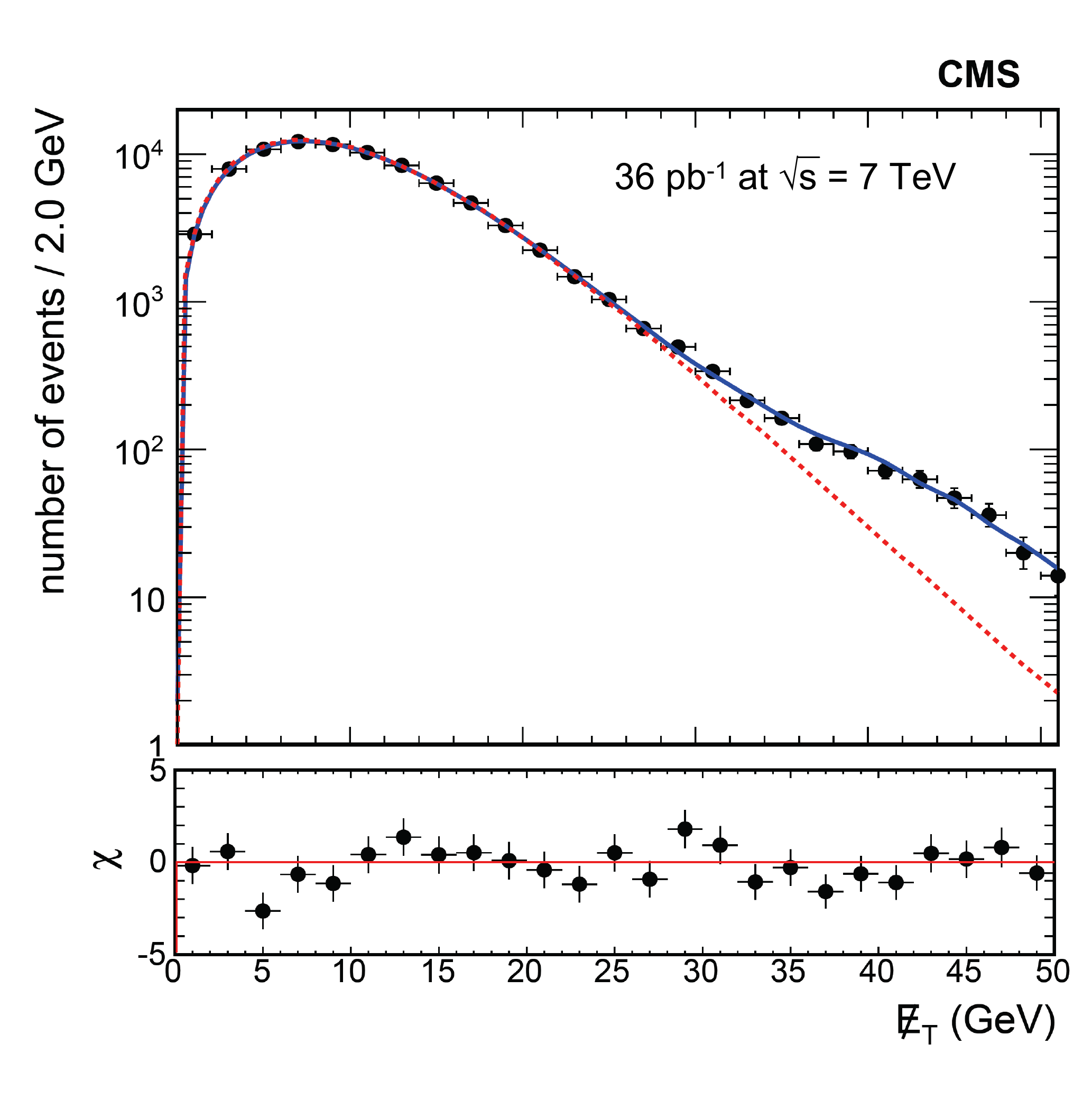}
\caption{Fit to the background-dominated control sample defined by inverting
the selection on the track-match variables,
while maintaining the rest of the signal selection.
The blue solid line represents the model used to fit the control data sample. This is a Rayleigh
function plus a floating-yield signal shape that accounts for the signal contamination in the
control region. The magenta dashed line shows the Rayleigh function alone with its parameters estimated
from the combined fit.
}\label{fig:e-inverted}
\end{center}
\end{figure}

The shape of the $\MET$ distribution for the QCD background is modeled by a parametric function (modified Rayleigh
distribution) whose expression is
\begin{equation}
f_{\mathrm{CQD}}(\MET) = \MET\exp\left(-\frac{\MET{^2}}{2(\sigma_0+\sigma_1 \MET)^{2}}\right)\,.
\label{eq:rayleigh}
\end{equation}
The fit to a control sample, defined by inverting the track-cluster matching selection
variables $\Delta\eta$, $\Delta\phi$, shown in Fig.~\ref{fig:e-inverted}, illustrates
the quality of the description of the background shape by the parameterized function,
including the region of the signal, at high \MET.
To study the systematic uncertainties associated with the background shape, the resolution term in
Eq.~(\ref{eq:rayleigh}) was changed by introducing an additional QCD shape parameter $\sigma_2$,
thus: $\sigma_0 + \sigma_1 \MET + \sigma_2 \MET^2$.

\begin{figure}[htbp]
\begin{center}
\includegraphics[width=0.48\textwidth]{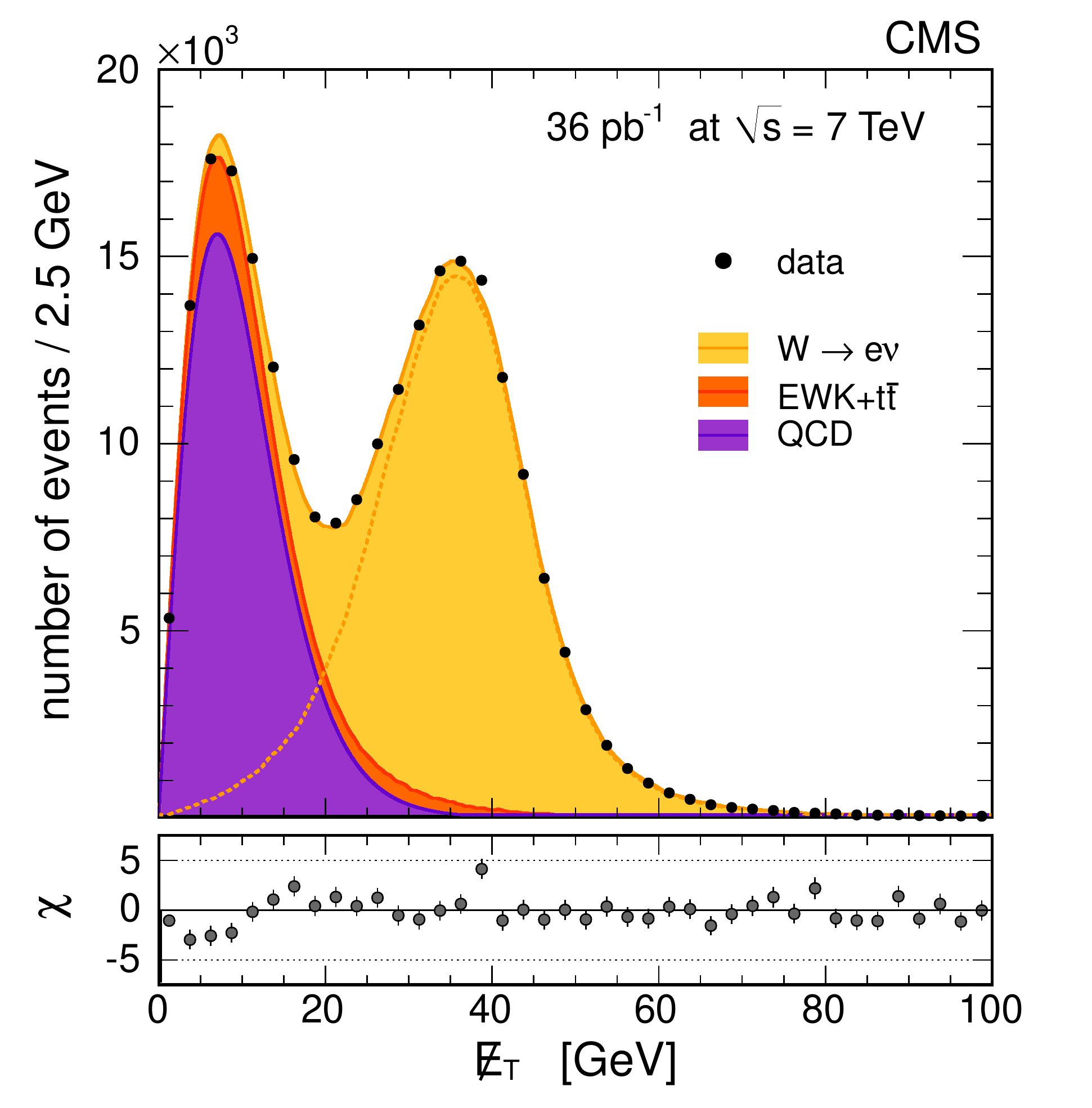}
\includegraphics[width=0.48\textwidth]{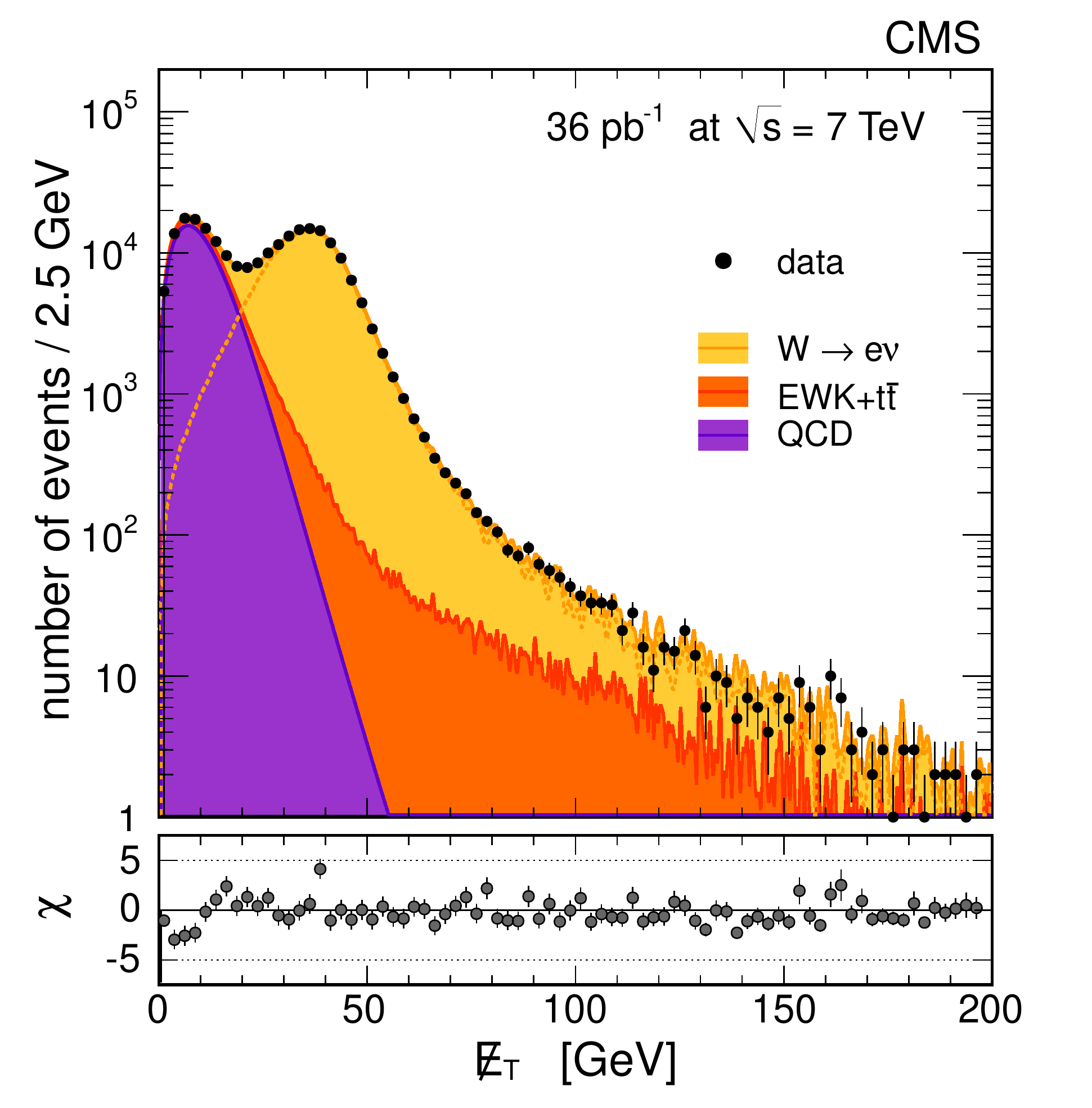}
\caption{ \label{fig:Wen}
The $\MET$ distribution for the selected $\Wen$ candidates on
a linear scale (left) and on a logarithmic scale (right).
The points with the error bars represent the data. Superimposed are the
contributions obtained with the fit
for QCD background (violet, dark histogram), all other backgrounds
(orange, medium histogram), and signal plus  background (yellow, light histogram).
The orange dashed line is the fitted signal contribution.
}
\end{center}
\end{figure}

\begin{figure}[htbp]
\begin{center}
\includegraphics[width=0.48\textwidth]{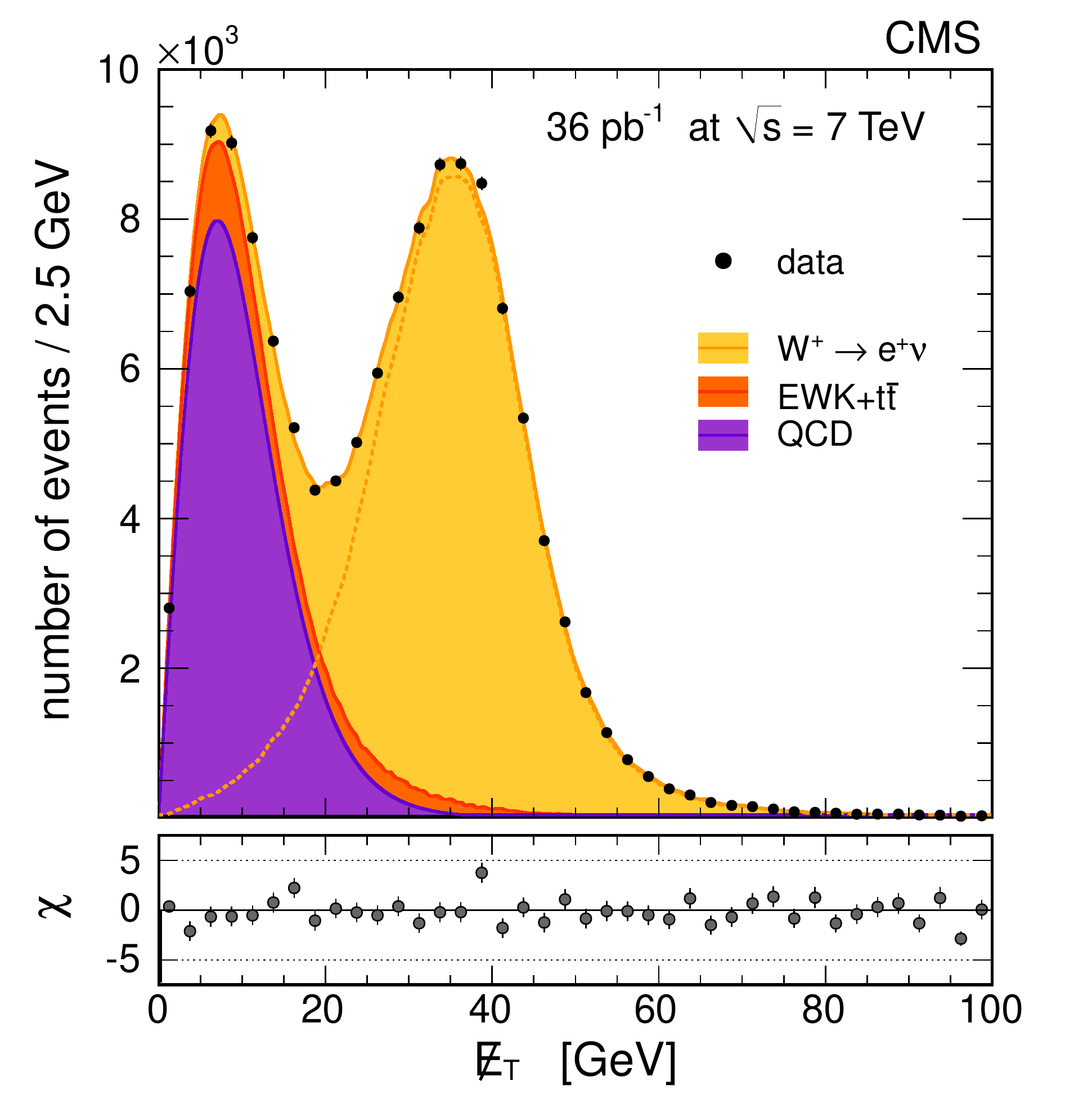}
\includegraphics[width=0.48\textwidth]{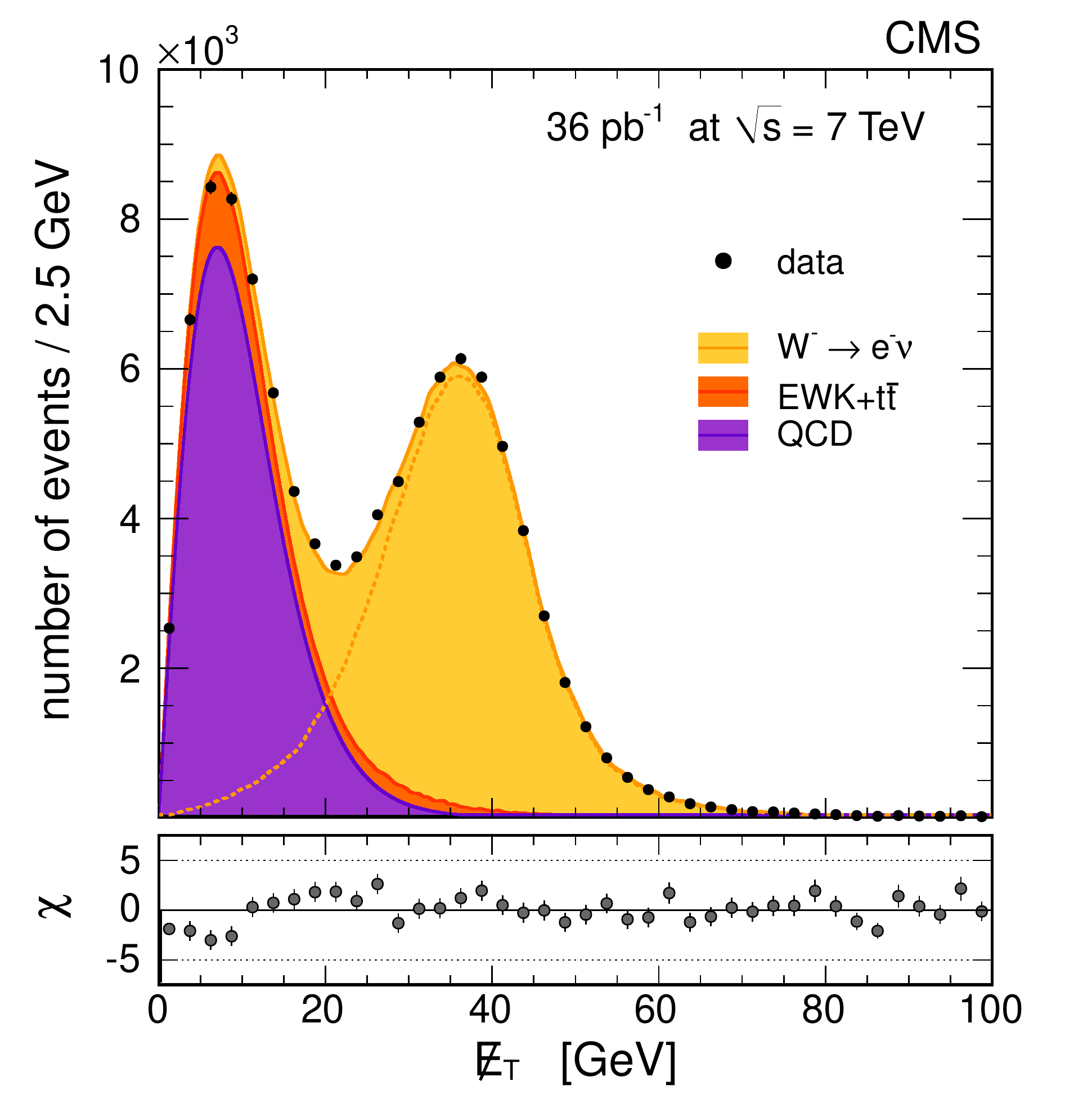}
\caption{ \label{fig:WenPM}
The $\MET$ distributions for the selected W$^+$ (left) and W$^-$ (right) candidates.
The points with the error bars represent the data. Superimposed are the contributions
obtained with the fit for QCD background (violet, dark histogram), all other backgrounds
(orange, medium histogram), and signal plus background (yellow, light histogram).
The orange dashed line is the fitted signal contribution.
}
\end{center}
\end{figure}

The free parameters of the UML fit are the QCD background yield,
the $\Wo$ signal yield, and the background shape
parameters $\sigma_0$ and $\sigma_1$.
The following signal yields are obtained:
$\WEIYIELD$ for the inclusive sample, $\WEPYIELD$ for the $\Wpen$ sample, and
$\WEMYIELD$ for the $\Wmen$ sample.
The fit to the inclusive $\Wen$ sample is displayed
in Fig.~\ref{fig:Wen}, while the fits for the charge-specific
channels are displayed in Fig.~\ref{fig:WenPM}.

The Kolmogorov--Smirnov probabilities for the fits to the charge-specific
channels are $\WEPKSPCOR$ for the $\Wp$ sample and
$\WEMKSPCOR$ for the $W^-$ sample.
\begin{figure}[t!]
\begin{center}
\includegraphics[width=0.48\textwidth]{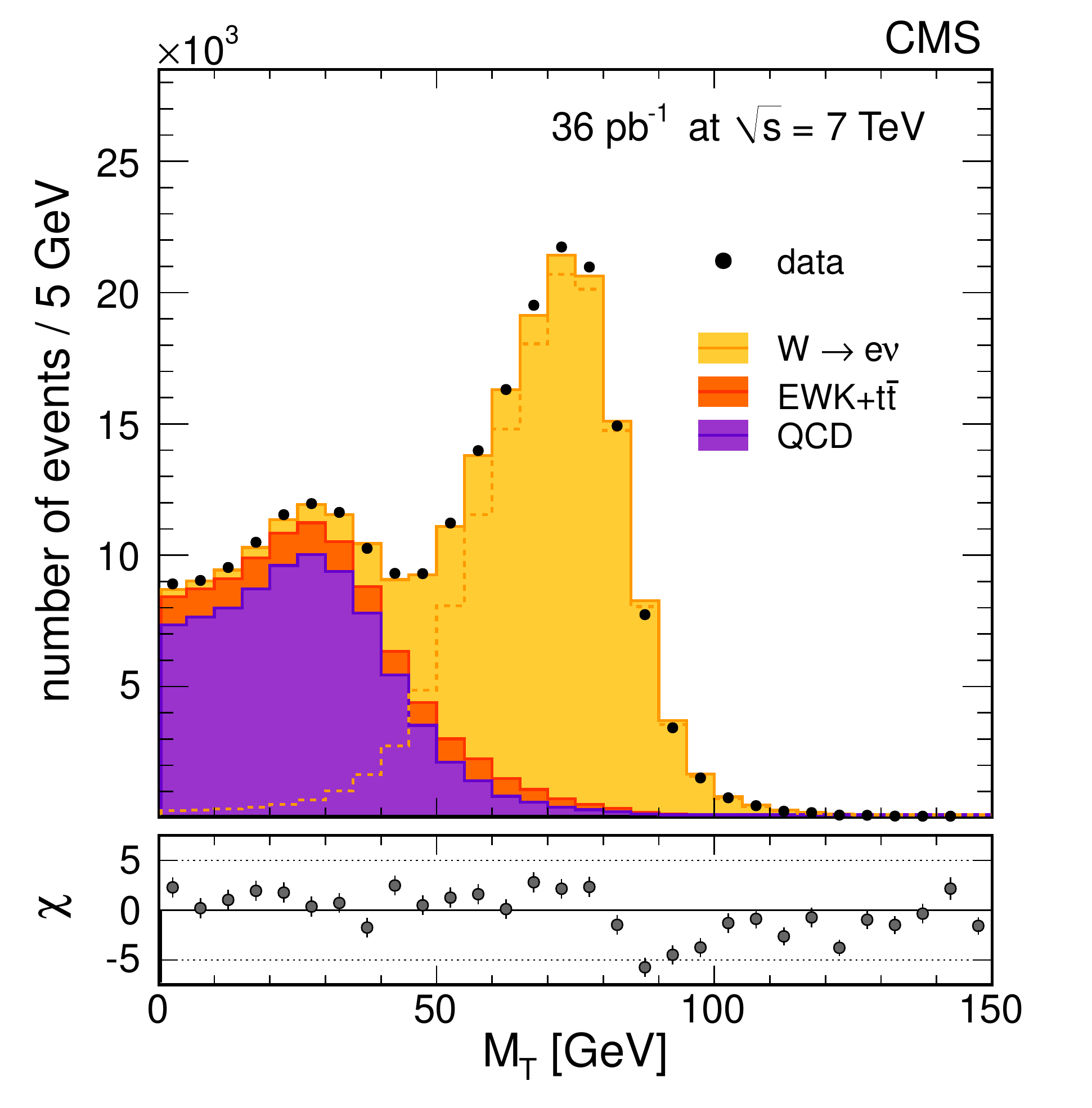}
\includegraphics[width=0.48\textwidth]{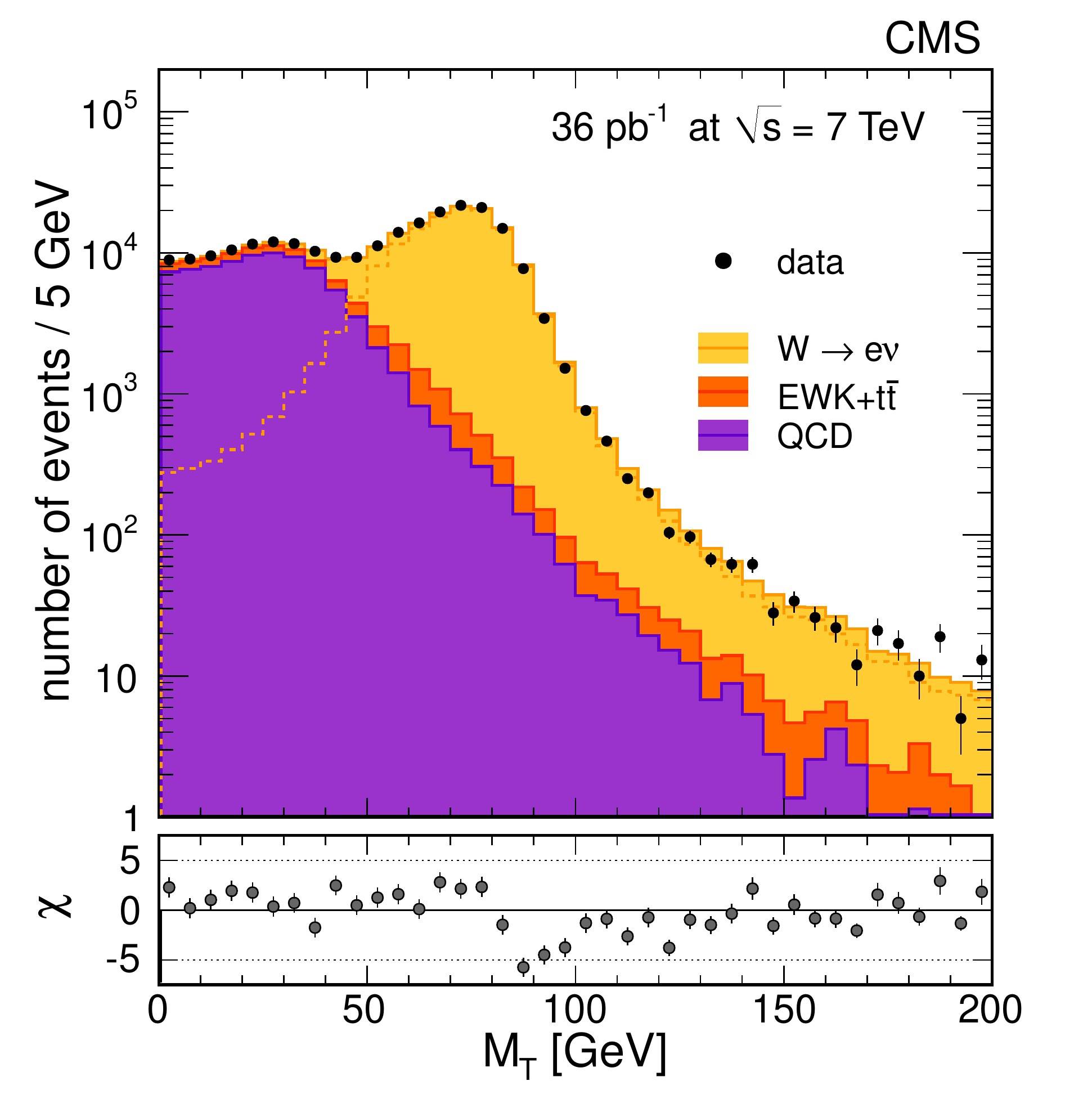}
\caption{ \label{fig:WenMT}
The $\MT$ distribution for the selected $\Wen$ candidates on
a linear scale (left) and on a logarithmic scale (right).
The points with the error bars represent the data. Superimposed are the
contributions obtained with the fit
for QCD background (violet, dark histogram), all other backgrounds
(orange, medium histogram), and signal plus  background (yellow, light histogram).
The orange dashed line is the fitted signal contribution.
}
\end{center}
\end{figure}
Figure~\ref{fig:WenMT} shows the distribution for the inclusive $\Wo$ sample
of the transverse mass, defined as
$\MT=\sqrt{2\Pt\MET (1-\cos(\Delta\phi_{\mathrm{l},\MET}))}$,
where $\Delta\phi_{\mathrm{l},\MET}$ is the azimuthal angle between the
lepton and the $\MET$ directions.

\subsubsection{Modeling the QCD Background Shape with a Fixed Distribution}
\label{sec:e-Wsigextr-FixedTemplate}

In this approach the QCD shape is extracted directly from data using a 
control sample obtained by inverting a subset of the requirements used to select the 
signal. After fixing the shape from data,
only the normalization is allowed to float in the fit.  

The advantage of this approach 
is that detector effects, such as anomalous signals in the ca\-lo\-ri\-me\-ters or 
dead ECAL towers, are automatically reproduced in the QCD shape, since these 
effects are not affected by the selection inversion used to define the control sample.
The track-cluster matching variable $\Delta\eta$ is found
to have the smallest correlation with $\MET$ and is therefore chosen as the one 
to invert in order to remove the signal and obtain the QCD control sample.  
Requirements on isolation and $H/E$ are the same as for the signal 
selection since these variables show significant correlation with \MET. 
\begin{figure}[h!]
  \begin{center}
    \includegraphics*[angle=-90,width=0.55\textwidth]{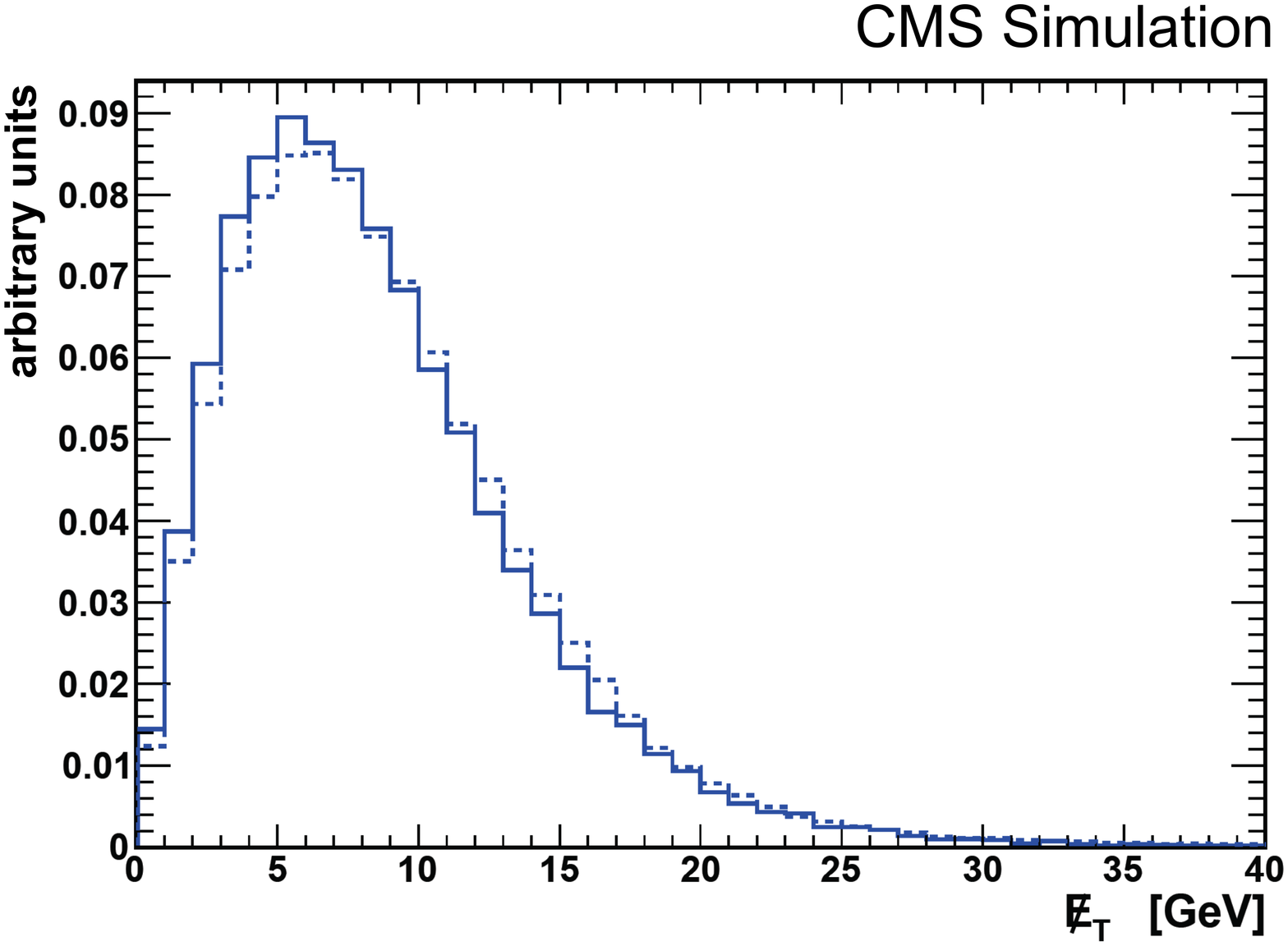}
    \caption{Normalised \MET distribution for QCD and $\gamma$+jet simulated 
events passing the signal selection (solid histogram) compared 
to the normalised distribution for events from all simulated samples passing
the same inverted selection criteria used to obtain the control sample in data
(dashed histogram).}
    \label{fig:antiselShape}
  \end{center}
\end{figure}

The shape of the $\MET$ distribution for QCD and $\gamma$+jet simulated 
events passing the signal selection is compared 
to the \MET distribution for a simulated control sample composed of all simulated samples (signal and 
all backgrounds, weighted according to the 
theoretical production cross sections), after applying the same anti-selection 
as in data (Fig.~\ref{fig:antiselShape}).

\begin{figure}[htb]
  \begin{center}
    \includegraphics*[width=0.5\textwidth]{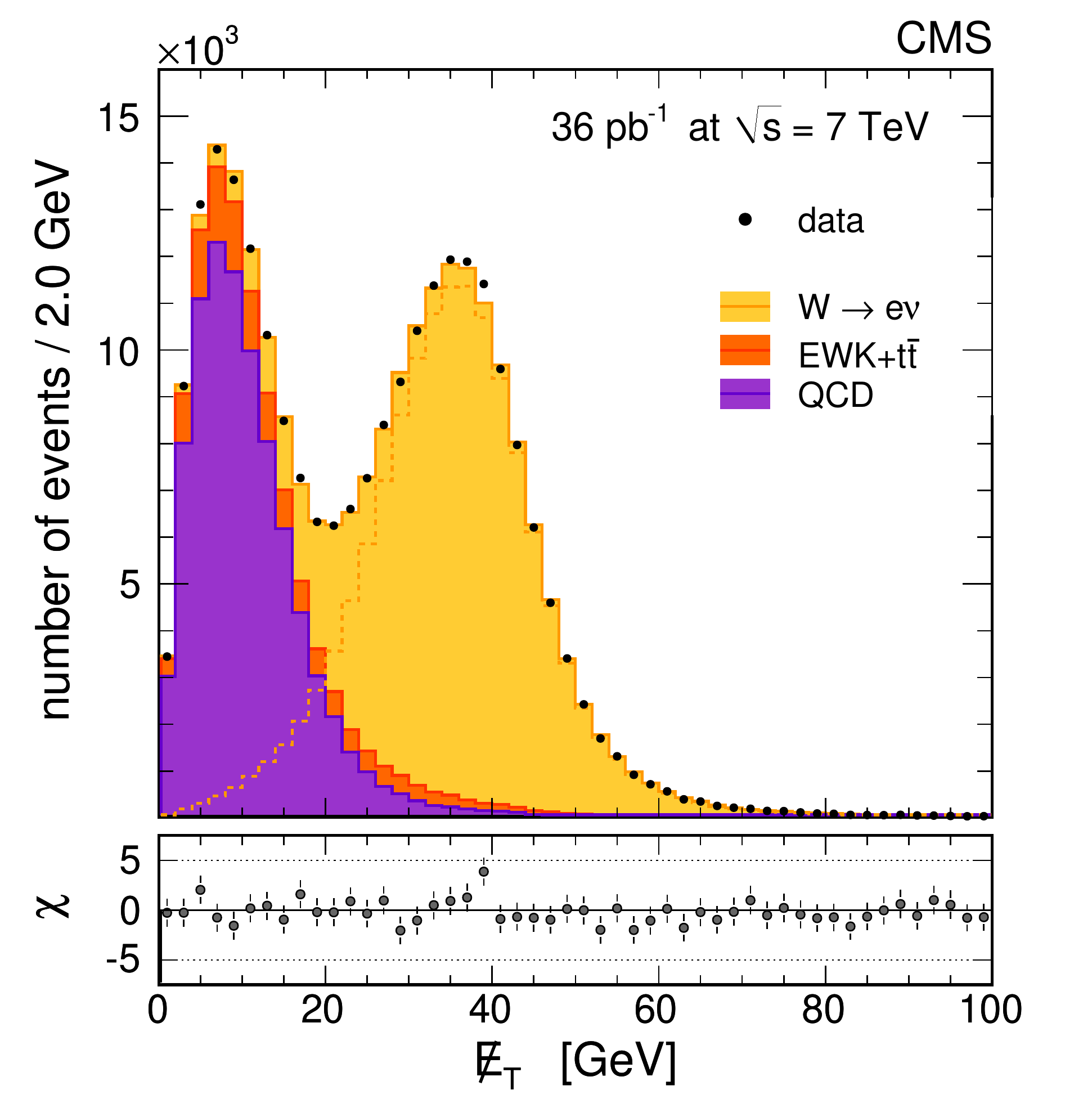}
    \caption{Result of the fixed-shape fit to the $\MET$ distribution for all W candidates.
The points with the error bars represent the data.   Superimposed are the results
of the maximum likelihood fit for QCD background (violet, dark histogram), other backgrounds
(orange, medium histogram), and signal plus  background (yellow, light histogram).
The orange dashed line (left plot) is the fit contribution from signal.
}
    \label{fig:resultAll}
  \end{center}
\end{figure}

The difference in the $\MET$ distributions from the signal and inverted selections
is found to be predominantly due to two effects, which can be reduced by applying corrections.
The first effect is due to a large difference in the distribution of the output of a multivariate 
analysis (MVA) used for electron identification in the PF algorithm, between the selected events 
and the control sample. The value of the MVA output determines whether an electron candidate 
is treated by the PF algorithm as a genuine electron, or as a superposition of a charged pion and 
a photon, with track momentum and cluster energy each contributing separately to $\MET$. The control 
sample contains a higher fraction of electron candidates in the latter category, resulting in a bias on 
the $\MET$ shape. A correction is derived to account for this.
The second effect comes from the signal 
contamination in the control sample. The size of the contamination (1.17$\%$) is 
measured from data, using the \TNP technique with $\Zee$  
events, by measuring the efficiency for a signal electron to pass the 
control sample selection.

The results of the inclusive fit to the $\MET$ distribution with the fixed QCD background shape
are shown in Fig.~\ref{fig:resultAll}; the only free parameters in the extended 
maximum likelihood fit are the QCD and signal yields.
By applying this second method the following yields are obtained:
$135\,982 \pm 388$ (stat.) for the inclusive sample, $81\,286 \pm 302$ (stat.) for the $\Wpen$ sample, and
$54\,703 \pm 249$ (stat.) for the $\Wmen$ sample.
The ratios of the inclusive, $\Wpen$, and $\Wmen$ yields between this 
method and the parameterized
QCD shape method are $0.997 \pm 0.005$, $0.997 \pm 0.005$, and $0.999 \pm 0.005$, respectively, 
considering only the uncorrelated systematic uncertainties between the two methods.

\label{sec:e-Wsigextr-ABCDE}

\subsubsection{The ABCD Method}

In this method the data are divided into four categories defined by boundaries on $\MET$ and the relative tracker 
isolation, $\ITRK/\ET$, of the electron candidate. The boundaries of the regions are chosen to minimize the overall 
statistical and systematic uncertainties on the signal yield.
Values of $\MET$ above and below the boundary of 25 GeV, together with $\ITRK/\ET$ values below 
the boundary of 0.04, define the regions A and B, respectively.
\begin{figure}[htbp]
\begin{center}
\includegraphics[width=0.35\textwidth]{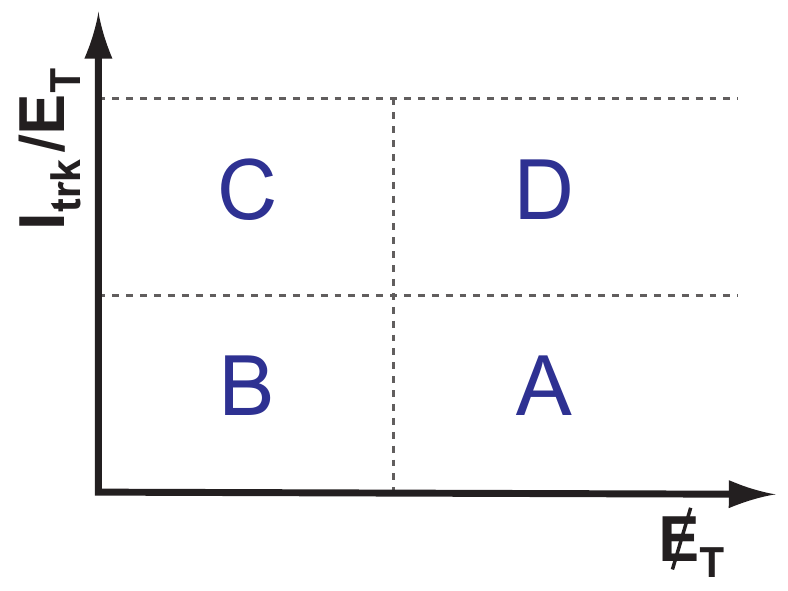}
\caption{The arrangement of the four categories of events used in the ABCD method. The vertical 
scale indicates increasing values of relative track isolation $\ITRK/\ET$
and the horizontal scale indicates increasing $\MET$.
}
\label{fig:abcde}
\end{center}
\end{figure}
Similarly, the regions above and below the $\MET$ boundary for $\ITRK/\ET$ values above 0.04, 
but below an upper $\ITRK/\ET$ bound of 0.2 (0.1) for electrons in the EB (EE), 
define the regions D and C, respectively. There is no upper bound for the $\MET$ values.
The different regions are shown graphically in Fig.~\ref{fig:abcde}, with region A 
having the greatest signal purity.
Combined regions are referred to as 'AB' (for A and B), for example.
The extracted signal corresponds to the entire ABCD region.

A system of equations is constructed relating the numbers of observed data events in each of the 
four regions to the numbers of background and signal events, with several parameters to be 
determined from auxiliary measurements or simulations.
In this formulation, two parameters, $f_\mathrm{A}$ and $f_\mathrm{D}$, relate to the QCD 
backgrounds and are defined as the ratios of events with a fake electron candidate in the A and D 
regions to the number in the AB and CD regions, respectively. The two parameters represent the 
efficiency with which misidentified electrons pass the boundary on $\MET$ dividing
AD from BC. If the efficiency for passing the $\MET$ boundary is largely independent of the choice of
the boundaries on $\ITRK/\ET$, then these two parameters will be approximately equal. Assuming $f_\mathrm{A}=f_\mathrm{D}$ holds exactly 
leads to a simplification of the system of equations such that all direct dependence of the signal
extraction on parameters related to the QCD backgrounds is eliminated. For this idealized case there would be
no uncertainty on the extracted signal yield arising from modeling of QCD backgrounds. Detailed studies of the
data suggest this assumption holds to a good degree. A residual bias in the extracted signal arising from this
assumption is estimated directly from the data by studying a control sample 
obtained with inverted quality requirements on the electron candidate, 
and an appropriate small correction to the yield is applied (${\approx}0.37\%$).
A systematic uncertainty on the signal yield is derived from the uncertainty on this bias correction.
This contribution is small and is dominated by the uncertainty on signal
contamination in the control sample.

Three other important parameters relate to signal efficiencies: $\epsilon_\mathrm{A}$ 
and $\epsilon_\mathrm{D}$, which
are the efficiencies for signal events in the AB and CD regions, respectively, to pass the 
$\MET$ boundary, and $\epsilon_\mathrm{P}$, which is the efficiency for the electron candidate of a signal event 
to pass the boundary on relative track isolation dividing the AB region from the CD region under the 
condition that this electron already lies in the ABCD region. 
The first two of these, $\epsilon_\mathrm{A}$ and $\epsilon_\mathrm{D}$, are
estimated from models of the $\MET$ in signal events using the 
methods described in Section~\ref{sec:WsignalMETtemplate}.
The third parameter, $\epsilon_\mathrm{P}$, is measured from data using the \TNP method, described in
Section~\ref{sec:ELEefficiencies}, and is one of the dominant sources of 
uncertainty on the $\Wo$ boson yield before
considering the final acceptance corrections.

Electroweak background contributions are estimated from MC samples
with an overall normalization scaled through an iterative method with the signal yield. 

The extracted yield with respect to the choice of boundaries in relative track isolation and $\MET$ is 
sensitive to biases in $\epsilon_\mathrm{P}$ and the QCD electron misidentification rate 
bias correction described above, respectively. 
The yield is very stable with respect to small changes in these selections, 
giving confidence that these 
important sources of systematic uncertainty are small.

The following signal yields are obtained:
$136\,003 \pm 498\,\mathrm{(stat.)}$ for the inclusive sample, $81\,525 \pm 385\,\mathrm{(stat.)}$ 
for the $\Wpen$ sample, and $54\,356 \pm 315\,\mathrm{(stat.)}$ for the $\Wmen$ sample.
The ratios of the inclusive, $\Wpen$, and $\Wmen$ yields between this method and the parameterized
QCD shape are $0.998 \pm 0.007$, $0.999 \pm 0.007$, and $0.993 \pm 0.007$, respectively, considering
only the uncorrelated systematic uncertainties between the two methods.

The results of the three signal extraction methods are summarised in Table~\ref{tab:WsignalCollection}.

\begin{table}[htbp] %
\begin{center}
   \caption[.]{ \label{tab:WsignalCollection}
Comparison of $\Wen$ signal extraction methods. The signal yield of each method is presented together with
its statistical uncertainty.
For the fixed shape and the ABCD methods, the ratios of the signal yields with the analytical function method
are also shown taking into account only the uncorrelated systematics between the methods used in the ratios.}
\begin {tabular} {|l|l|c|c|c|}
\hline
\multicolumn{2}{|l|}{Source}        & $\Wen$           & $\Wpen$           & $\Wmen$            \\
\hline\hline
Analytical fun. &yield    & $\WEIYIELD$      &  $\WEPYIELD$      &  $\WEMYIELD$      \\
\hline
\multirow{2}{*}{Fixed shape} & yield                         & $\WEIftYIELD$    &  $\WEPftYIELD$    &  $\WEMftYIELD$      \\
 & ratio & $\rWEIftYIELD$   &  $\rWEPftYIELD$   &  $\rWEMftYIELD$      \\
\hline
\multirow{2}{*}{ABCD} & yield                                & $\WEIabYIELD$    & $\WEPabYIELD$     & $\WEMabYIELD$    \\
 & ratio       & $\rWEIabYIELD$   & $\rWEPabYIELD$    & $\rWEMabYIELD$    \\
\hline
\end {tabular}
\end{center}
\end{table}

\subsection{\texorpdfstring{Modeling of the QCD Background and $\Wmn$ Signal Yield}{Modeling of the QCD Background the W->mu nu Signal Yield}}
\label{sec:Wmunu}

The $\Wmn$ analysis is performed using fixed distributions for the $\MET$ shapes
obtained from data for the QCD background component and from simulations,
after applying proper corrections, for the signal and the remaining background components.

Different approaches to signal extraction are considered for $\Wmn$, as for $\Wen$.
The alternative methods do not demonstrate better performance than the use of fixed shapes
in the W signal fit. Given the lower backgrounds in the muon channel with respect to the
electron channel, the alternative strategies are not pursued at the same level of detail
as in the electron case.

The $\MET$ shape of the QCD background component is obtained from a high-purity QCD sample
of events that pass the signal selection, except that the isolation requirement
is inverted and set to $\IRelComb > 0.2$ (Fig.~\ref{figure:Wmunu_iso}).

\begin{figure}[htbp] {\centering
    \includegraphics[width=7.5cm]{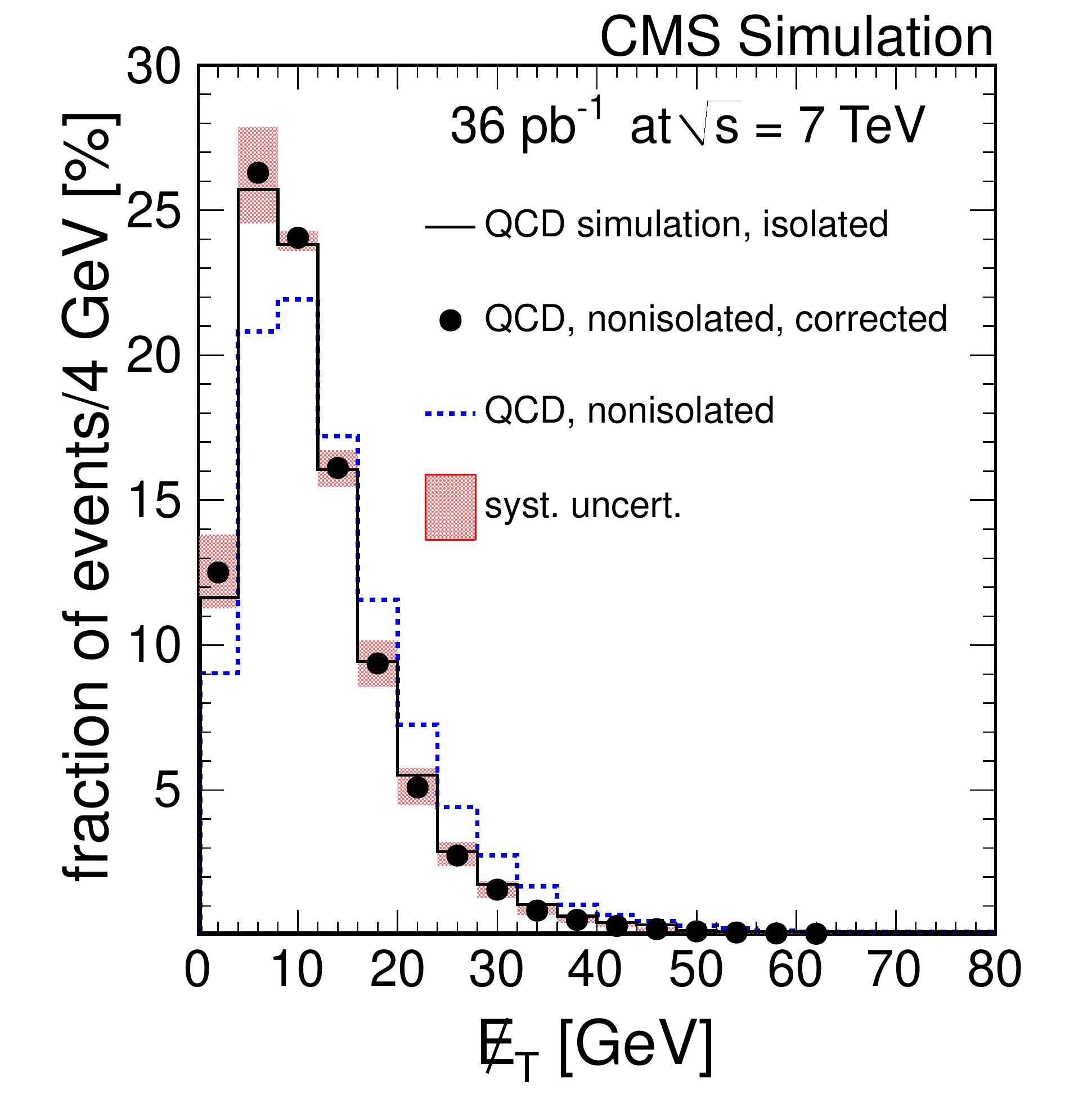}
    \includegraphics[width=7.5cm]{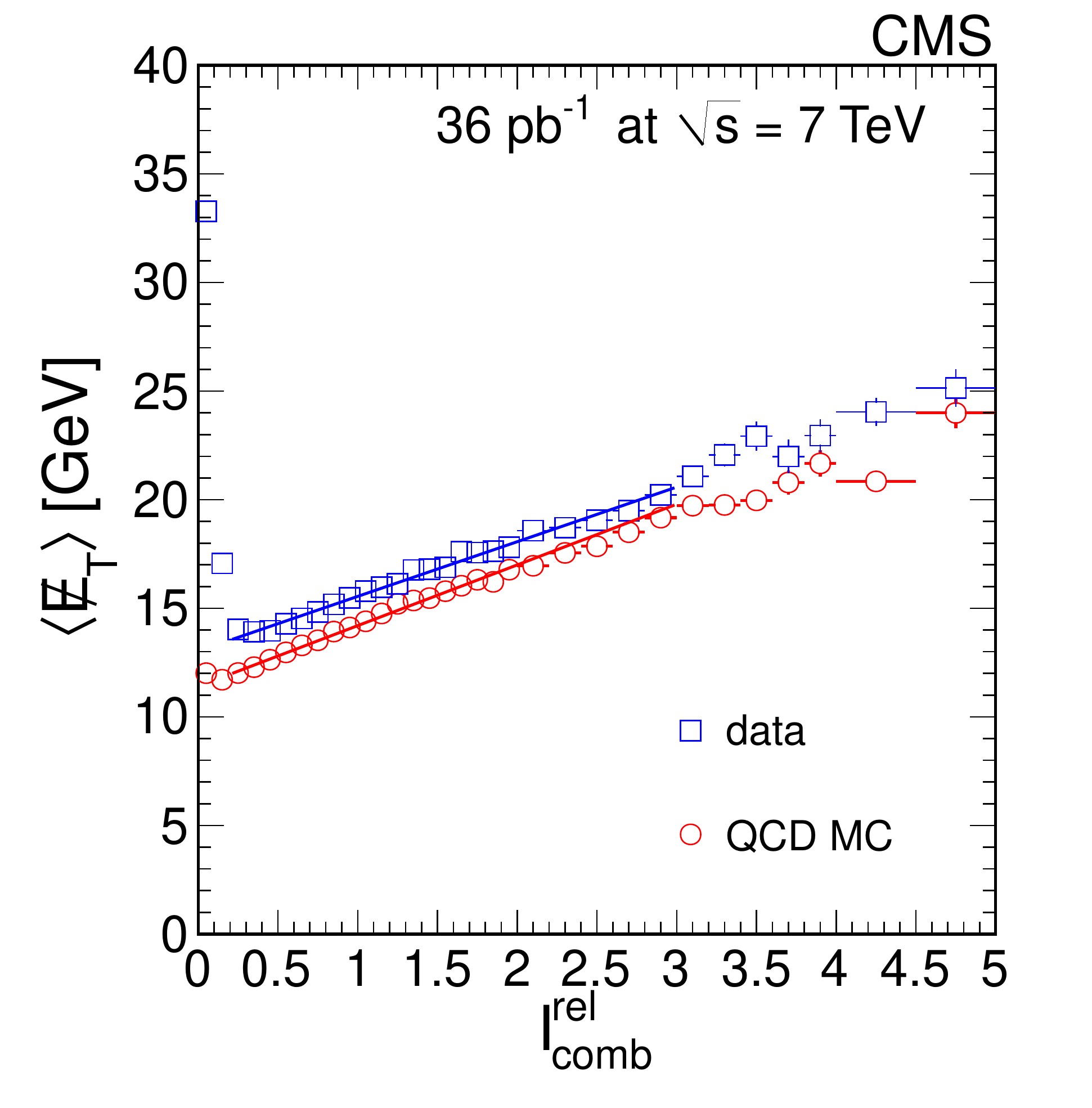}
    \caption{Left: distribution of the corrected $\MET$ for selected events
with a non isolated muon (black points) superimposed
on the distribution of uncorrected $\MET$ for the same events (blue, dashed line) and
$\MET$ for events with an isolated muon (black, solid histogram). All distributions are from simulated QCD events.
The shaded area represents the systematic uncertainty due to corrections
with factors $\alpha \pm \Delta \alpha$, for $\Delta \alpha = 0.08$.
Right: distribution of the average $\MET$ versus $\IRelComb$
for simulated QCD events (red circles) and
for data (blue squares).
The high values of $\MET$ in the first two bins in $\IRelComb$
are due to the presence of the W signal events. The superimposed lines are linear fits
in the range $[0.2, 3.0]$ of $\IRelComb$.
    \label{figure:Wmunu_QCD}}
}
\end{figure}

Simulation studies indicate that this distribution does not accurately
reproduce the $\MET$ shape when muon isolation is required.
This is shown in Fig.~\ref{figure:Wmunu_QCD} (left),
where the solid line represents the shape for events with an isolated muon and the dashed line
the shape obtained by inverting the isolation requirement. 

A positive correlation between the isolation variable $\IRelComb$ and $\MET$ is shown
in Fig.~\ref{figure:Wmunu_QCD} (right, red open circles). This behavior can be
parameterized in terms of a linear function
$\MET\propto (1+\alpha\,\IRelComb)$, as shown in the same figure.
A compensation for the correlation is subsequently made
by applying a correction of the kind of $\MET^\prime = \MET/(1+\alpha\,\IRelComb)$ to
the events selected by inverting the isolation requirement and a new corrected shape is obtained.
The agreement of this new shape (black points in Fig.~\ref{figure:Wmunu_QCD}, left)
with the prediction from events with an isolated muon is considerably improved.
It is also observed that a maximal variation in the correction factor of $\Delta \alpha =  0.08$ successfully
covers the simulation prediction for events with an isolated muon over the whole $\MET$ interval (shaded area in Fig.~\ref{figure:Wmunu_QCD}, left).
\begin{figure}[htbp]
{\centering
    \includegraphics[width=7.5cm]{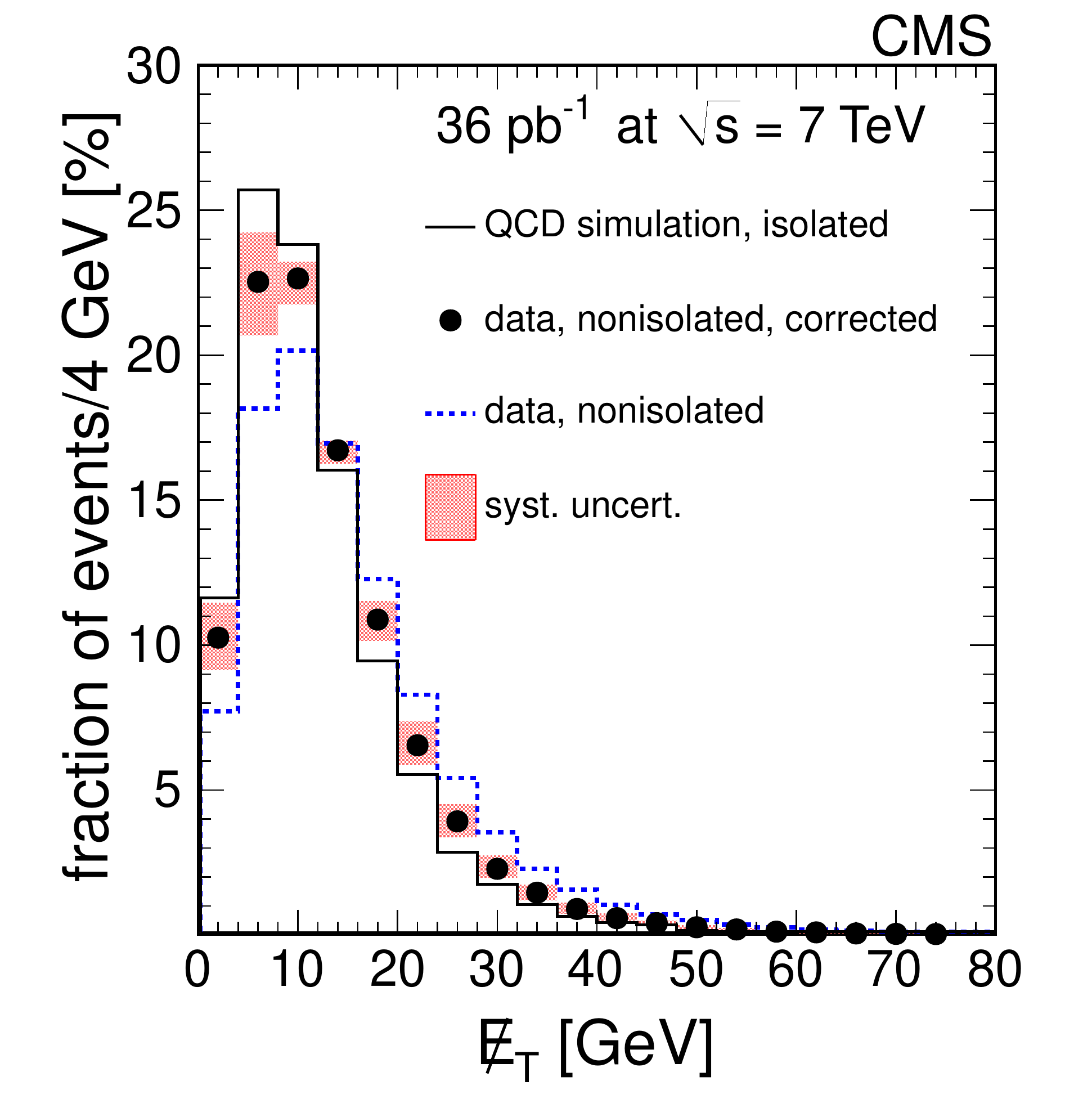}
    \caption{
Distribution of the corrected $\MET$ for selected events
with a nonisolated muon in data (black points) superimposed
on the uncorrected $\MET$ distributions for data (blue dashed line) and
simulated QCD events (black, solid histogram, same as
the black, solid histogram in Fig.~\ref{figure:Wmunu_QCD}).
The shaded area represents the systematic uncertainty due to corrections with factors
$\alpha \pm \Delta \alpha$ for $\Delta \alpha = 0.08$.
    \label{figure:Wmunu_QCD_data}}
}
\end{figure}

The same positive correlation between $\MET$ and $\IRelComb$ is observed in the data
(blue squares in Fig.~\ref{figure:Wmunu_QCD}, right).
A correction $\MET^\prime = \MET/(1+\alpha\,\IRelComb)$, with $\alpha \approx 0.2$,
was applied.
The shapes obtained in data are shown in Fig.~\ref{figure:Wmunu_QCD_data} where
the uncorrected and corrected data shapes from events selected by inverting the isolation requirement, together with the
simulation expectation for events with an isolated muon, are shown.
The shaded area in Fig.~\ref{figure:Wmunu_QCD_data} is bounded by the two distributions, obtained using two extreme correction
parameters $\alpha \pm \Delta \alpha$, with $\Delta \alpha = 0.08$, as evaluated in simulations.
This area is taken as a systematic uncertainty on the QCD background shape.

Several parameterizations for the correction are considered,
but the impact on the corrected distribution and therefore on the final result is small.
Associated uncertainties on the cross section and ratios are evaluated as the differences between
the fit results obtained with the optimal $\alpha$ value and
two extreme cases, $\alpha \pm \Delta \alpha$.

The following signal yields are obtained: $140\,757 \pm 383$ for the inclusive sample,
$56\,666\pm240$ for the $\Wmmn$ sample, and $84\,091\pm291$ for the $\Wpmn$ sample.

\begin{figure}[!ht] {\centering
   \includegraphics[width=7cm]{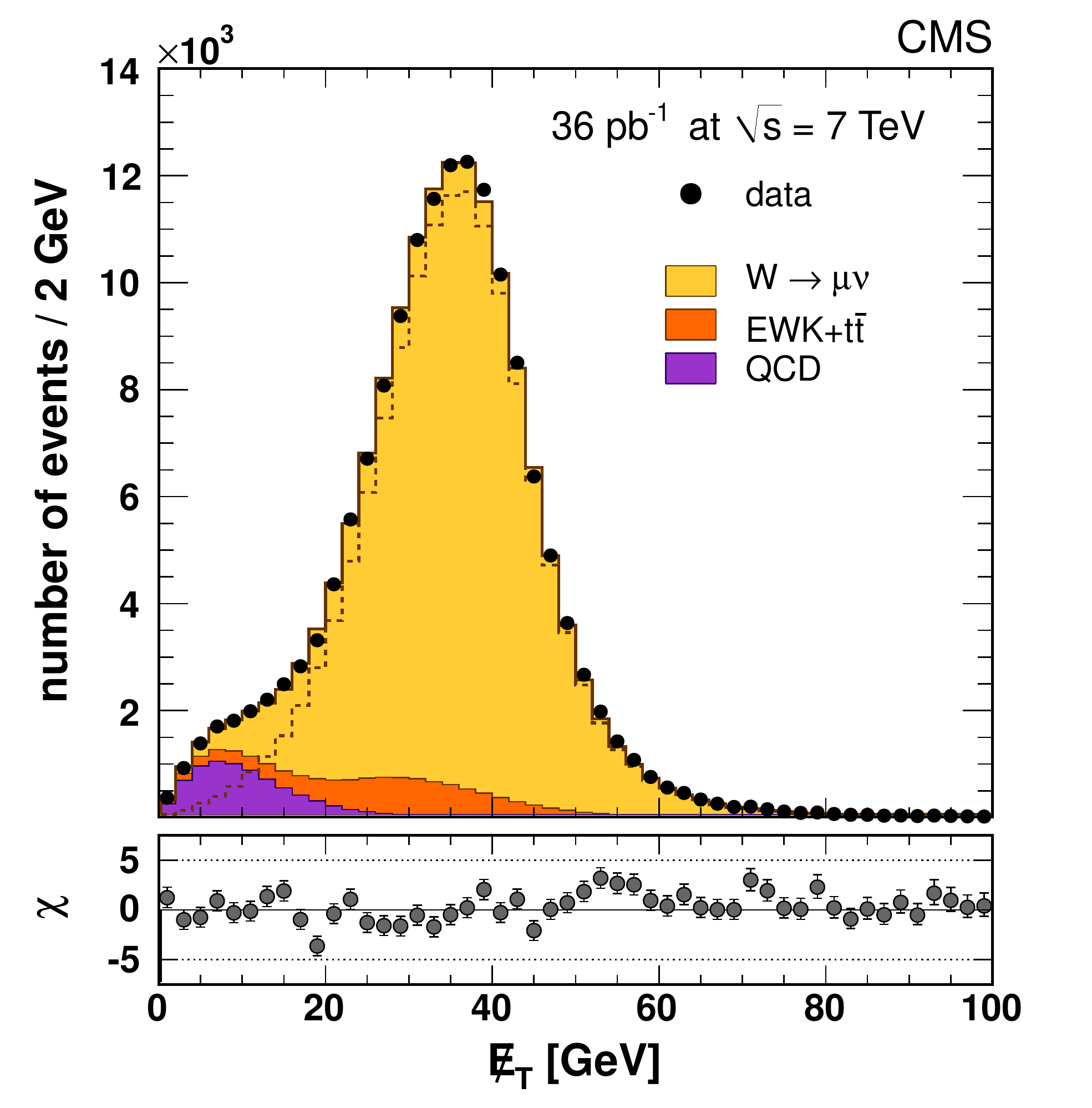}
   \includegraphics[width=7cm]{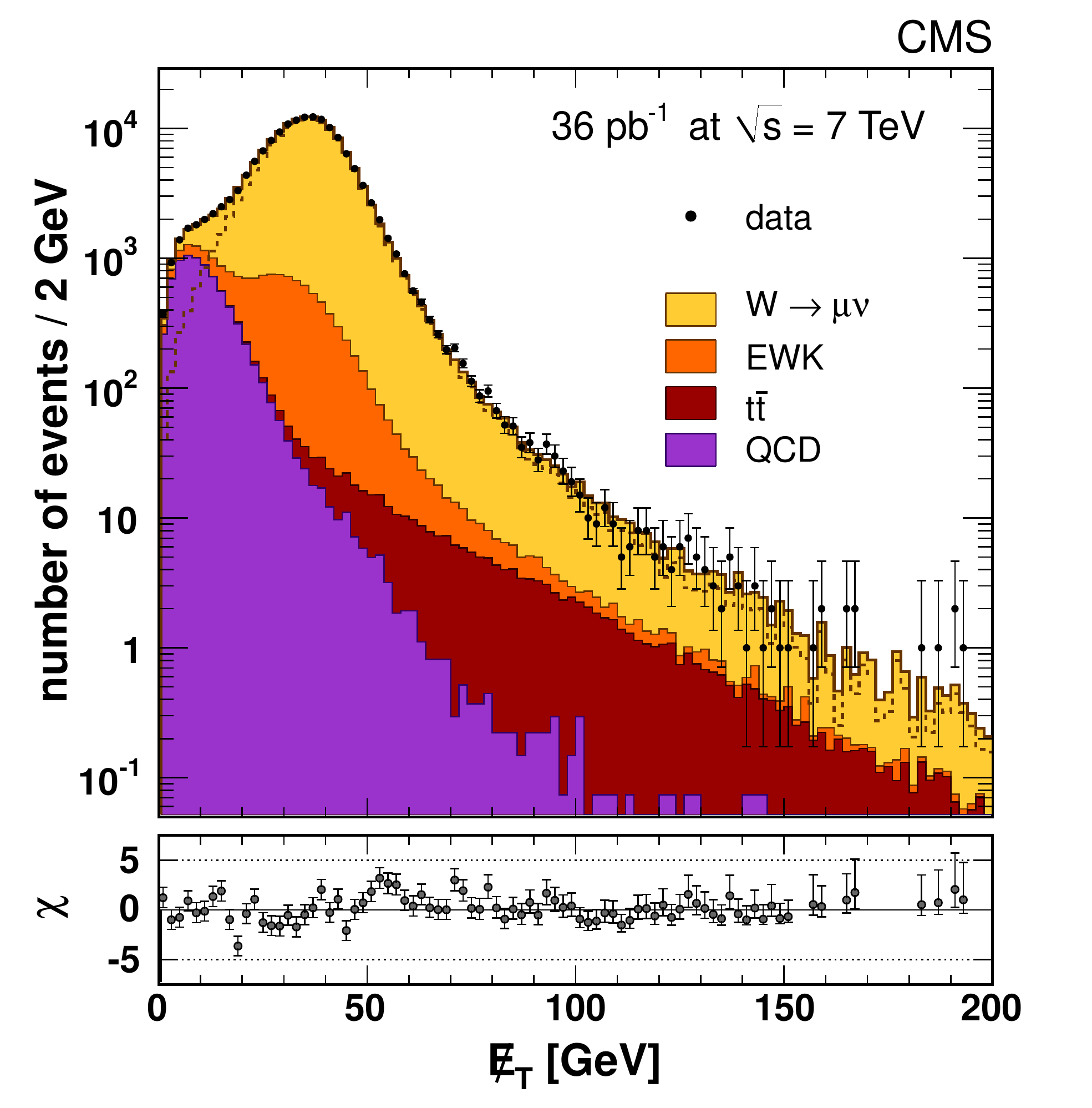}
     \caption{
The $\MET$ distribution for the selected $\Wmn$ candidates on
a linear scale (left) and on a logarithmic scale (right).
The points with the error bars represent the data. Superimposed are the
contributions obtained with the fit
for QCD background (violet, dark histogram), all other backgrounds
(orange, medium histogram), and signal plus  background (yellow, light histogram).
The black dashed line is the fitted signal contribution.
     \label{figure:Wmunu_exp_fit}}
}
 \end{figure}

\begin{figure}[!ht] {\centering
   \includegraphics[width=7cm]{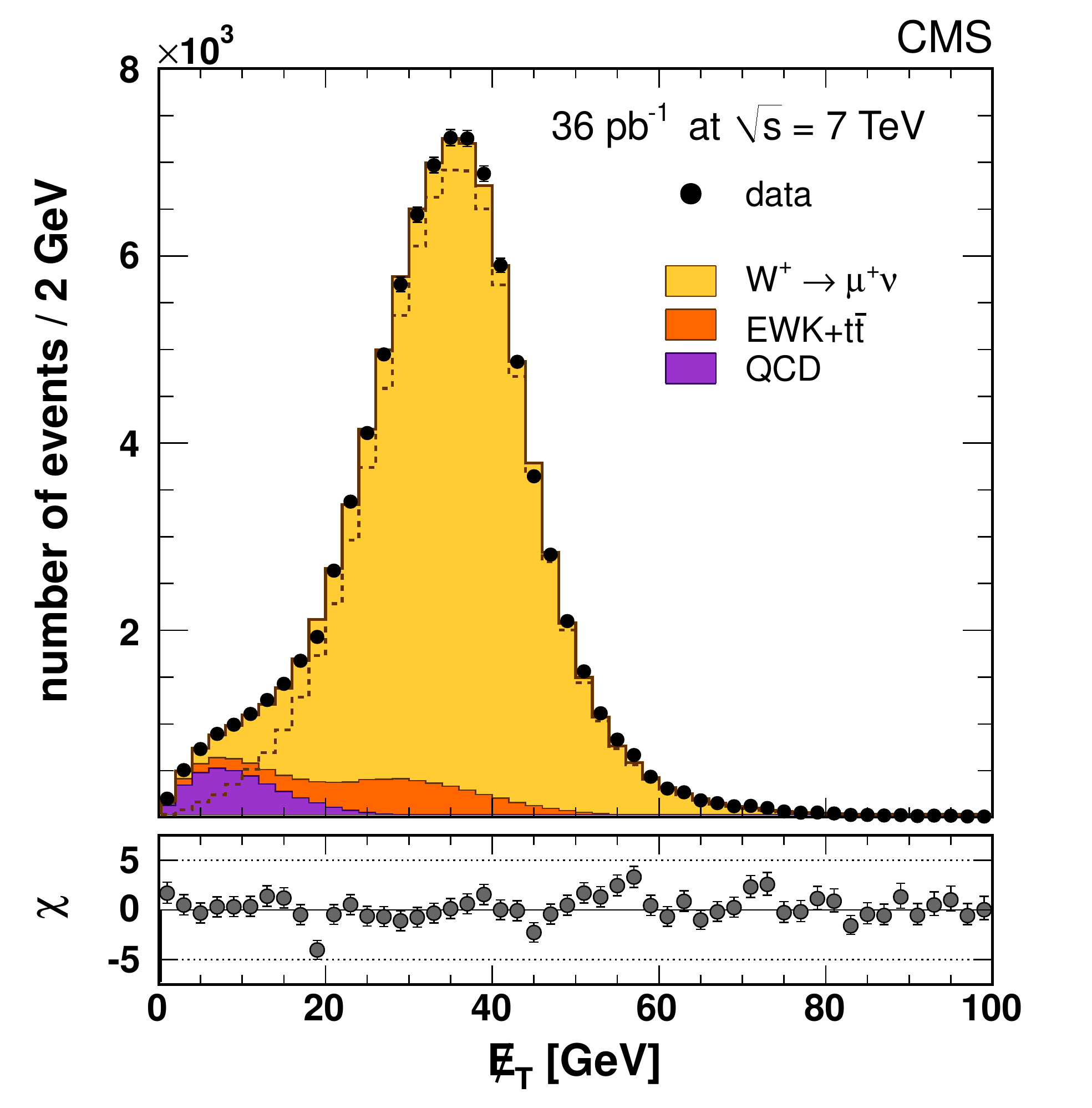}
   \includegraphics[width=7cm]{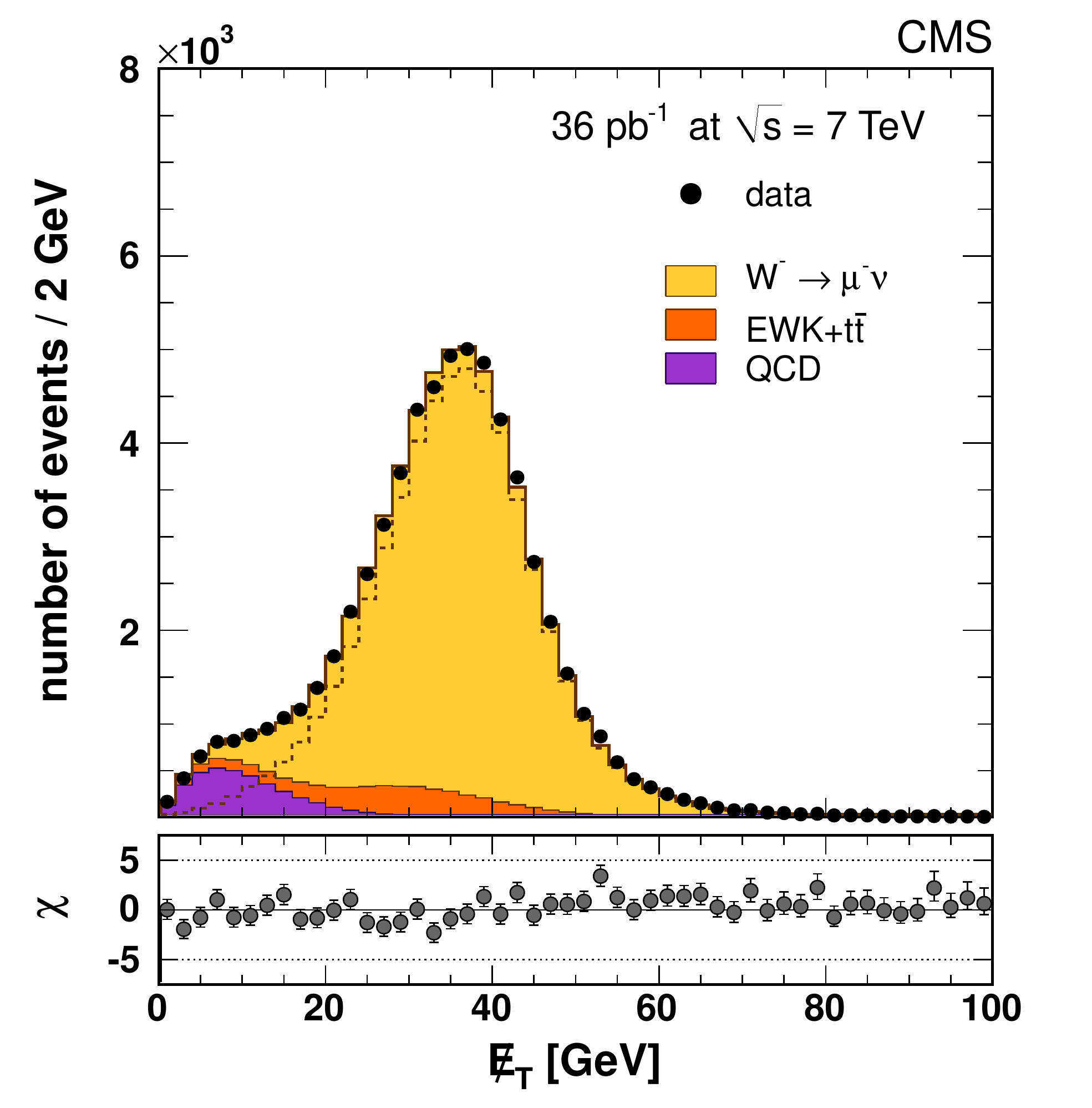}
    \caption{
The $\MET$ distributions for the selected W$^+$ (left) and W$^-$ (right) candidates.
The points with the error bars represent the data. Superimposed are the contributions
obtained with the fit for QCD background (violet, dark histogram), all other backgrounds
(orange, medium histogram), and signal plus background (yellow, light histogram).
The black dashed line is the fitted signal contribution.
    \label{figure:Wmn_PlusMinus}}
}
\end{figure}

\begin{figure}[!ht] {\centering
   \includegraphics[width=7cm]{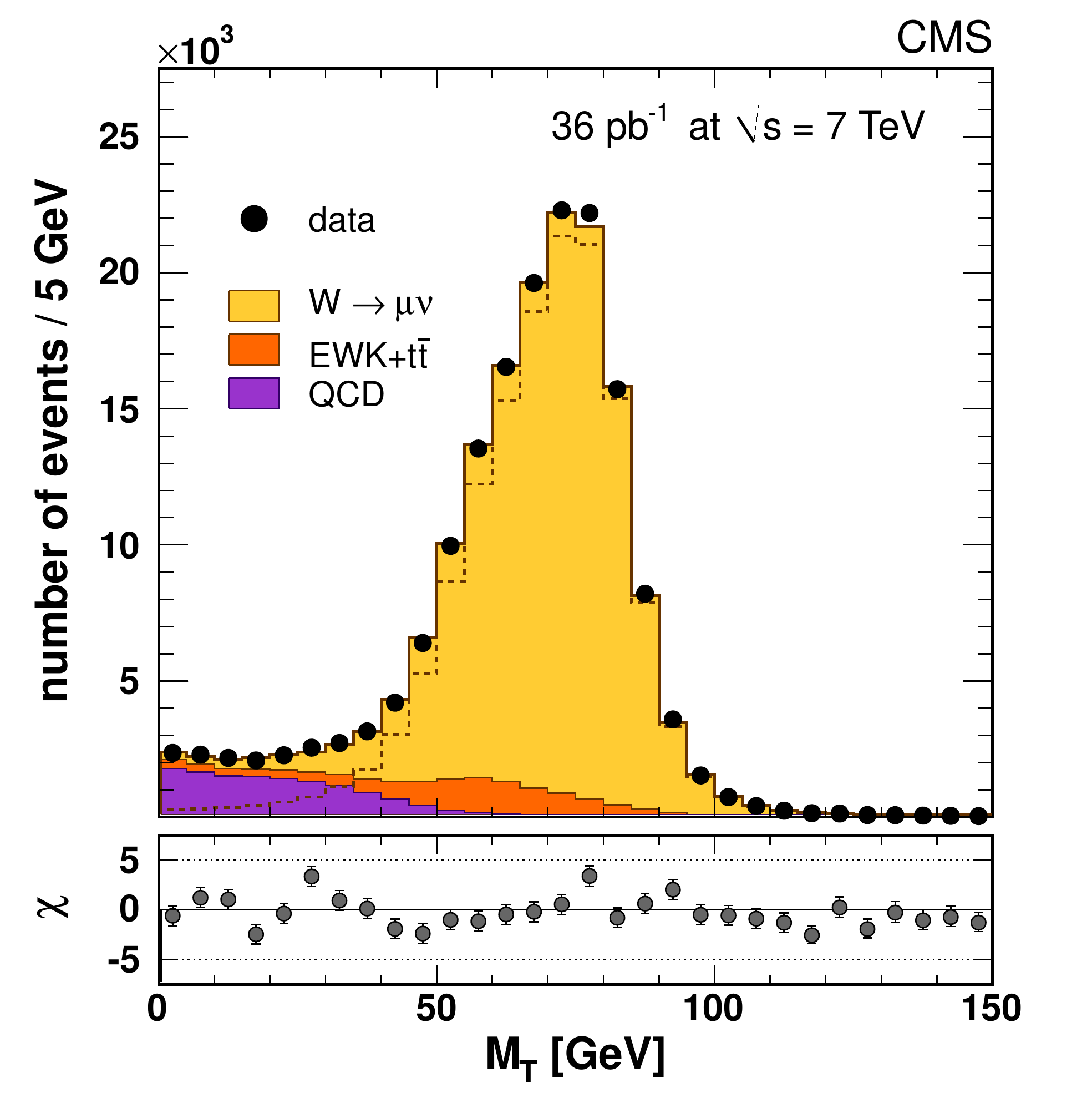}
   \includegraphics[width=7cm]{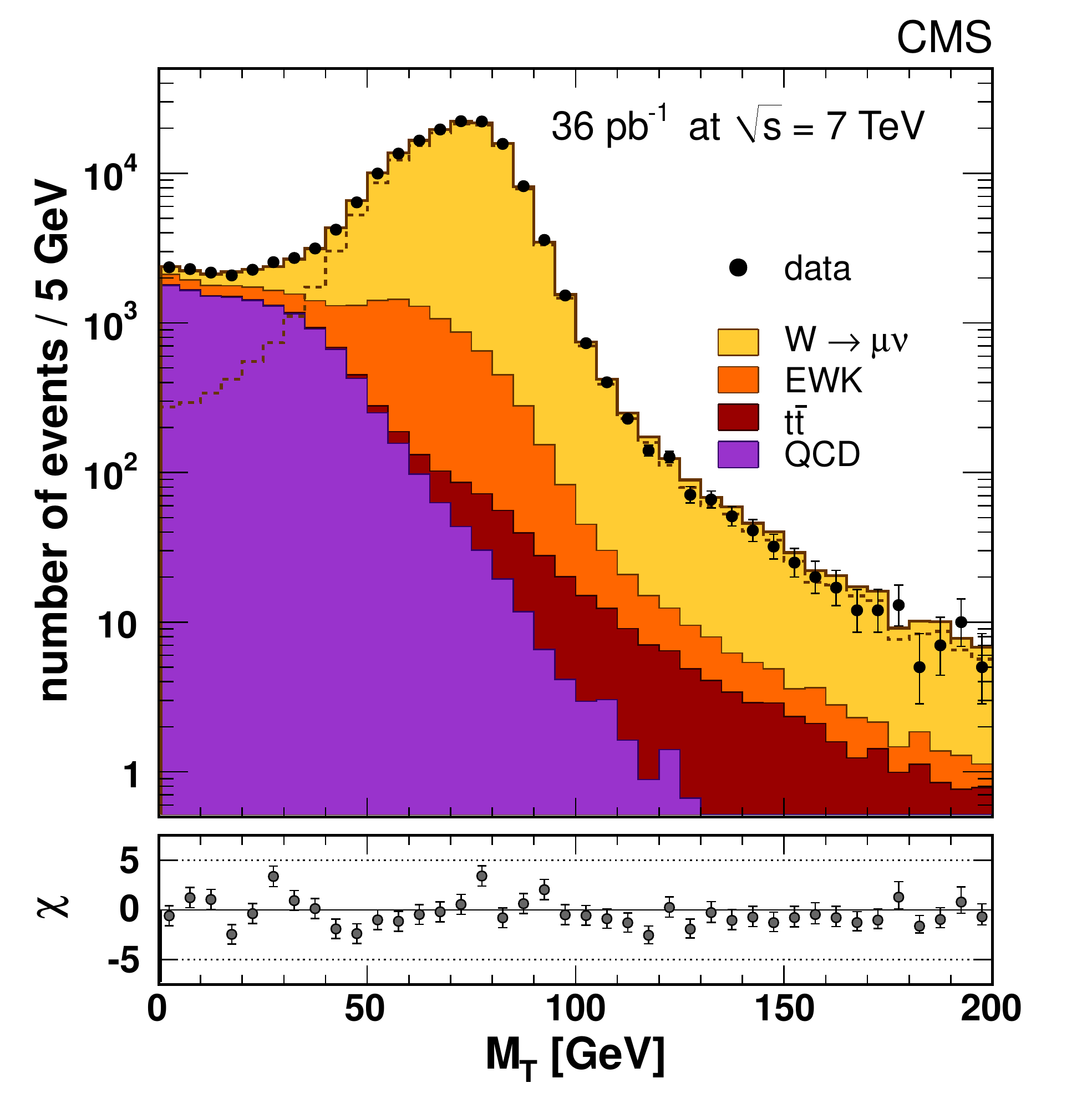}
     \caption{
The $\MT$ distribution for the selected $\Wmn$ candidates on
a linear scale (left) and on a logarithmic scale (right).
The points with the error bars represent the data. Superimposed are the
contributions obtained with the fit
for QCD background (violet, dark histogram), all other backgrounds
(orange, medium histogram), and signal plus background (yellow, light histogram).
The black dashed line is the fitted signal contribution.
     \label{figure:Wmunu_exp_fit_mt}}
}
 \end{figure}

The $\MET$ distributions are presented in
Fig.~\ref{figure:Wmunu_exp_fit} (full sample) and Fig.~\ref{figure:Wmn_PlusMinus}
 (samples selected by the muon charge) superimposed on the individual fitted
contributions of the W signal and the EWK and QCD backgrounds.
Figures~\ref{figure:Wmunu_exp_fit} and~\ref{figure:Wmn_PlusMinus}
show the $\MET$ distributions for data and fitted signal, plus background components.
 Figure~\ref{figure:Wmunu_exp_fit_mt}
shows the $\MT$ distributions for data and signal, plus background components, fitted
from the $\MET$ spectra.

\section{\texorpdfstring{The $\Zll$ Signal Extraction}{The Z-> ll Signal Extraction}}
\label{sec:ZsignalExtraction}

The $\Zll$ yield can be obtained by counting the
number of selected candidates after subtracting the residual background.
The $\Zll$ yield and lepton efficiencies are also determined using a simultaneous fit
to the invariant mass spectra of multiple dilepton categories.
The simultaneous fit deals correctly with correlations
in determining the lepton efficiencies and the $\Zo$ yield
from the same sample.
The Z yield extracted in this way does not need to be corrected for efficiency effects
in order to determine the cross section, and
the statistical uncertainty on the $\Zo$ yield absorbs the uncertainties
on the determination of lepton efficiencies that would be propagated as systematic
uncertainties in the counting analysis.
Both methods were performed for the $\Zee$ analysis, while only
the simultaneous fit was used for the $\Zmm$ analysis
after taking into account the results from the previous studies~\cite{WZCMS:2010}.

\subsection{EWK and QCD Backgrounds}
\label{sec:bkgZll}

For the $\Zee$ analysis the background contributions from EWK processes $\Ztt$, $\ttbar$,
and diboson production are estimated from the yields of events selected in NLO MC samples
normalized to the NNLO cross sections and scaled to the considered integrated luminosity.
They amount to \ZEEEWKBKG events, where the uncertainty combines the NNLO
and luminosity uncertainties. Data are used to estimate the background
originating from W+jets, $\gamma$+jets, and QCD multijet events where the
selected electrons come from misidentified jets or photons
(referred to as 'QCD background').
This background contribution is estimated using the distribution
of the relative track isolation, $\ITRK/\Et$,
and amounts to $4.9 \pm 8.4\, \textrm{(stat.)} \pm 8.4\, \textrm{(syst.)}$ events.
As a cross-check, the ``same-sign/opposite-sign'' method was used,
which is based on the signs of the charges of the two electron candidates, the measured
charge misidentification for electrons that pass the nominal selection criteria, and the
hypothesis that the QCD background is charge-symmetric.
The QCD background estimate with this method is $59 \pm 17
\textrm{(stat.)} \pm 160\, \textrm{(syst.)}$ events.
The two methods are consistent with the presence of negligible QCD background in our sample.

Backgrounds in the $\Zmm$ analysis containing two isolated global muons
have been estimated with simulations to be very small.
This category of dimuon events is defined as the ``golden'' category.
The simulation prediction of the smallness of the $\ttbar$ and QCD backgrounds was validated with data.
First, the selected dimuon sample was enriched with $\ttbar$ events by applying
a requirement on $\MET$, because of the presence of neutrinos in $\ttbar$ events,
and an agreement between data and the simulation prediction was found
with the dimuon invariant mass requirement inverted,
where the residual Z signal is negligible.
The QCD component has been checked using the same-sign dimuon events and dimuon events with
both muons failing the isolation requirement, and was found to be in agreement with the simulation predictions.
The conclusion from the maximum amount of measured data-simulation discrepancy
was that the uncertainty in the residual background subtraction
has a negligible effect on the $\Zmm$ measured yield.
The backgrounds to the $\Zmm$ categories having one global and one looser muon
are significantly larger than in the golden category.
Simulation estimates in this case are not used for such backgrounds and
fits to the dimuon invariant mass
distributions are performed including parameterized background components, as
described in Section~\ref{sec:Zmumu}.

Backgrounds estimates in the $\Zee$ and $\Zmm$ analyses are summarized in Table~\ref{tab:ZllBG}.
\begin{table} %
\begin{center}
\caption[.]{\label{tab:ZllBG}
Estimated background-to-signal ratios in the $\Zee$ and $\Zmm$ (only for candidates
in the golden category) channels.
The QCD background for the $\Zee$ channel has been estimated with data,
while all other estimates are based on MC simulations, and their corresponding uncertainties
are statistical only.}
\begin{tabular}{|l|c|c|}
\hline
Processes & $\Zee$ sel. & $\Zmm$ sel. \\
\hline\hline
Diboson production   & $(0.157\pm 0.001)\%$ & $(0.158\pm 0.001)\%$ \\
$\ttbar$             & $(0.117\pm 0.008)\%$ & $(0.141\pm 0.014)\%$ \\
$\Ztt$               & $(0.080\pm 0.006)\%$ & $(0.124\pm 0.005)\%$ \\
W+jets               & $(0.010\pm 0.002)\%$ & $(0.008\pm 0.002)\%$ \\
\hline
Total EWK plus $\ttbar$  & $(0.365\pm 0.010)\%$ & $(0.430\pm 0.015)\%$ \\
\hline
QCD            & $(0.06\pm 0.14)\%$ &  $(0.013\pm 0.001)\%$ \\
\hline
Total background                & $(0.42\pm 0.14)\%$ & $(0.444\pm 0.015)\%$ \\
\hline
\end{tabular}
\end{center}
\end{table}

\subsection{\texorpdfstring{The $\Zee$ Signal Extraction}{The Z->ee Signal Extraction}}

In the following sections the use of a pure $\Zee$ sample
for the determination of the residual energy-scale and resolution corrections is first discussed.
Then the signal extraction with the counting analysis and the
simultaneous fit methods are presented.

\subsubsection{Electron Energy Scale}
\label{sec:e-escale}

The lead tungstate crystals of the ECAL are subject to transparency loss
during irradiation, followed by recovery in periods with no
irradiation. The magnitude of the changes to the energy response is
dependent on instantaneous luminosity and was, at the end of the 2010
data taking period, up to 1$\%$ in the barrel region, and 4$\%$ or more in
parts of the endcap. The changes are monitored continuously by injecting
laser light and recording the response. The corrections derived from
this monitoring are validated by studying the variation of the $\pi^0$ mass
peak as a function of time for different regions of the ECAL (using $\pi^0$ data
collected in a special calibration stream), and by studying the overall $\Zee$ mass peak and width.
With the current corrections, residual variations of the energy scale with time are
at the level of 0.3\% in the barrel and less than 1\% in the endcaps.

\par
The remaining mean scale correction factors to be applied to the data and the
resolution corrections (smearing) to be applied to the simulated sample
are estimated from $\Zee$ events. Invariant mass distributions for electrons
in several $\eta$ bins in the EB and EE are derived
from simulations and compared to data. A simultaneous fit of a Breit--Wigner convolved with a
Crystal-Ball function to each $\Zee$ mass distribution is performed  in order
to determine the energy scale correction factors for the data and the resolution
smearing corrections for the simulated samples. The energy scale correction
factors are below 1$\%$ while the resolution smearing corrections are below 1$\%$
everywhere, with the exception of the transition region between the EB and the EE,
where they reach 2$\%$.
Those corrections are propagated in the analysis and proper systematic uncertainties
for the cross section measurements are estimated as discussed in Section~\ref{subsec:ELEsystematics}.

\par
\subsubsection{Counting Analysis}
After energy scale corrections, applied to electron ECAL clusters before
any threshold requirement, 10 fewer events ($-0.12\%$) were selected compared to the number of selected
events before the application of the energy scale corrections.
This brings the final $\Zee$ sample to $\ZEESAMPLEN$ and, after
background subtraction, the $\Zo$ yield is $\ZEEYIELD$ events.
This yield is used for the cross section estimation.

\par
\begin{figure}
  \begin{center}
   \includegraphics[width=0.48\textwidth]{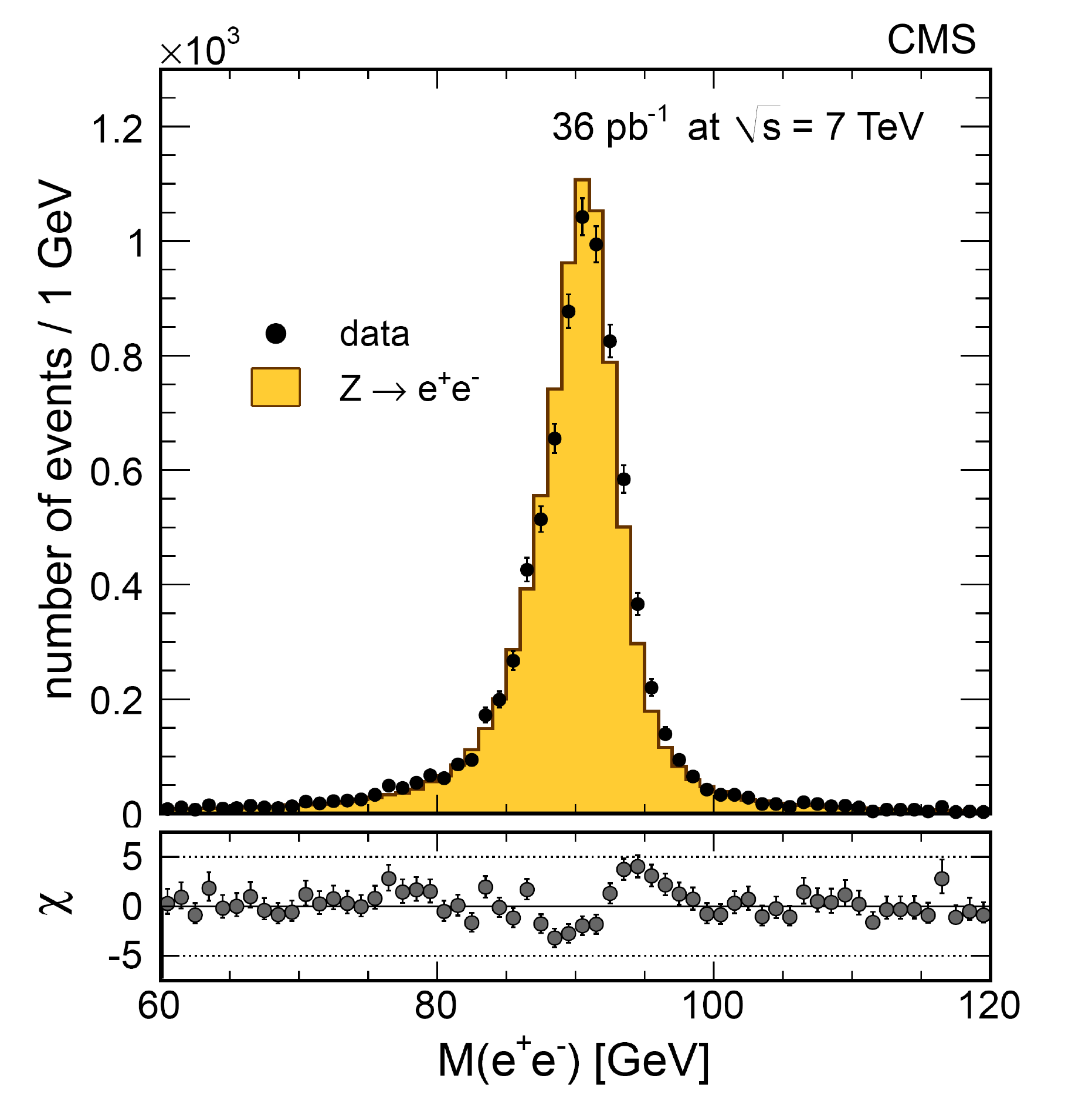}
   \includegraphics[width=0.48\textwidth]{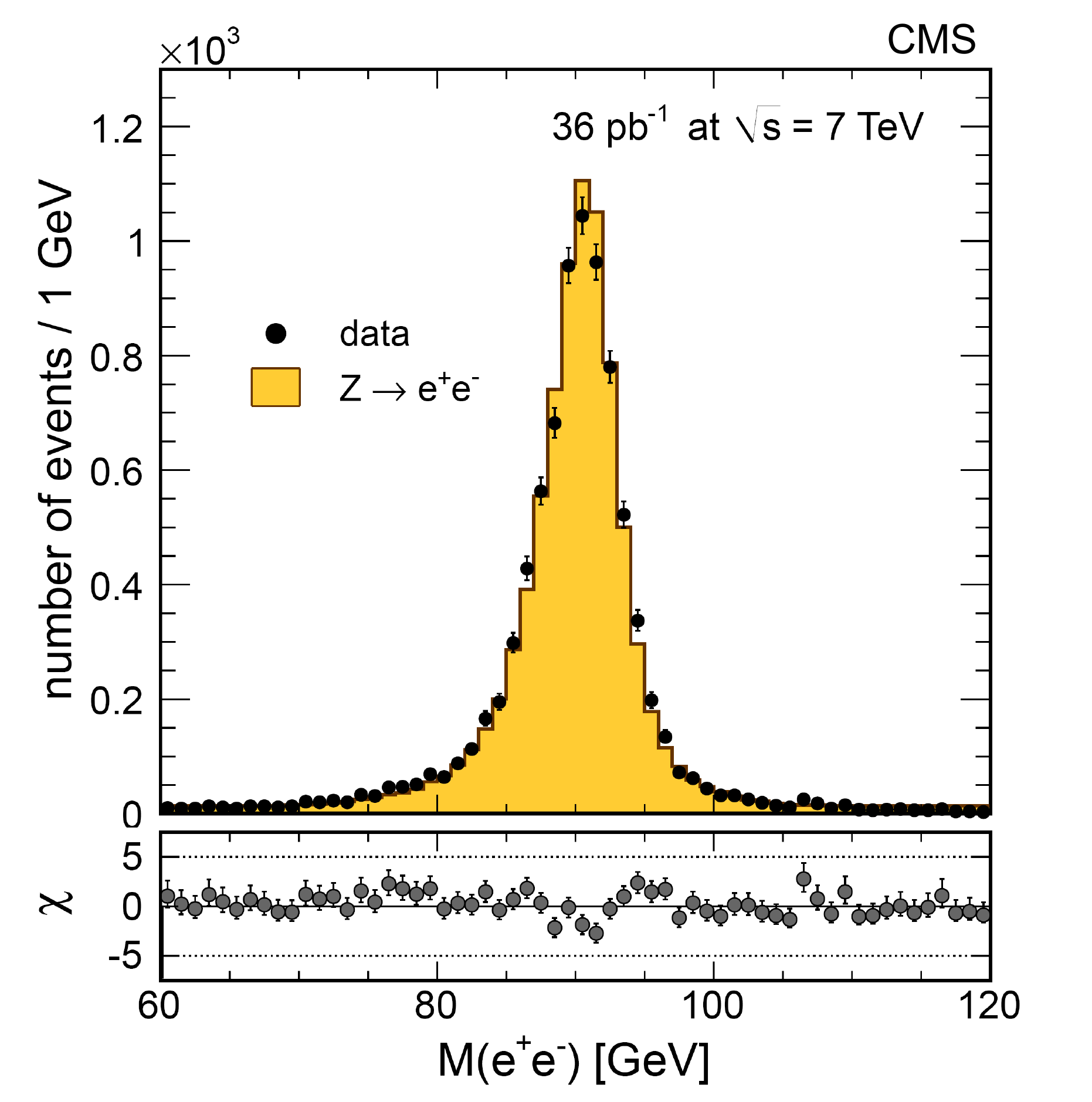}
   \includegraphics[width=0.48\textwidth]{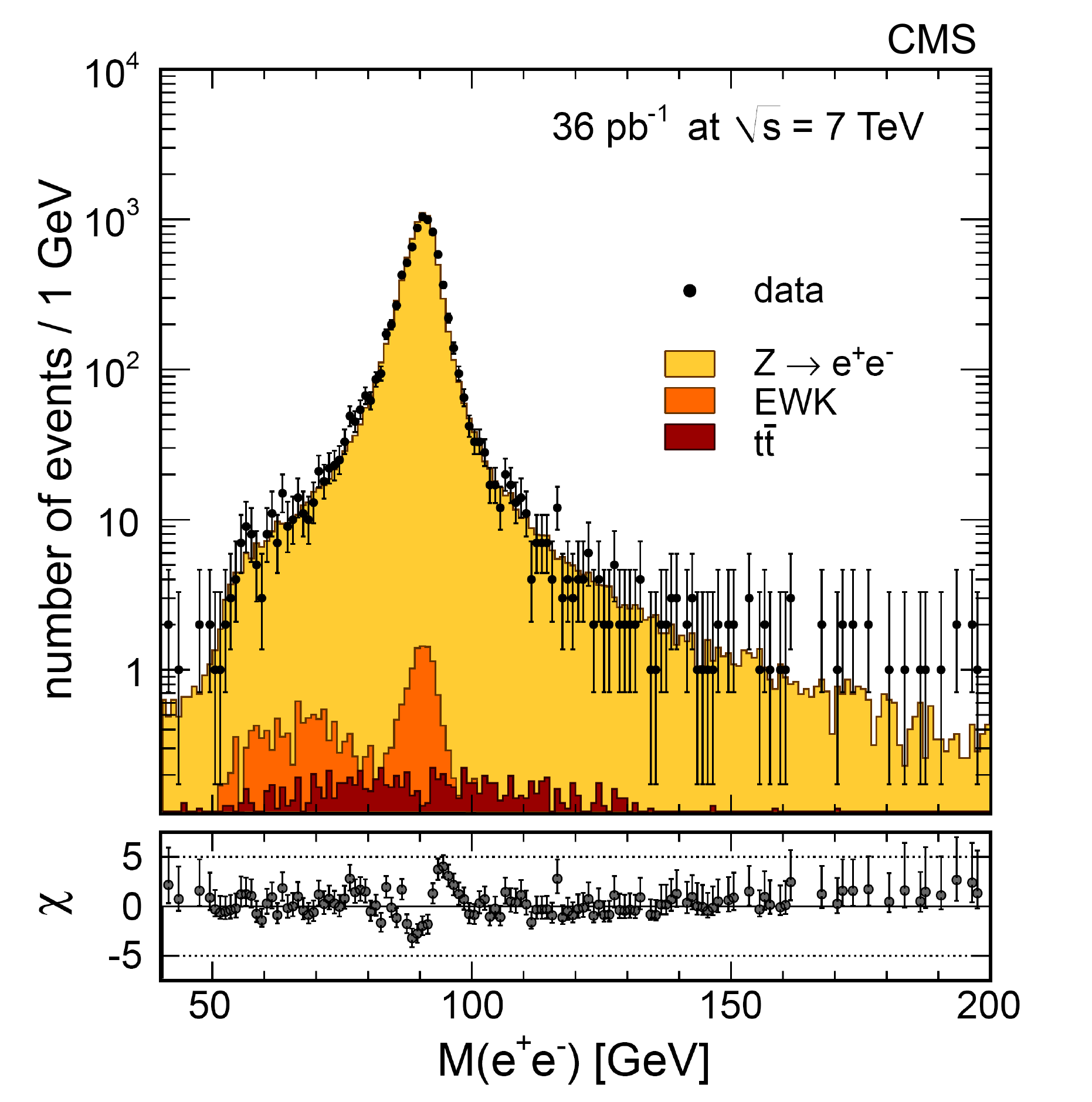}
   \includegraphics[width=0.48\textwidth]{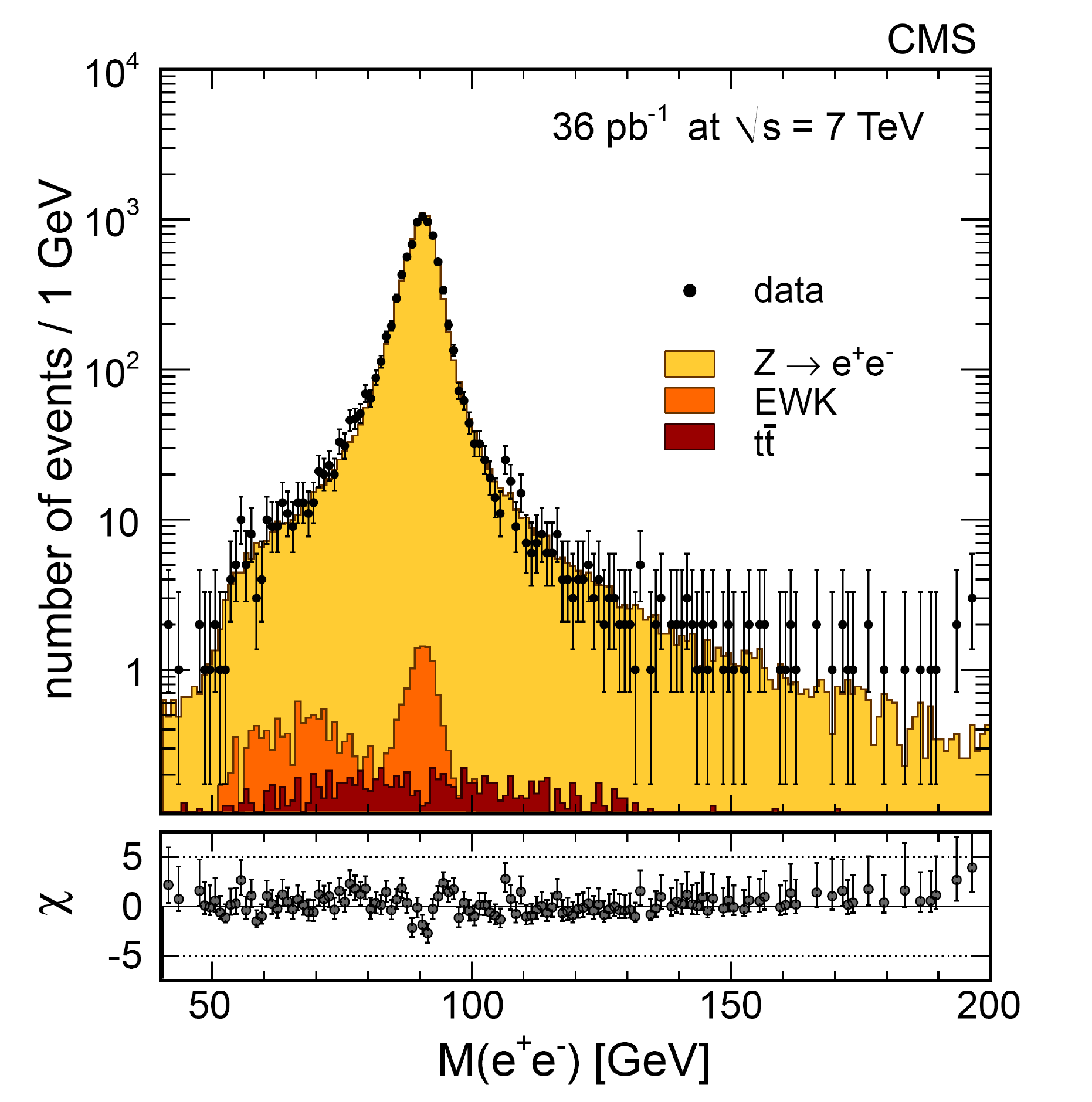}
   \caption{ \label{fig:Zee}
Distributions of the dielectron invariant mass for the selected $\Zee$ candidates on
a linear scale (top) and on a logarithmic scale (bottom) before (left)
and after (right) applying energy-scale correction factors.
The points with the error bars represent the data.
Superimposed are the expected distributions from simulations, normalized
to an integrated luminosity of $36$~pb$^{-1}$. The expected distributions are
the Z signal (yellow, light histogram), other EWK processes (orange, medium histogram),
and $\ttbar$ background (red, dark histogram).
Backgrounds are negligible and cannot be seen on the linear-scale plots.
}
  \end{center}
\end{figure}

The dielectron invariant mass spectra for the selected sample
with the tight selection before and after the application of the corrections are
shown in Fig.~\ref{fig:Zee} along with the predicted distributions.
The data and simulation distributions are normalized to account for the difference in selection
efficiency.

\subsubsection{Simultaneous Fit}

The $\Zo$ event yield and the electron efficiencies can be extracted from
a simultaneous fit. Two categories of events are
considered: events where both electrons satisfy
the tight selection with $\Et>25\GeV$, and events that consist of
one electron with $\Et>25\GeV$ that passes the tight selection, and one
ECAL cluster with $\Et>25\GeV$ that fails
the selection, either at the reconstruction or electron identification level.

In each category, a signal-plus-background function is fitted to the observed mass spectrum.
The signal shape is taken from signal samples simulated with POWHEG at the NLO generator level,
and is convolved with a Crystal-Ball function modified to include an extra
Gaussian on the high end tail with floating mean and width.
In the first category, the nearly vanishing background is fixed to the
value reported in Table~\ref{tab:ZllBG}. In the second
category of events, the background is modeled by an exponential distribution.

The estimated cross section is $988 \pm 10\, \mathrm{(stat.)} \pm 4\mathrm{(syst.)}\,\mathrm{pb}$.
The cross section is in good agreement with the counting analysis estimate of
 $992 \pm 11\, \mathrm{(stat.)}\,\mathrm{pb}$, considering only the statistical uncertainty.
Both techniques give equivalent results. The counting analysis estimate is used for the
cross section measurement in the $\Zee$ channel.

\subsection{\texorpdfstring{The $\Zmm$ Signal Extraction}{The Z->mu mu Signal Extraction}}
\label{sec:Zmumu}

The yield of the $\Zmm$ events is determined from a fit simultaneously with the
average muon reconstruction efficiencies in the tracker and in
the muon detector, the muon trigger efficiency,
as well as the efficiency of the applied isolation requirement.
\Zmm candidates are obtained as pairs of muon candidates of different types
and organized into categories according to different requirements:
\begin{itemize}
\item $\Zmumu$: a pair of isolated global muons, further split into two samples:
\begin{itemize}
\item $\ZmumuTwoHlt$: each muons associated with an HLT trigger muon;
\item $\ZmumuOneHlt$: only one of the two muons associated with an HLT trigger muon;
\end{itemize}
\item $\Zmus$: one isolated global muon and one isolated
  stand-alone muon;
\item $\Zmut$: one isolated global muon and one isolated tracker track;
\item $\ZmumuNonIso$: a pair of global muons, of which one is isolated and the
other is nonisolated.
\end{itemize}

With the exception of the $\ZmumuOneHlt$ category, each global muon must correspond to an HLT trigger muon.
The five categories are explicitly forced to be mutually exclusive in the event
selection: if one event falls into the first category it is excluded from the second;
if it does not fall into the first category and falls into the second, it is excluded
from the third, and so on. In this way non-overlapping, hence statistically
independent, event samples are defined. The expected number of events in which more than
one dimuon combination is selected is almost negligible.
In those few cases all possible combinations are considered.

The five unknowns, the Z yield and four efficiency terms, can be written in terms of
the five signal yields in each category as follows:
\begin{eqnarray}
 \label{eqNmumuTwoHlt}
   \NmumuTwoHlt & = & \NZtomumu \effHlt^2 \effIso^2 \effTrk^2 \effSa^2,  \\
  \label{eqNmumuOneHlt}
   \NmumuOneHlt & = & 2 \NZtomumu \effHlt (1 - \effHlt) \effIso^2 \effTrk^2 \effSa^2,  \\
  \label{eqNmus}
   \Nmus & = & 2 \NZtomumu \effHlt \effIso^2 \effTrk (1 - \effTrk) \effSa^2,  \\
  \label{eqNmut}
   \Nmut & = & 2 \NZtomumu \effHlt \effIso^2 \effTrk^2 \effSa(1 -\effSa), \\
  \label{eqNmumuNonIso}
   \NmumuNonIso & = & 2 \NZtomumu \effHlt^2  \effIso (1 - \effIso)  \effTrk^2 \effSa^2.
\end{eqnarray}
The various efficiency terms in Eqs.~(\ref{eqNmumuTwoHlt}) to~(\ref{eqNmumuNonIso}),
the average efficiencies
of muon reconstruction in the tracker, $\effTrk$, in the muon detector as
a stand-alone muon, $\effSa$, the average efficiency of the isolation requirement,
$\effIso$, and the average trigger efficiency, $\effHlt$,
can be factorized because the muon selection
factorizes the requirements on the tracker and muon detector quantities separately.
Neither selection on $\chi^2$ per degree of freedom nor requirement of the muon reconstruction through the
tracker-muon algorithm is applied in order to
avoid efficiency terms that cannot be described as a product of contributions from the tracker and
the muon detector.

The dimuon invariant mass spectra for the five categories are divided into
bins of different sizes, depending on the number of observed events.
The distributions of the dimuon invariant mass for the different categories can be written
as the sum of a signal peak plus a background component.

\begin{figure}[hbtp]
    \begin{minipage}{73mm}
      \begin{center}
        \resizebox{1.0\textwidth}{!}{{\includegraphics{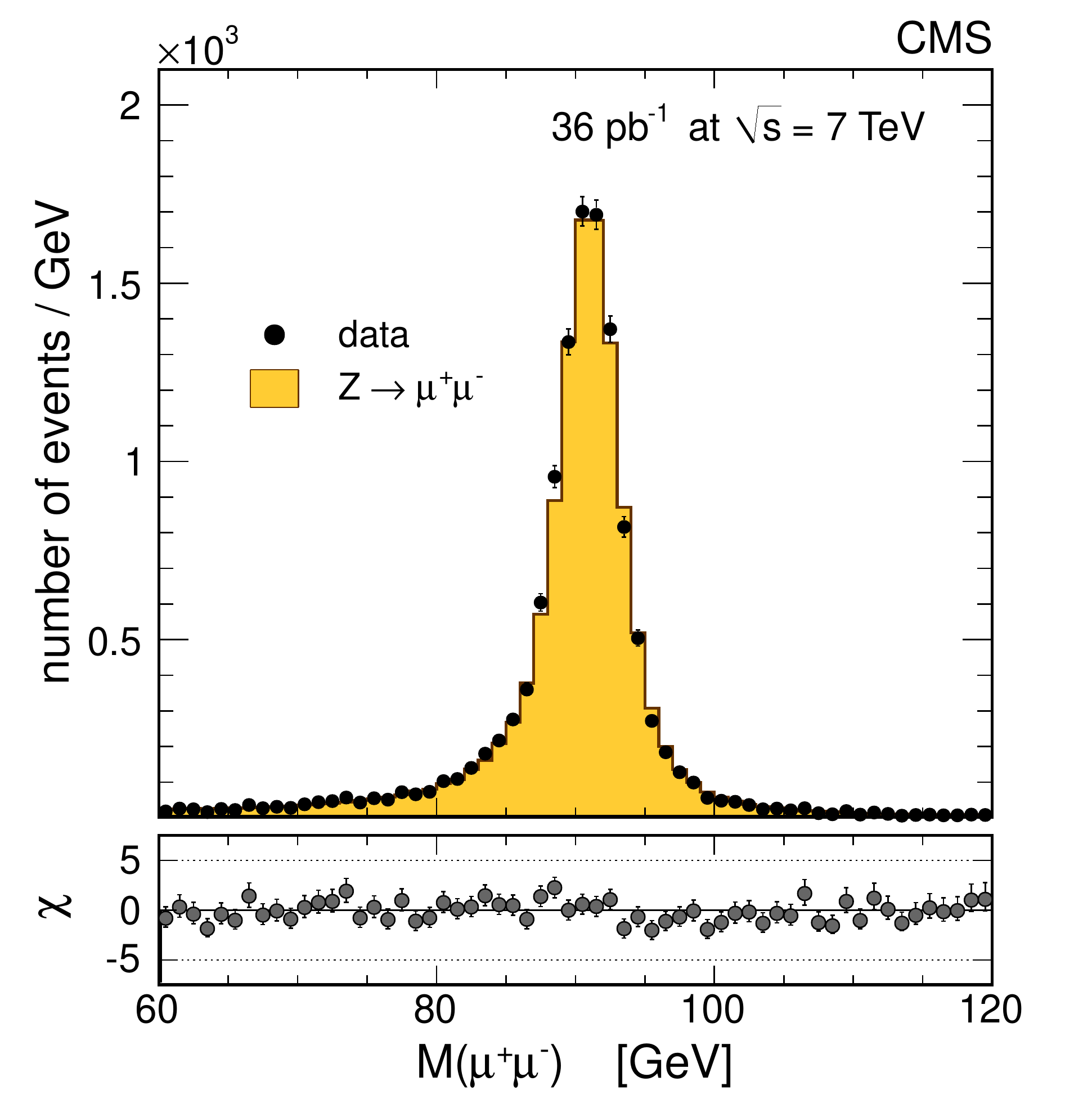}}}
      \end{center}
    \end{minipage}
    \begin{minipage}{73mm}
       \begin{center}
       \resizebox{!}{1.0\textwidth}{\includegraphics{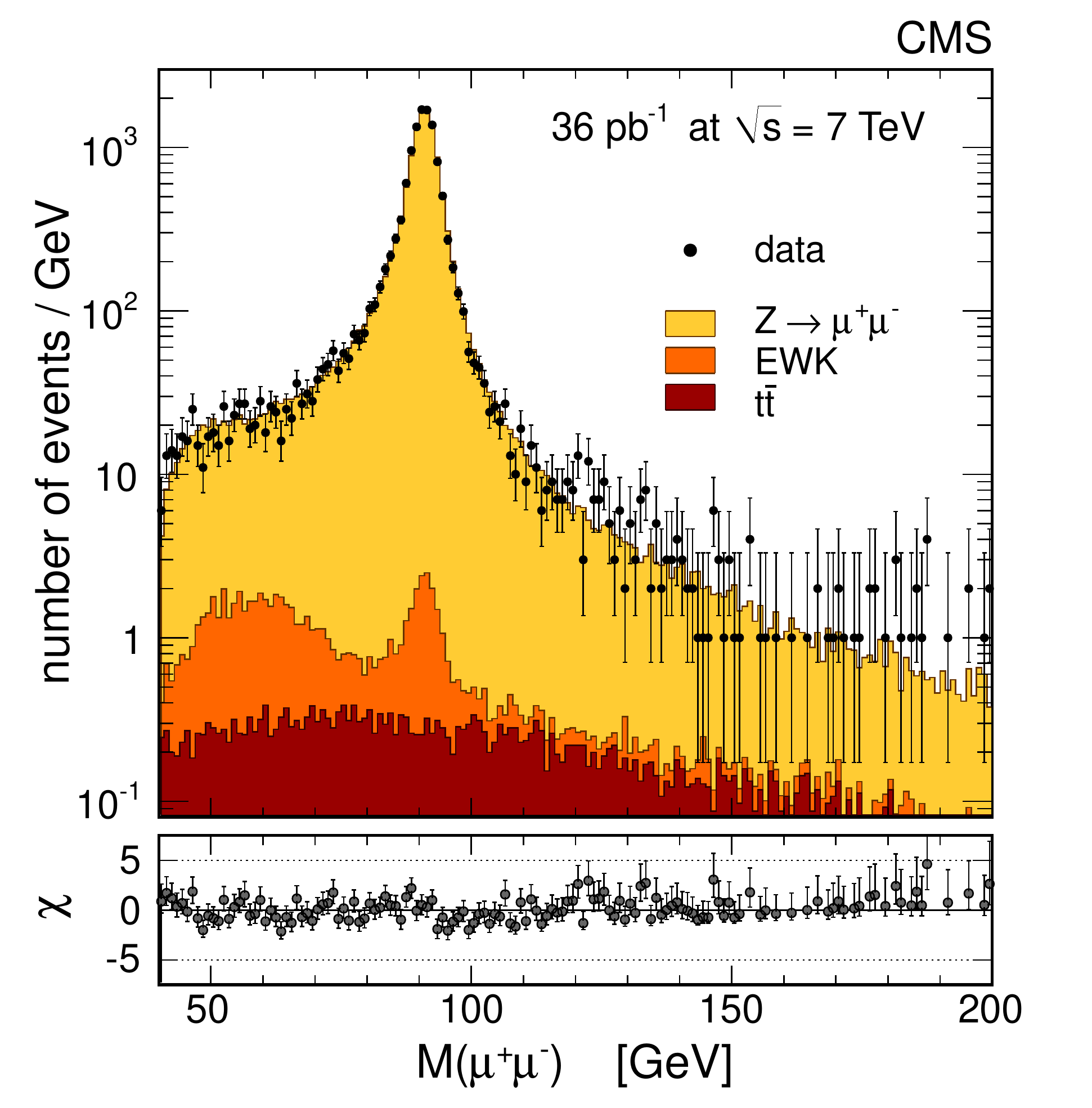}}
      \end{center}
    \end{minipage}
\caption{
Distributions of the dimuon invariant mass for the selected $\Zmm$ golden candidates on
a linear scale (left) and on a logarithmic scale (right).
The points with the error bars represent the data.
Superimposed are the expected distributions from simulations, normalized
to an integrated luminosity of $36$~pb$^{-1}$. The expected distributions are
the Z signal (yellow, light histogram), other EWK processes (orange, medium histogram),
and $\ttbar$ background (red, dark histogram).
Backgrounds are negligible and cannot be seen on the linear-scale plots.
}
\label{fig:zGolden36pb}
\end{figure}

\begin{figure}[hbtp]
    \begin{minipage}{73mm}
      \begin{center}
        \resizebox{1.0\textwidth}{!}{{\includegraphics{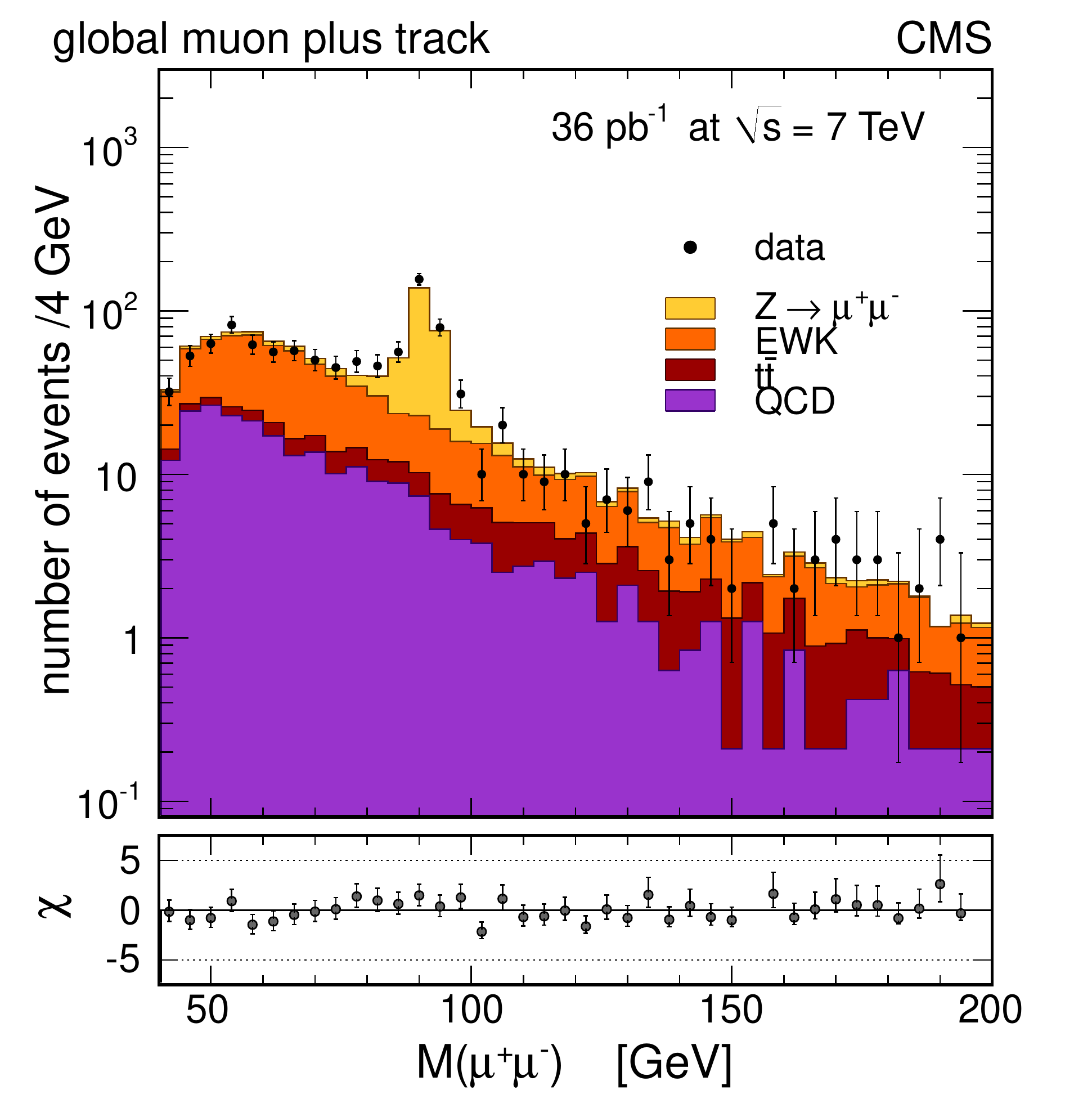}}}
      \end{center}
    \end{minipage}
    \begin{minipage}{73mm}
       \begin{center}
       \resizebox{!}{1.0\textwidth}{\includegraphics{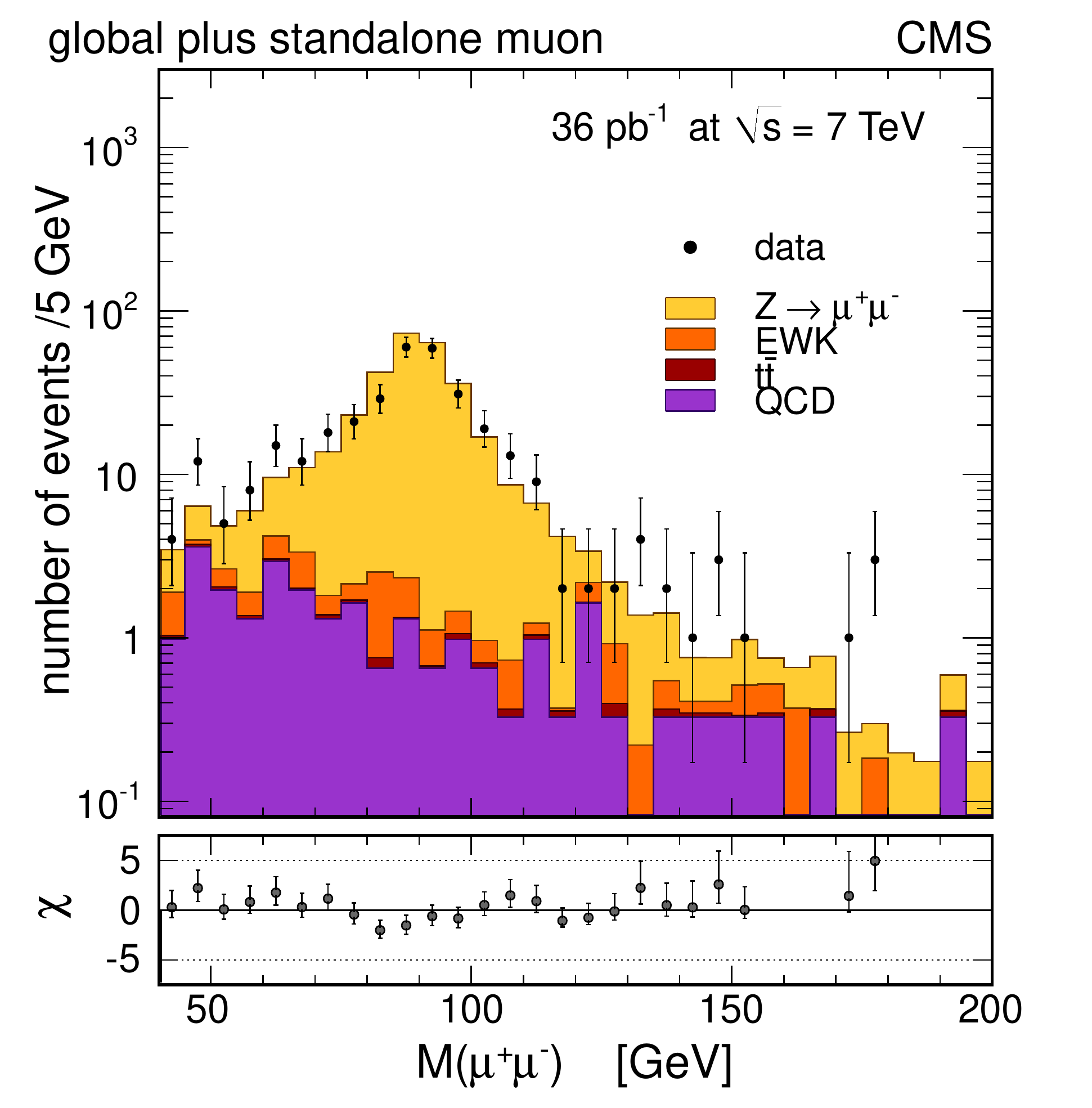}}
      \end{center}
    \end{minipage}
\caption{Distributions of the dimuon invariant mass for the selected
$\Zmut$ (left) and $\Zmus$ (right) candidates.
The points with the error bars represent the data.
Superimposed are the expected distributions from simulations, normalized
to an integrated luminosity of $36$~pb$^{-1}$. The expected distributions are
the Z signal (yellow, light histogram), other EWK processes (orange, medium histogram),
$\ttbar$ background (red, dark histogram) and QCD background (violet, black histogram).
}
\label{fig:zNoGold1}
\end{figure}

\begin{figure}[hbtp]
    \begin{center}
     \begin{minipage}{73mm}
       \begin{center}
        \resizebox{!}{1.0\textwidth}{\includegraphics{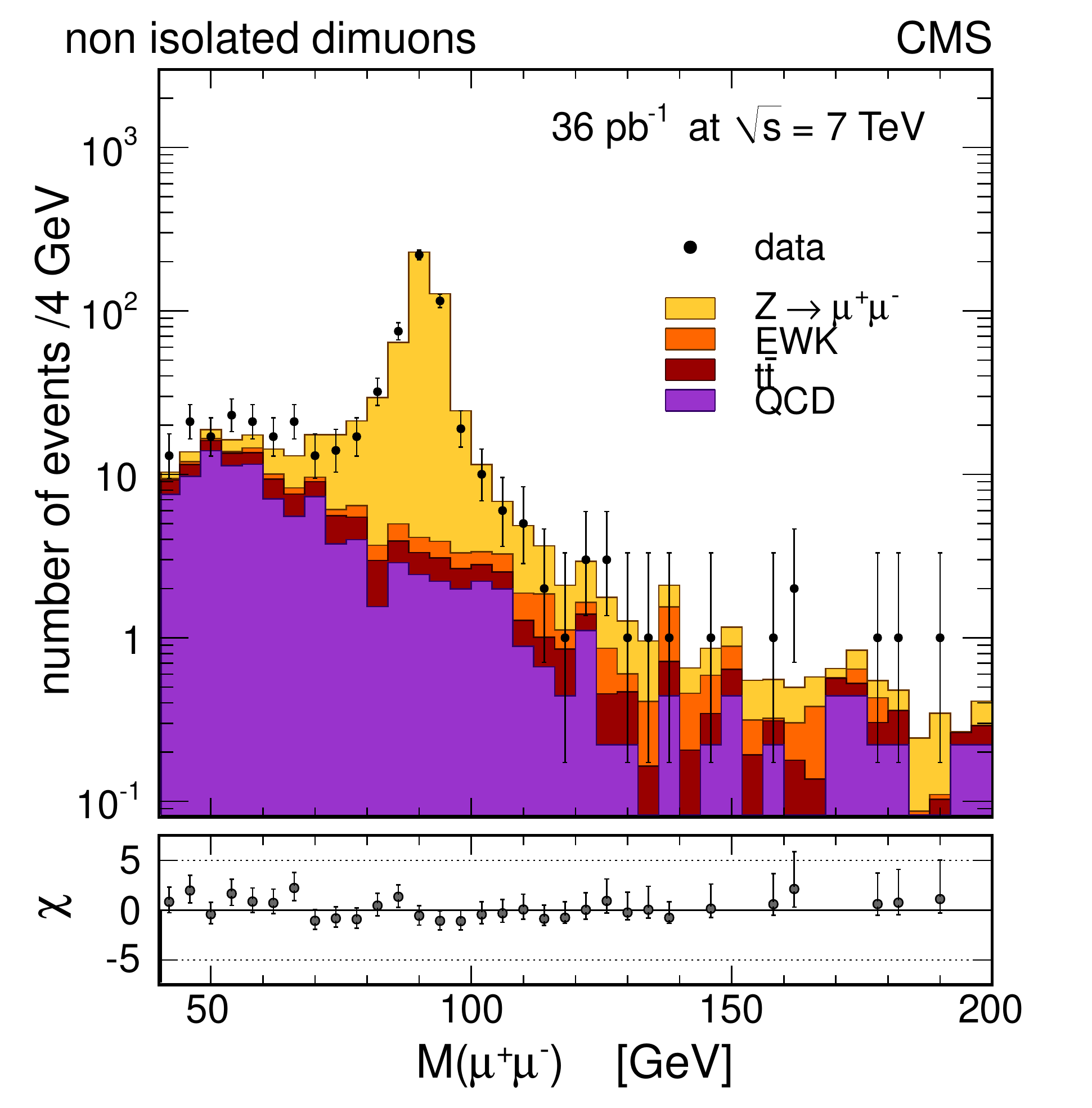}}
       \end{center}
     \end{minipage}
   \end{center}
\caption{Distributions of the dimuon invariant mass for the selected
$\ZmumuNonIso$ candidates.
The points with the error bars represent the data.
Superimposed are the expected distributions from simulations, normalized
to an integrated luminosity of $36$~pb$^{-1}$. The expected distributions are
the Z signal (yellow, light histogram), other EWK processes (orange, medium histogram),
$\ttbar$ background (red, dark histogram), and QCD background (violet, black histogram).
}
\label{fig:zNoGold2}
\end{figure}

Figure~\ref{fig:zGolden36pb} shows the dimuon invariant mass spectrum for the $\Zmm$ golden events
on both a linear scale and a logarithmic scale, and Figs.~\ref{fig:zNoGold1}
and~\ref{fig:zNoGold2} show the
invariant mass distributions for the remaining categories.
The spectra are in agreement with the simulation.


The signal-peak distribution can be considered to be identical in the categories
$\Zmumu$ and $\Zmut$  because the momentum resolution in CMS is determined predominantly
by the tracker measurement for muons with $\Pt \leq 200$ GeV.
The binned spectrum of the dimuon invariant mass in the
$\Zmumu$ category, which has the most events of all categories,
is taken as shape model for all categories but $\Zmus$.
The large size of the golden sample ensures that the statistical
uncertainty of the invariant mass distribution has a negligible effect on the cross
section measurement.
The small presence of background is neglected in this distribution.
The uncertainty due to this approximation has been evaluated and
taken as the systematic uncertainty as described in Section~\ref{sec:muonSyst}.

Because only tracker isolation is used, the shape obtained from golden events
can also be used to model the $\ZmumuNonIso$ peak distribution.
A requirement on calorimetric isolation would have distorted the dimuon invariant mass
distribution of events with one nonisolated muon because of FSR,
as has been observed both in simulation and data.

The model of the invariant mass shape for the $\Zmus$ category is also derived from golden dimuon events.
The three-momentum for one of the two muons is taken from only the muon detector track fit,
in order to emulate a stand-alone muon.
To avoid using the same event twice in forming the $\Zmus$ shape model,
the higher-$\Pt$ (lower-$\Pt$) muon is chosen for even (odd) event numbers.

Background shapes are modeled as products of an exponential times a polynomial
whose degree depends on the category.
Different background models and different binning sizes are considered for the
categories other than $\Zmumu$ and a systematic uncertainty related to the fitting
procedure is determined accordingly.

A simultaneous binned fit based on a Poissonian likelihood~\cite{PoisLR} is performed
for the different categories.
Table~\ref{fig:fitRes_36pb} reports the signal yield and single-muon efficiencies determined from
the simultaneous fit and the ratios of the fitted to simulation efficiencies.
A goodness-of-fit test gives a probability ($p$-value) of 0.36 for this fit.

\begin{table}[htbp] 
\begin{center}
\caption{Signal yield and efficiencies determined from data with the simultaneous fit, and
  ratios of efficiencies determined from the fit and the simulation.}
\label{fig:fitRes_36pb}
\begin{tabular}{|c|c|c|}
\hline
Quantity & Fit results from data & Data/simulation\\
\hline\hline
$\NZtomumu$ &13\,728 $\pm$ 121   & \\
\hline
$\effHlt$ & 0.9203 $\pm$ 0.0019  &0.9672 $\pm$ 0.0020   \\
$\effIso$ & 0.9813 $\pm$ 0.0010& 0.9962 $\pm$  0.0011 \\
$\effSa$ & 0.9762  $\pm$ 0.0012 & 0.9964 $\pm$ 0.0013 \\
$\effTrk$ & 0.9890 $\pm$ 0.0006  & 0.9949 $\pm$ 0.0007  \\
\hline
\end{tabular}
\end{center}
\end{table}

The background in the $\Zmumu$ golden category (of the order of few per mille)
was neglected in the fit. In order to correct the fitted yield $\NZtomumu$ for
the presence of this background, we subtract the small estimated irreducible
background fraction.

A $(1.0\pm 0.5)\%$ overall efficiency correction due to the loss of muon events because of trigger
prefiring is also applied (Section~\ref{sec:muonEff}).

The estimated cross section is $968\pm 8 \mathrm{(stat.)}$~pb.

\section{Systematic Uncertainties}
\label{sec:systematics}

The largest uncertainty contribution on the measured cross sections is related to the
integrated luminosity~\cite{lumiPAS}, and amounts to $\LUMISYST\%$.
\par
The next most important source of systematic uncertainty is due to the
lepton efficiency correction factors obtained
from the \TNP method.
In the $\Zmm$ analysis, the efficiency uncertainties are 
absorbed in the statistical uncertainty of the measurement, via the simultaneous 
fit to the yield and efficiencies.

Table~\ref{tab:syst} shows a summary of systematic uncertainties
for the $\Wo$ and $\Zo$ cross section measurements.
Tables~\ref{tab:systEL} and~\ref{tab:systMU}
show a summary of systematic uncertainties
for the individual cross sections ($\Wp, \Wm$) and the ratios ($\Wp/\Wm$, $\Wo/\Zo$).
Details of systematic uncertainties for the muon and electron channels are described in the
following subsections.

\begin{table}[htbp] %
\begin{center}
   \caption[.]{ \label{tab:syst}
Systematic uncertainties in percent for inclusive W and Z cross sections.
The ``n/a'' entry means that the source does not apply.
A common luminosity uncertainty of 4\% applies to all channels.}
\begin {tabular} {|l|c|c|c|c|}
\hline
Source                                   & $\Wen$         & $\Wmn$           & $\Zee$         & $\Zmm$ \\
\hline\hline
Lepton reconstruction \& identification  & \WEITNPSYST    & \WMIEFFSYST      & \ZEETNPSYST    &  n/a \\
Trigger prefiring                      & n/a            & \WMIEFFPRET      & n/a            & \ZMMEFFPRET \\
Energy/momentum scale \& resolution             & \WEIESCALESYST & \WMISCALESYST    & \ZEEESCALESYST & \ZMMSCALESYST \\
$\MET$ scale \& resolution               & \WEIMETSYST    & \WMIMETSYST      &  n/a           &  n/a   \\
Background subtraction / modeling        & \WEIBKGSYST    & \WMIQCDSHAPESYST & \ZEEBKGSYST    & \ZMMBKGTOTSYST  \\
Trigger changes throughout 2010          & n/a            & n/a              & n/a            & \ZMMTRIGABSYST \\
\hline
Total experimental                       & \WEEXPSYST    & \WMEXPSYST       & \ZEEEXPSYST    & \ZMMEXPSYST \\
\hline
PDF uncertainty for acceptance           & \WEIPDFACCSYST & \WMIPDFACCSYST   & \ZEEPDFACCSYST & \ZMMPDFACCSYST \\
Other theoretical uncertainties          & \WEITHSYST     & \WMITHSYST       & \ZEETHSYST     & \ZMMTHSYST  \\
\hline
Total theoretical                        & \WEITOTTHSYST & \WMITOTTHSYST & \ZEETOTTHSYST & \ZMMTOTTHSYST \\
\hline
Total (excluding luminosity)                      & \WEITOTSYST    & \WMITOTSYST      & \ZEETOTSYST    & \ZMMTOTSYST \\
\hline
\end {tabular}
\end{center}
\end{table}

\begin{table}[htbp] %
\begin{center}
   \caption[.]{ \label{tab:systEL}
Systematic uncertainties in percent for individual W cross sections and the ratios in the
electron channel.
A common luminosity uncertainty of 4\% applies to all cross sections. }
\begin {tabular} {|l|c|c|c|c|}
\hline
Source       & $\Wp$ (e) & $\Wm$ (e) & $\Wp/\Wm$ (e) & $W/Z$ (e) \\
         \hline\hline
Lepton reconstruction \& identification  & 1.5 & 1.5 & 1.5 & 1.1 \\
Energy scale \& resolution             & 0.5 & 0.6 & 0.1 & 0.2 \\
$\MET$ scale \& resolution               & 0.3 & 0.3 & 0.1 & 0.3 \\
Background subtraction / modeling        & 0.3 & 0.5 & 0.4 & 0.3 \\
\hline
Total experimental                       & 1.6 & 1.7 & 1.6 & 1.2 \\
\hline
PDF uncertainty for acceptance           & 0.7 & 1.2 & 1.6 & 0.6 \\
Other theoretical uncertainties          & 1.0 & 0.7 & 1.2 & 1.2 \\
\hline
Total theoretical                        & 1.2 & 1.4 & 2.0 & 1.4 \\
\hline
Total (excluding luminosity)                                   & 2.1 & 2.2 & 2.6 & 1.8 \\
\hline
\end {tabular}
\end{center}
\end{table}

\begin{table}[htbp] %
\begin{center}
   \caption[.]{ \label{tab:systMU}
Systematic uncertainties in percent for individual W cross sections and ratios in the
muon channel.
A common luminosity uncertainty of 4\% applies to all cross sections. }
\begin {tabular} {|l|c|c|c|c|}
\hline
Source       & $\Wp$ ($\mu$) & $\Wm$ ($\mu$) & $\Wp/\Wm$ ($\mu$) & $W/Z$ ($\mu$) \\
         \hline\hline
Lepton reconstruction \& identification  & 0.9 & 0.9 & 1.3 & 0.9 \\
Trigger prefiring                       & 0.5 & 0.5 & 0   & 0   \\
Momentum scale \& resolution             & 0.19 & 0.25 & 0.06 & 0.35 \\
$\MET$ scale \& resolution               & 0.2 & 0.2 & 0.0   & 0.2 \\
Background subtraction / modeling        & 0.4 & 0.5 & 0.2 & 0.4 \\
\hline
Total experimental                       & 1.1 & 1.2 & 1.3 & 1.1 \\
\hline
PDF uncertainty for acceptance           & 0.9 & 1.5 & 1.9 & 0.9 \\
Other theoretical uncertainties          & 0.9 & 0.8 & 0.8 & 1.4 \\
\hline
Total theoretical                        & 1.3 & 1.7 & 2.1 & 1.6 \\
\hline
Total (excluding luminosity)                             & 1.7 & 2.1 & 2.5 & 2.0 \\
\hline
\end {tabular}
\end{center}
\end{table}

\subsection{Electron Channels}
\label{subsec:ELEsystematics}

\par
The propagation of statistical and systematic uncertainties on the data/simulation
efficiency correction factors ($\rhoeff$)
from the \TNP method (reconstruction, identification, and trigger)
results in uncertainties of $\WEITNPSYST\%$ and $\ZEETNPSYST\%$ for the $\Wen$ and $\Zee$ analyses,
respectively. The uncertainties on the $\Wp$ and $\Wm$ cross sections are larger than that for
the inclusive $\Wo$ because of the larger statistical uncertainty when efficiencies are estimated
per charge. The systematic uncertainty, which depends on the efficiency
under study, is determined by considering alternative signal and background models. 
The size of the systematic uncertainty is 0.3$\%$ for the electron selection efficiencies 
and 1.0$\%$ for the electron reconstruction efficiency. The estimation of the 
trigger efficiency is considered to be background-free so there is no need to 
perform a fit for the signal estimation. Theoretical uncertainties on the 
corrected efficiencies related to the PDF uncertainties and the PDF choice were 
found to be negligible.

\par
The electron energy scale has an impact on the \ET distribution
for the signal. To study this effect, the
energy-scale corrections obtained
from the shift of the $\Zo$ mass peak
(Section~\ref{sec:e-escale}) are applied to electrons in the EB and EE
in simulation (before the $\Et$ requirement)
and  the missing $\Et$ is recomputed.
The obtained variations on the signal yield from the UML fit
are $\WEIESCALESYST\%$ for the inclusive $\Wo$, $\WEPESCALESYST\%$ for the $\Wp$, and
$\WEMESCALESYST\%$ for the $\Wm$ samples and $\WERESCALESYST\%$ on the
$\Wp/\Wm$ ratio.  All the charge-related studies (determination of individual $\Wp$ and $\Wm$
yields and $\Wp/\Wm$ ratio and associated systematic uncertainties)
include data/simulation
charge misidentification scale factors, estimated from
the fraction of same-sign events in the $\Zee$ data and simulated samples.
\par
The energy scale of electrons has an impact on the $\Zo$ yield because of 
the $\Et>25~\GeV$ requirement on the two electrons and the mass window requirement.
Applying the energy-scale corrections
mentioned above to the EB and EE electrons and reprocessing the data, 
the $\Zo$~yield is decreased by 10~events ($\ZEESAMPLE \to \ZEESAMPLEN$).
A systematic uncertainty equal to this decrease of $\ZEEESCALESYST\%$
is assigned to the $\Zo$ signal yield.
The energy-scale uncertainty for the $\Wo$ selection is included in the systematic uncertainty 
described in the previous paragraph. There, the systematic uncertainty 
is larger than that for the $\Zo$ selection because the energy scale also affects the 
$\MET$ shape used for the signal extraction.
The $\Wo$ selection itself is affected by the energy scale at the level of $0.12\%$.
\par
The \MET shape used in the $\Wo$ fits is also distorted
by energy resolution uncertainties; this induces a change in the $\Wo$ signal yield
by 0.02\%.

\par

The \MET energy scale is affected by our limited knowledge of the intrinsic hadronic
recoil response. From the discrepancies found in the data/simulation
comparisons (Section~\ref{sec:WsignalMETtemplate}), uncertainties due to
the \MET energy scale are estimated to be 0.3\% for inclusive $\Wo$,
$\Wp$, and $\Wm$ yields, and 0.1\% for the $\Wp/\Wm$ ratio.

\par
The systematic uncertainties on the background subtraction
address the possible difference between the true background distribution
and the modified Rayleigh function that is used in the UML fit.
We make the assumption that any such difference can be accounted
for by an additional $\sigma_2$ parameter (defined in Section~\ref{sec:WQCDbkg}),
which affects the resolution at large values of \MET (below the signal).
The value of $\sigma_2$ is first determined for three samples: the control sample
in the data, the control sample in the QCD simulation, and the
selected sample in the QCD simulation. The values obtained are $\sigma_2=0.0009~\GeV^{-1}$,
$0.0010~\GeV^{-1}$, and $0.0007~\GeV^{-1}$, respectively for $\Wp$ and
$\sigma_2=0.0007~\GeV^{-1}$,
$0.0009~\GeV^{-1}$, and $0.0008~\GeV^{-1}$ for $\Wm$.
The three values of $\sigma_2$ are then fixed in turn, and $\sigma_0$ and $\sigma_1$
are set to their values from data to generate distributions (of the size
of our sample) with
the three-parameter function, which  we then fit with our nominal two-parameter
function. The maximal relative difference in the yields
is quoted as the systematic uncertainty on background subtraction: $\WEIBKGSYST\%$ for inclusive $\Wo$,
$\WEPBKGSYST\%$ for $\Wp$, $\WEMBKGSYST\%$ for $\Wm$, and $\WERBKGSYST\%$ for the ratio.

\par
In the following paragraphs we discuss the systematic uncertainties of 
the fixed shape and the ABCD methods which were also explored in order 
to cross check the extraction of the $\Wen$ signal. 
These uncertainties correspond to the specific methods and are not propagated
to the final cross section measurement reported in this paper. 

The systematic uncertainties on the background subtraction using fixed-shape 
distributions are summarized in Table~\ref{tab:FixedTempSyst}.  
The total uncertainty is taken as the sum in quadrature of 
the values in the table, giving 0.40\%.  
The total uncertainty is dominated by the uncertainty of the correction of the 
signal contamination in the control sample. This uncertainty is evaluated by 
propagating the uncertainty on the measured contamination using the \TNP technique. 
The statistical uncertainty of this evaluation is calculated using 
a large number of shapes in which the number of events is generated 
from a Poisson distribution with the mean equal to the number of events in the nominal shape. 
The signal yield under variation of the requirements used to define the control 
sample was also studied and found to be very stable with an RMS spread  
of 0.12\% for the range of selections considered. In order to take into 
account the observation of small residual correlations that are not corrected for,
an additional systematic uncertainty of 0.35$\%$ is assigned as a conservative estimate of their size.

\begin{table}[htbp] %
  \begin{center}
    \caption{Summary of systematic uncertainties for background modeling using 
fixed-shape distributions.}
    \label{tab:FixedTempSyst}
    \begin{tabular}{|l|c|}
      \hline
      Source of systematic uncertainty & Value \\
      \hline\hline
      MVA Correction & 0.05\% \\
      Signal contamination & 0.15\% \\
      Statistical fluctuations & 0.12\% \\
      Residual correlations & 0.34\% \\
      \hline
      Total & 0.40\% \\
      \hline
    \end{tabular}
  \end{center}
\end{table}

\par
The systematic uncertainties on the signal extraction using the ABCD method
are summarized in Table~\ref{tab:ABCDEsyst}. The total uncertainty is
taken as the sum in quadrature of the individual components listed in the table,
and corresponds to 0.7$\%$.
The two most important sources of systematic uncertainty arise from the 
modeling of the signal shape. The largest of these (0.53$\%$) comes from 
the uncertainty on $\varepsilon_\mathrm{P}$, dominated by the statistical uncertainties 
in the \TNP method. Uncertainties on the electron energy scale, and hadronic 
recoil response and resolution affect the modeling of the $\MET$ distribution 
for the signal and together give rise to the second largest uncertainty (0.4$\%$).
The uncertainty coming from the modeling of electroweak backgrounds is 
estimated to be 0.2$\%$. The assumption that the fake electron efficiency to 
pass the $\MET$ boundary is independent of the relative track isolation leads 
to a small bias for which a correction is applied. The uncertainty on this 
correction gives rise to a very small error on the yield of 0.07$\%$. This is dominated 
by the uncertainty on the signal contamination of the anti-selected sample used 
to estimate the correction.

\begin{table}[hbtp] %
  \begin{center}
    \caption{Summary of systematic uncertainties for the ABCD method.}
    \label{tab:ABCDEsyst}
    \begin{tabular}{|l|c|}
      \hline
      Source of systematic uncertainty & Value \\
      \hline\hline
      Signal contamination in bias correction & 0.07\% \\
      EWK backgrounds & 0.20\% \\
      Tag-and-probe  & 0.53\% \\
      $\MET$ modeling  & 0.40\% \\
      \hline
      Total & 0.70\% \\
      \hline
    \end{tabular}
  \end{center}
\end{table}

The QCD background in the $\Zee$ channel is estimated, as discussed earlier, 
using the shape information of the relative track isolation distribution. 
The relative uncertainty (approximately 0.14$\%$) of the 
total Z yield is used as the systematic uncertainty.

\subsection{Muon Channels}
\label{sec:muonSyst}

The total uncertainty of 0.9\%  (statistical plus systematic) on the 
correction factors $\rhoeff$ is used as the systematic uncertainty due to muon efficiency 
(reconstruction, identification, selection, isolation,
and trigger) for the $\Wmn$ yield. 
The systematic uncertainty assigned to the efficiencies is evaluated using
a large simulated sample including the Z signal and all potential backgrounds. 
Additional uncertainties are evaluated by varying the initial Z preselection criteria and
the mass window to perform the background subtraction fit, and by using alternative parameterizations to model the 
background. The statistical uncertainties on the fit parameters describing the background correction 
are also included. The effect of the uncertainties due to the choice of PDFs used in the Z simulation is also studied and found to 
be negligible.

The full difference in correction factors for the positively and negatively charged muons, 1.3\%, 
is propagated as a systematic uncertainty in the measurement of the $\Wp/\Wm$ cross-section ratio.

A conservative systematic uncertainty of 0.5\%, due to the correction for the trigger prefiring inefficiency 
(Section~\ref{sec:muonEff}), is assigned to both the $\Zmm$ and $\Wmn$ cross-section estimates.

Dedicated studies comparing the peak position and width of the observed Z distribution
with the expected one indicate a muon momentum scale effect of ${\sim} 0.25\%$ for 40 GeV muons.
In order to evaluate the impact on the W cross-section measurement,
the fitting procedure with a new signal distribution where the muon $\Pt$ in the simulations
is modified according to the observed effect, is performed. The difference with respect to the value
quoted above is $\WMISCALESYST$\% for the inclusive W sample, 0.19\% for $\Wp$, and 0.25\% for $\Wm$,
and for the $\Wp/\Wm$ ratio it reduces to 0.06\%.
Muon momentum scale and resolution affect the measurement of the $\Zmm$ cross section 
with a $\ZMMSCALESYST\%$ uncertainty.

The QCD background shape for the W analysis is tested by applying fits to the $\MET$ spectrum
with the two extreme $\MET$ shapes, corresponding to the maximal variations of the correction 
factor, $\alpha$. The variation in the signal yield with respect to that obtained using the reference 
distribution is $\WMIQCDSHAPESYST\%$  for the inclusive W sample, 0.4\% for $\Wp$, 0.5\% for $\Wm$, 
and 0.2\% for the $\Wp/\Wm$ ratio.

The recoil modeling in the signal shape is also a potential source of
uncertainty. This uncertainty is estimated by applying the signal shape predicted by the simulation
to the fits of the $\MET$ distribution. The variation in the signal
yield with respect to the reference result is $\WMIMETSYST\%$.

The systematic uncertainty on the $\Zmm$ signal extraction procedure 
 has been evaluated as follows. The
uncertainty of the fit model is estimated by varying in different ways the background models 
and changing the dimuon mass binning of the various dimuon categories.
Half of the difference between the maximum and minimum fitted yields across all the tested variations
is taken as a systematic uncertainty. This amounts to $\ZMMFITSYST\%$.

The signal shape has been determined assuming that the golden
samples are background-free.
A flat distribution is added as background contribution to the signal shapes
and this produces a relative change in the fitted Z yield 
equal to one third of the introduced background fraction.
An irreducible contamination is known to be present from simulation with the given selection.
It amounts to less than 0.5\%, so a conservative estimate of $\ZMMBKGSYST\%$
systematic uncertainty due to neglecting the background
in the signal shapes used for the fit is assigned.
Adding those two contributions in quadrature, a total 
systematic uncertainty due to the fit method of
$\ZMMBKGTOTSYST\%$ is assigned. 

The stability of the measured Z yields was also checked in 
the two run periods with different trigger thresholds and
the corresponding variation of the signal yield of 0.1\% is taken as
a conservative systematic uncertainty.

\subsection{Theoretical Uncertainties}
\label{sec:theory}

The main theoretical uncertainty on the cross section estimation arises from the computation of the
geometrical and kinematic acceptance of the detector.  Uncertainty due to
the PDF choice, and uncertainties in the PDFs themselves are
studied using the full PDF eigenvector set and comparing among PDFs
provided by the CTEQ, MSTW, and NNPDF groups. For the estimation of 
the acceptance uncertainties, we followed the recipe prescribed by the 
PDF4LHC working group~\cite{PDF4LHC}.

Systematic uncertainties on the acceptances due to the PDF choice are
reported in Table~\ref{tab:pdfSyst}.
Here $\Delta_{i}$ denotes the uncertainty (68\% confidence level (CL)) within a given set 
$i$ ($i=$ CT10~\cite{CTEQ10}, MSTW08NLO~\cite{Martin:2009iq}, NNPDF2.1~\cite{NNPDF21}). 
The quantity $\Delta_{\mathrm{sets}}$ corresponds to half of the maximum difference between the central values of any pair of sets. 
The final systematic uncertainty (last column) considers half of the maximum
difference between the extreme values (central values plus positive or minus negative 
uncertainties), again for any pair of the three 
sets, plus the remaining $\alpha_S$ uncertainties.
As can be seen from Table~\ref{tab:pdfSyst}, the $\Wm$ acceptance uncertainties 
are larger than the $\Wp$ ones. This is true for each 
PDF set as well as for the total assigned acceptance uncertainty and reflects the 
larger d-quark PDF uncertainties with respect to those for the u quark.
The acceptance estimates obtained using the different PDF sets are summarized in
Table~\ref{tab:pdfAcc}.

\begin{table}[htb] %
\begin{center}
\caption{Systematic uncertainties from the PDF choice on estimated 
acceptances and acceptance correction factors after the analysis selections. }
\label{tab:pdfSyst}
\begin{tabular}{| l | c | c | c | l | c |}
\hline
\centering Quantity & $\Delta_{\mathrm{CTEQ}}$ (\%) & $\Delta_{\mathrm{MSTW}}$ (\%) &  $\Delta_{\mathrm{NNPDF}}$ (\%) & $\Delta_{\mathrm{sets}}$ (\%) & Syst. (\%) \\
\hline
\hline
$\Wp$ acceptance (e) & $\pm 0.5$ & $\pm 0.3$ &  $\pm 0.4$ & $0.2$~{\tiny (NNPDF-MSTW)} & $0.7$ \\
$\Wm$ acceptance (e) & $\pm 0.9$ & $\pm 0.5$ &  $\pm 0.7$ & $0.5$~{\tiny (NNPDF-MSTW)} & $1.2$ \\
$\Wo$ acceptance (e)   & $\pm 0.5$ & $\pm 0.3$ &  $\pm 0.4$ & $0.2$~{\tiny (MSTW-CTEQ)} & $0.6$ \\
$\Zo$ acceptance (e)   & $\pm 0.7$ & $\pm 0.4$ &  $\pm 0.6$ & $0.3$~{\tiny (NNPDF-MSTW)} & $0.9$ \\
$\Wp/\Wm$ correction (e) & $\pm 1.6$ & $\pm 0.5$ &  $\pm 0.7$ & $0.7$~{\tiny (NNPDF-MSTW)} & $1.6$ \\
$\Wo/\Zo$ correction (e) & $\pm 0.6$ & $\pm 0.2$ &  $\pm 0.3$ & $0.2$~{\tiny (NNPDF-MSTW)} & $0.6$ \\
\hline
$\Wp$ acceptance ($\mu$)   & $\pm 0.7$ & $\pm 0.4$ &  $\pm 0.6$ & $0.3$~{\tiny (NNPDF-MSTW)} & $0.9$ \\
$\Wm$ acceptance ($\mu$) & $\pm 1.1$ & $\pm 0.6$ &  $\pm 0.9$ & $0.5$~{\tiny (MSTW-CTEQ)} & $1.5$ \\
$\Wo$ acceptance ($\mu$) & $\pm 0.7$ & $\pm 0.4$ &  $\pm 0.6$ & $0.2$~{\tiny (MSTW-CTEQ)} & $0.8$ \\
$\Zo$ acceptance ($\mu$)   & $\pm 1.0$ & $\pm 0.6$ &  $\pm 0.9$ & $0.2$~{\tiny (NNPDF-MSTW)} & $1.1$ \\
$\Wp/\Wm$ correction ($\mu$) & $\pm 1.9$ & $\pm 0.6$ &  $\pm 0.9$ & $0.8$~{\tiny (NNPDF-MSTW)} & $1.9$ \\
$\Wo/\Zo$ correction ($\mu$) & $\pm 0.8$ & $\pm 0.2$ &  $\pm 0.3$ & $0.2$~{\tiny (NNPDF-CTEQ)} & $0.9$ \\
\hline
\end{tabular}
\end{center}
\end{table}

\begin{table}[htb] %
\begin{center}
\caption{Predictions of the central values of the acceptances and the ratios of 
acceptances for various PDF sets. }
\label{tab:pdfAcc}
\begin{tabular}{| l | c | c | c |}
\hline
\centering Quantity & CTEQ & MSTW &  NNPDF  \\
\hline
\hline
$A_\Wp(\mathrm{e})$ & 0.5017 & 0.5016 & 0.5036 \\
$A_\Wm(\mathrm{e})$ & 0.4808 & 0.4855 & 0.4804 \\
$A_\Wo(\mathrm{e})$ & 0.4933 & 0.4951 & 0.4942 \\
$A_\Zo(\mathrm{e})$ & 0.3876 & 0.3892 & 0.3872 \\
$A_\Wp(\mathrm{e})/A_\Wm(\mathrm{e})$ & 0.9583 & 0.9488 & 0.9626 \\
$A_\Wo(\mathrm{e})/A_\Zo(\mathrm{e})$ & 0.7857 & 0.7853 & 0.7880 \\
\hline
$A_\Wp(\mu)$ & 0.4594 & 0.4587 & 0.4617 \\
$A_\Wm(\mu)$ & 0.4471 & 0.4519 & 0.4472 \\
$A_\Wo(\mu)$ & 0.4543 & 0.4559 & 0.4557 \\
$A_\Zo(\mu)$ & 0.3978 & 0.3990 & 0.3973 \\
$A_\Wp(\mu)/A_\Wm(\mu)$ & 0.9732 & 0.9614 & 0.9778 \\
$A_\Wo(\mu)/A_\Zo(\mu)$ & 0.8756 & 0.8761 & 0.8796 \\
\hline
\end{tabular}
\end{center}
\end{table}

\begin{table}[!ht] %
\begin{center}
\caption{Uncertainties on acceptances due to theoretical assumptions. The different contributions are due to
ISR plus NNLO effects, factorization and renormalization scales, PDF uncertainties, FSR modeling, and EWK corrections.}
\label{tab:th_results}
\begin{tabular}{|l|ccccc|c|}
	\hline
	Quantity & ISR+NNLO & $\mu_R$,$\mu_F$ Scales & PDF & FSR & EWK & Total \\
	\hline\hline
	$\Wp$ acceptance (e)     & $0.63\%$ & $0.77\%$ & $0.7\%$ & $0.17\%$ & $0.14\%$ & $1.2\%$ \\
	$\Wm$ acceptance (e)     & $0.31\%$ & $0.50\%$ & $1.2\%$ & $0.20\%$ & $0.29\%$ & $1.4\%$ \\
	$\Wo$ acceptance (e)       & $0.53\%$ & $0.34\%$ & $0.6\%$ & $0.13\%$ & $0.14\%$ & $0.9\%$ \\
	$\Zo$ acceptance (e)       & $0.84\%$ & $0.39\%$ & $0.9\%$ & $0.54\%$ & $0.84\%$ & $1.6\%$ \\
	$\Wp/\Wm$ correction (e) & $0.32\%$ & $1.14\%$ & $1.6\%$ & $0.26\%$ & $0.25\%$ & $2.0\%$ \\
	$\Wo/\Zo$ correction (e)     & $0.31\%$ & $0.48\%$ & $0.6\%$ & $0.44\%$ & $1.00\%$ & $1.4\%$ \\
	\hline
	$\Wp$ acceptance $(\mu)$     & $0.72\%$ & $0.49\%$ & $0.9\%$ & $0.34\%$ & $0.14\%$ & $1.3\%$ \\
	$\Wm$ acceptance $(\mu)$     & $0.50\%$ & $0.37\%$ & $1.5\%$ & $0.16\%$ & $0.39\%$ & $1.7\%$ \\
	$\Wo$ acceptance $(\mu)$       & $0.65\%$ & $0.44\%$ & $0.8\%$ & $0.21\%$ & $0.13\%$ & $1.1\%$ \\
	$\Zo$ acceptance $(\mu)$       & $1.08\%$ & $0.20\%$ & $1.1\%$ & $0.25\%$ & $1.08\%$ & $1.9\%$ \\
	$\Wp/\Wm$ correction $(\mu)$ & $0.23\%$ & $0.61\%$ & $1.9\%$ & $0.31\%$ & $0.43\%$ & $2.1\%$ \\
	$\Wo/\Zo$ correction $(\mu)$     & $0.43\%$ & $0.38\%$ & $0.9\%$ & $0.27\%$ & $1.22\%$ & $1.6\%$ \\
	\hline
\end{tabular}
\end{center}
\end{table}

Table~\ref{tab:th_results} summarizes the different theoretical uncertainties on the acceptance
due to ISR and NNLO, higher order effects, PDFs, FSR, and missing
EWK contributions.

The baseline MC generator used to simulate the W and Z signals, {\sc POWHEG}, is
accurate up to the NLO in perturbative QCD, and up to
the leading-logarithmic (LL) order  for soft, nonperturbative QCD effects.
A description accurate to just beyond the next to next to LL (NNLL)
can be attained with a resummation procedure~\cite{Collins-1, Collins-2}.
The {\sc ResBos} generator~\cite{ResBos} implements both the resummation and NNLO calculations,
which are missing in the baseline generator, and its predictions for the W boson $\PT$ spectrum 
show remarkable agreement with $\mathrm{p}\bar{\mathrm{p}}$ data at 
$\sqrt{s} = 1.96$~TeV~\cite{ResBosComp}. Final state radiation is incorporated 
in {\sc ResBos} via {\sc PHOTOS}~\cite{PHOTOS}.
The effect of soft nonperturbative effects, hard higher-order effects, and initial-state radiation (ISR),
which are not accounted for in the baseline generator, is studied by comparing {\sc ResBos} 
results with {\sc POWHEG},
and the difference is taken as a systematic uncertainty (second column in Table~\ref{tab:th_results}).

Fixed-order cross section calculations depend on the renormalization ($\mu_R$) and factorization
 ($\mu_F$) scales. 
Higher-order virtual 
processes influence the W and Z boson momentum and rapidity distributions. 
{\sc ResBos} fixes $\mu_R$ and $\mu_F$ to the boson mass, so 
{\sc FEWZ}~\cite{Melnikov:2006kv, Melnikov:2006di} code is used to
estimate the effect of scale dependence of NNLO calculations that is quoted as a systematic uncertainty.
The acceptance is computed by varying up and down the renormalization 
and factorization scales within a factor of two, keeping $\mu_R =\mu_F$.
Half of the maximum excursion range due to this variation is taken as a systematic
uncertainty (third column in Table~\ref{tab:th_results}). The PDF uncertainties
from Table~\ref{tab:pdfSyst} are reported in the fourth column of Table~\ref{tab:th_results}
and added in quadrature to the other contributions to determine the total theoretical
uncertainties, shown in the last column.

On top of higher-order QCD corrections, the effect of EWK corrections, not fully 
implemented in our baseline MC samples, is estimated using the {\sc HORACE} generator~\cite{HORACE-1, HORACE-2, HORACE-3, HORACE-4},
which implements both FSR and virtual and nonvirtual corrections. 
Individual effects are separated and the final-state effects
are then compared to the {\sc PYTHIA} results, as {\sc PYTHIA} is used for FSR in the 
{\sc POWHEG} event generation. While {\sc PYTHIA} only partially accounts for NLO EWK corrections 
by generating QED ISR and FSR with a parton shower approximation, HORACE also implements one-loop
virtual corrections and photon emission from W boson. The difference between the two generators is taken
as a systematic uncertainty (fifth column in Table~\ref{tab:th_results}). Moreover, FSR is simulated beyond the
single-photon emission in HORACE using the parton shower method. The difference due to FSR in HORACE and
PYTHIA is taken as a contribution to the systematic uncertainty (sixth column in Table~\ref{tab:th_results}).

\section{Results}
\label{sec:results}

The results for the electron and muon channels are presented separately.
Assuming lepton universality, we combine our measurements in the different lepton
decay modes. The electron and muon channels are combined by calculating an
average value weighted by the combined statistical and systematic uncertainties,
taking into account the correlated uncertainties.
For the cross-section measurements, correlations are only numerically
relevant for theoretical uncertainties,
including the PDF uncertainties on the acceptance values. For the cross
section ratio measurements, the correlations
of lepton efficiencies are taken into account in each lepton channel.
In the combination of lepton channels, fully
correlated theoretical uncertainties are assumed for the acceptance factor,
with other uncertainties assumed uncorrelated. The luminosity uncertainty
cancels exactly in the cross-section ratios.

The NNLO predictions of the total cross sections and their ratios were estimated
using {\sc fewz} and the MSTW 2008 PDF. The uncertainties, at 68$\%$ CL, include
contributions from the strong coupling $\alpha_S$~\cite{Martin:2009bu,GWatt}, the choice of heavy quark masses
(charm and bottom quarks)~\cite{Martin:2009hq} as well as neglected higher-order corrections
beyond NNLO, by allowing the renormalization and factorization scales to vary in a similar way to that
described in Section~\ref{sec:theory}.

The following cross sections for inclusive $\Wo$ production are measured:
{\footnotesize
\begin{eqnarray*}
 \WEISIGBR , \\
 \WMISIGBR , \\
 \WLISIGBR .
\end{eqnarray*}
}%
The corresponding NNLO prediction is $\THEORYSIGBRWI$.
The results for charge-specific $\Wo$ production are
{\footnotesize
\begin{eqnarray*}
 \WEPSIGBR , \\
 \WMPSIGBR , \\
 \WLPSIGBR ,
\end{eqnarray*}
}%
and
{\footnotesize
\begin{eqnarray*}
 \WEMSIGBR , \\
 \WMMSIGBR , \\
 \WLMSIGBR .
\end{eqnarray*}
}%
The NNLO predictions for these cross sections are $\THEORYSIGBRWP$
for~$\Wp$ and $\THEORYSIGBRWM$ for~$\Wm$.
The following cross sections for inclusive $\Zo$ production are measured:
{\footnotesize
\begin{eqnarray*}
 \ZEESIGBR , \\
 \ZMMSIGBR , \\
 \ZLLSIGBR .
\end{eqnarray*}
}%
The reported $\Zo$ cross sections correspond to the
invariant mass range $60 < m_{\ell^+\ell^-} < 120~\GeV$, and
are corrected for the kinematic acceptance but not for $\gamma^*$ exchange.
The NNLO prediction for $\Zo$ production is $\THEORYSIGBRZ$.

\par
The ratio of cross sections for $\Wo$ and $\Zo$ production is
\begin{displaymath}
  \frac{\sigma_{\Wo}}{\sigma_{\Zo}} =
  \frac{N_{\Wo}}{N_{\Zo}} \,
  \frac{\epsilon_{\Zo}}{\epsilon_{\Wo}} \,
  \frac{A_{\Zo}}{A_{\Wo}} \,,
\end{displaymath}
where $A_{\Zo}$ and $A_{\Wo}$ are the acceptances for Z and W selections, respectively.
The two different decay channels
are combined by assuming fully correlated uncertainties for the
acceptance factors, with other uncertainties assumed uncorrelated.
The resulting ratios are:
\begin{eqnarray*}
  \RESRATWZE , \\
  \RESRATWZM , \\
  \RESRATWZL . \\
\end{eqnarray*}
The NNLO prediction for this ratio is $\THEORYRATIOWZ$,
in good agreement with the measured value.

The ratio of cross sections for $\Wp$ and $\Wm$ production is given by
\begin{displaymath}
  \frac{\sigma_{\Wp}}{\sigma_{\Wm}} =
  \frac{N_{\Wp}}{N_{\Wm}} \,
  \frac{\epsilon_{\Wm}}{\epsilon_{\Wp}} \,
  \frac{A_{\Wm}}{A_{\Wp}} \,,
\end{displaymath}
where $A_{\Wp}$ and $A_{\Wm}$ are the acceptances for $\Wp$ and $\Wm$, respectively.
The two different decay channels are combined by assuming fully correlated uncertainties for the acceptance
factors, with other uncertainties assumed uncorrelated.  This results
in the measurements:
\begin{eqnarray*}
  \RESRATWWE , \\
  \RESRATWWM , \\
  \RESRATWWL . \\
\end{eqnarray*}
The NNLO prediction for this ratio is $\THEORYRATIOWW$, which agrees
with the presented measurement.

\begin{figure}[htbp]
\begin{center}
  \includegraphics[width=0.495\textwidth]{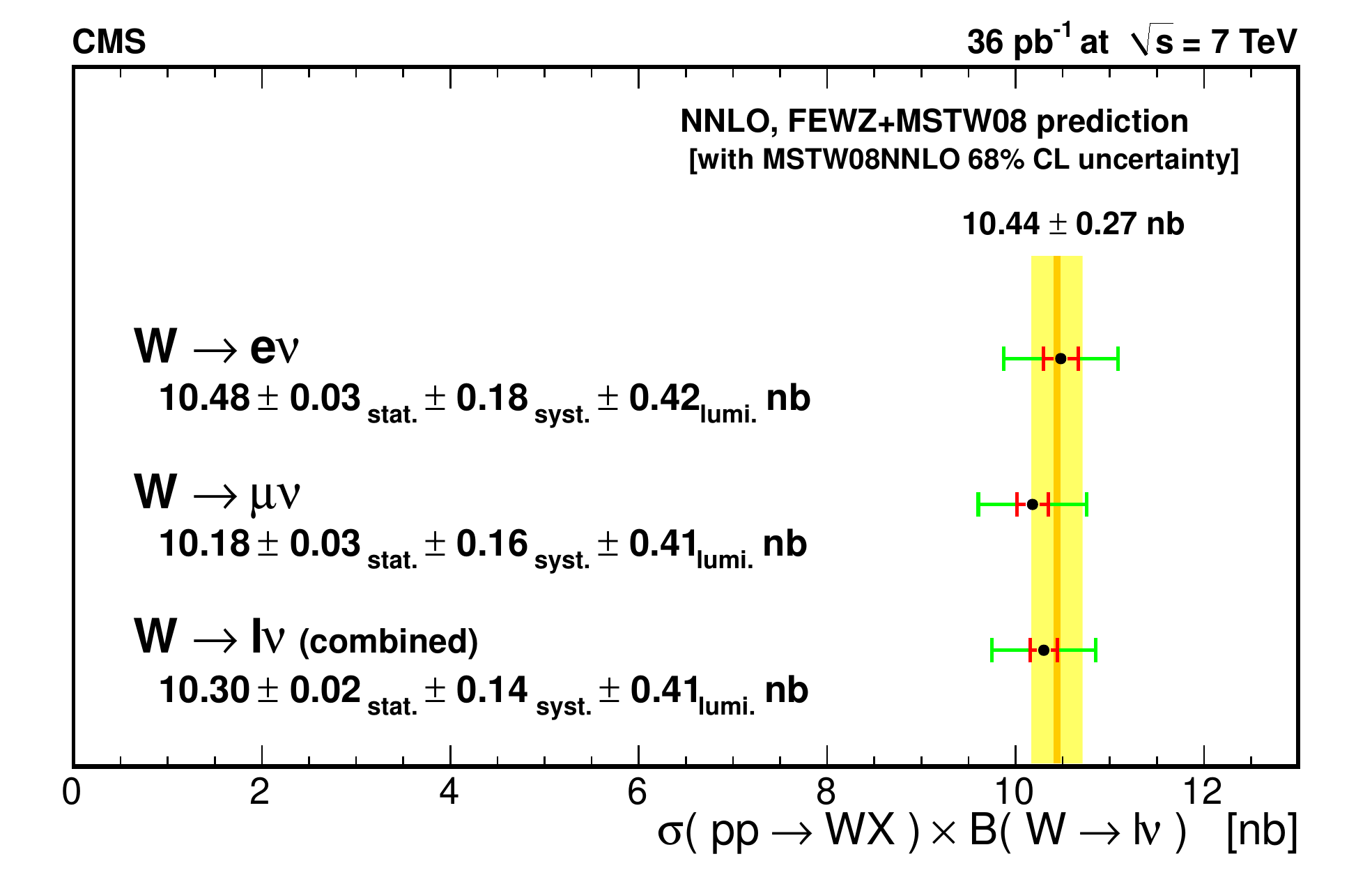}
  \includegraphics[width=0.495\textwidth]{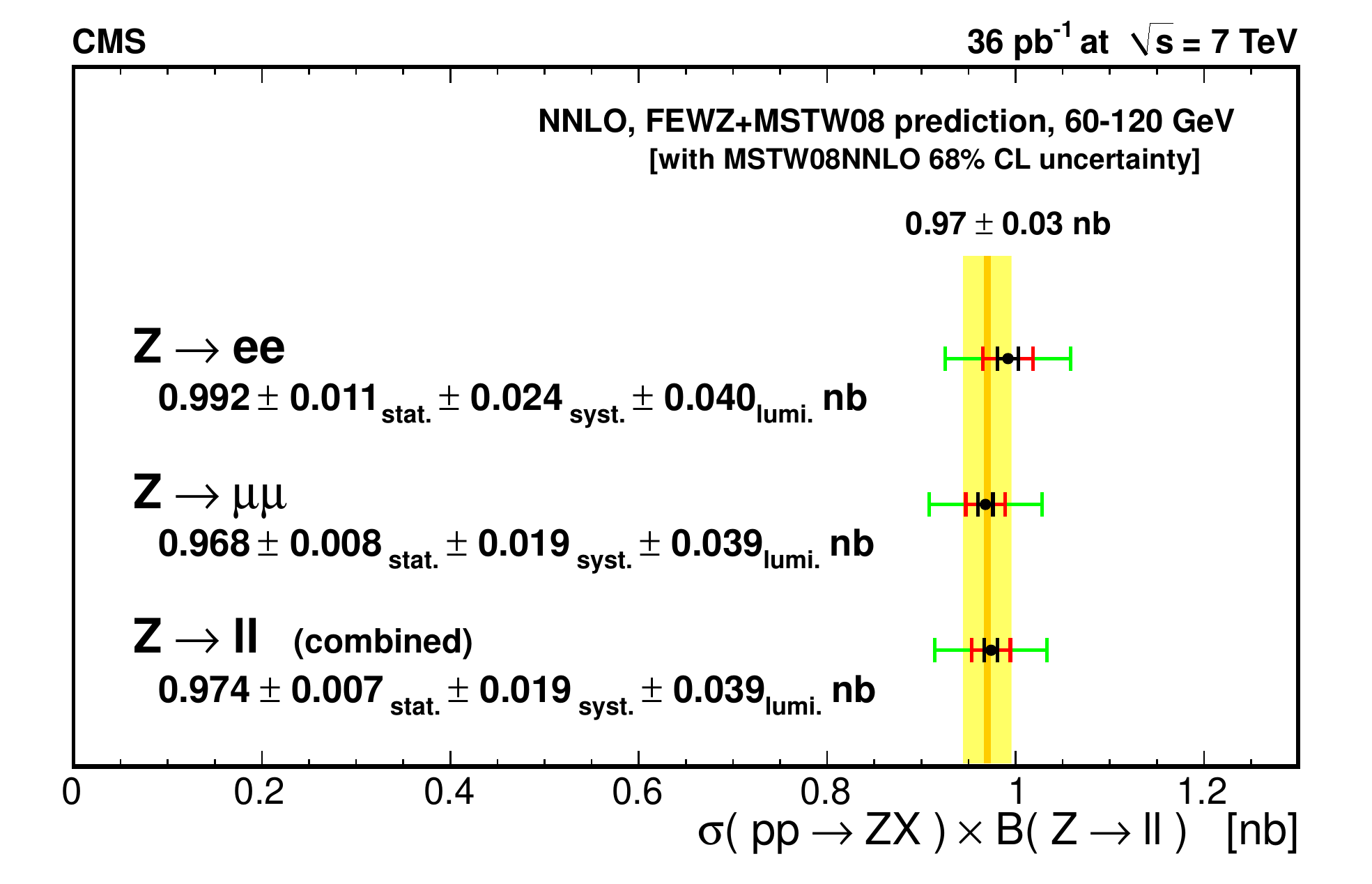}
\caption[.]{\label{fig:WZ_LEPstylePlots}
Summary of the $\Wo$ and $\Zo$ production cross section times branching ratio measurements.
Measurements in the electron and muon channels, and combined, are compared to the
theoretical predictions (yellow band) computed at the NNLO in QCD with recent PDF sets.
Statistical uncertainties are represented as a black error bars, while the red error bars also include systematic uncertainties,
and the green error bars also include luminosity uncertainties.
}
\end{center}
\end{figure}

\begin{figure}[htbp]
\begin{center}
  \includegraphics[width=0.495\textwidth]{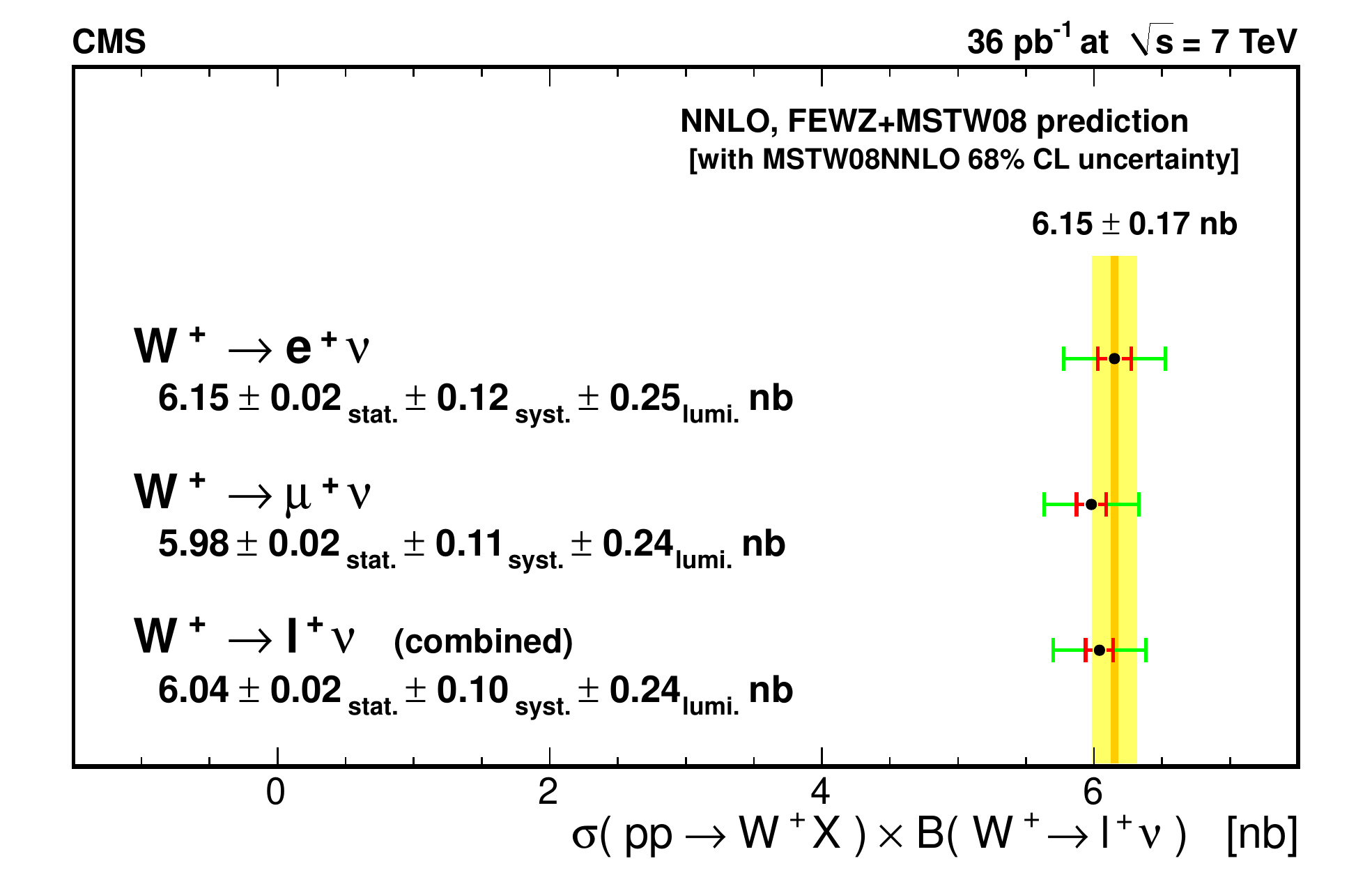}
  \includegraphics[width=0.495\textwidth]{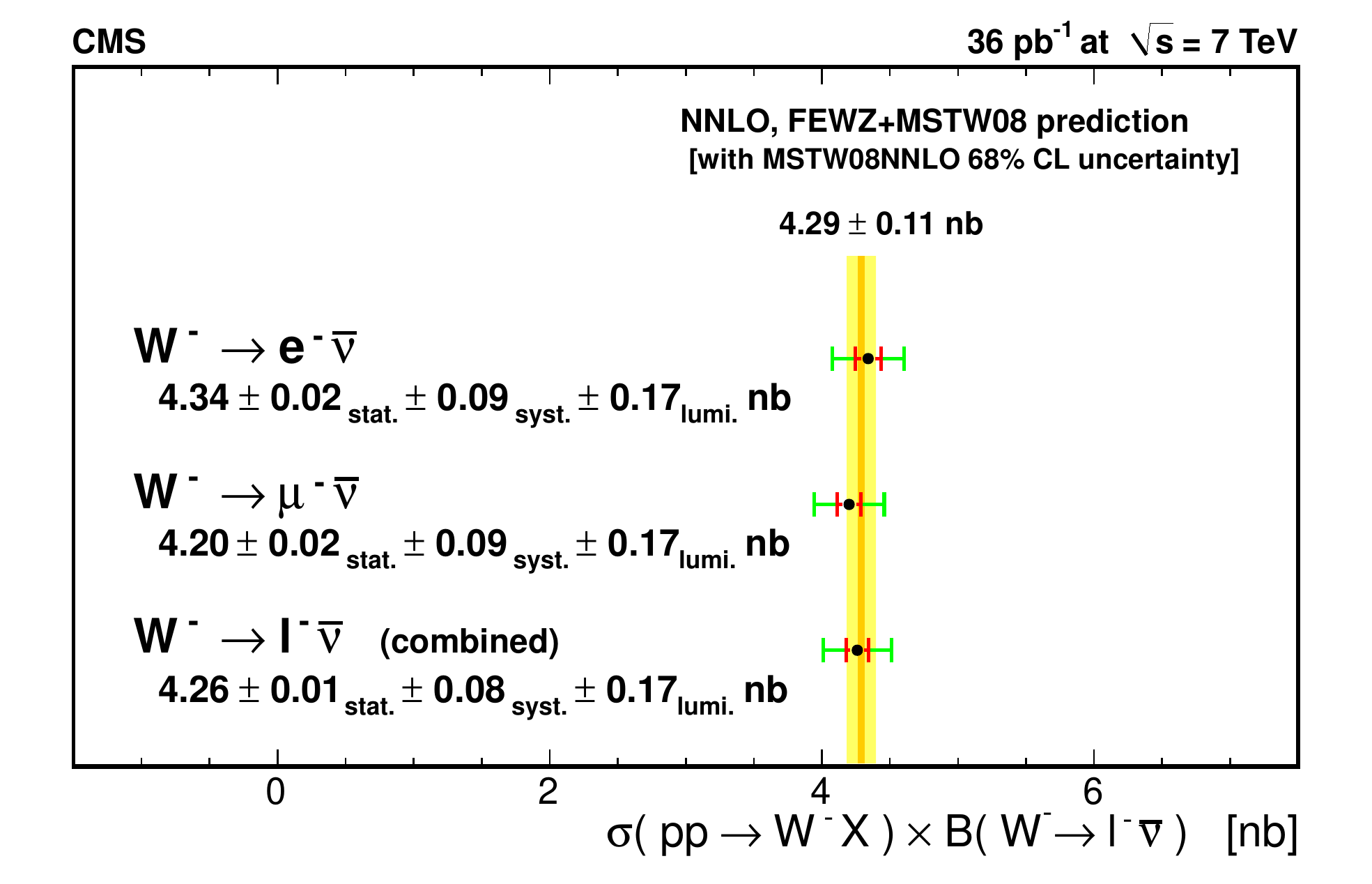}
\caption[.]{\label{fig:WPM_LEPstylePlots}
Summary of the $\Wp$ and $\Wm$ production cross section times branching ratio measurements.
Measurements in the electron and muon channels, and combined, are compared to the
theoretical predictions computed at the NNLO in QCD with recent PDF sets.
Statistical uncertainties are negligible in this plot; the red error bars represent systematic uncertainties,
and the green error bars also include luminosity uncertainties.
}
\end{center}
\end{figure}

\begin{figure}[htbp]
\begin{center}
  \includegraphics[width=0.495\textwidth]{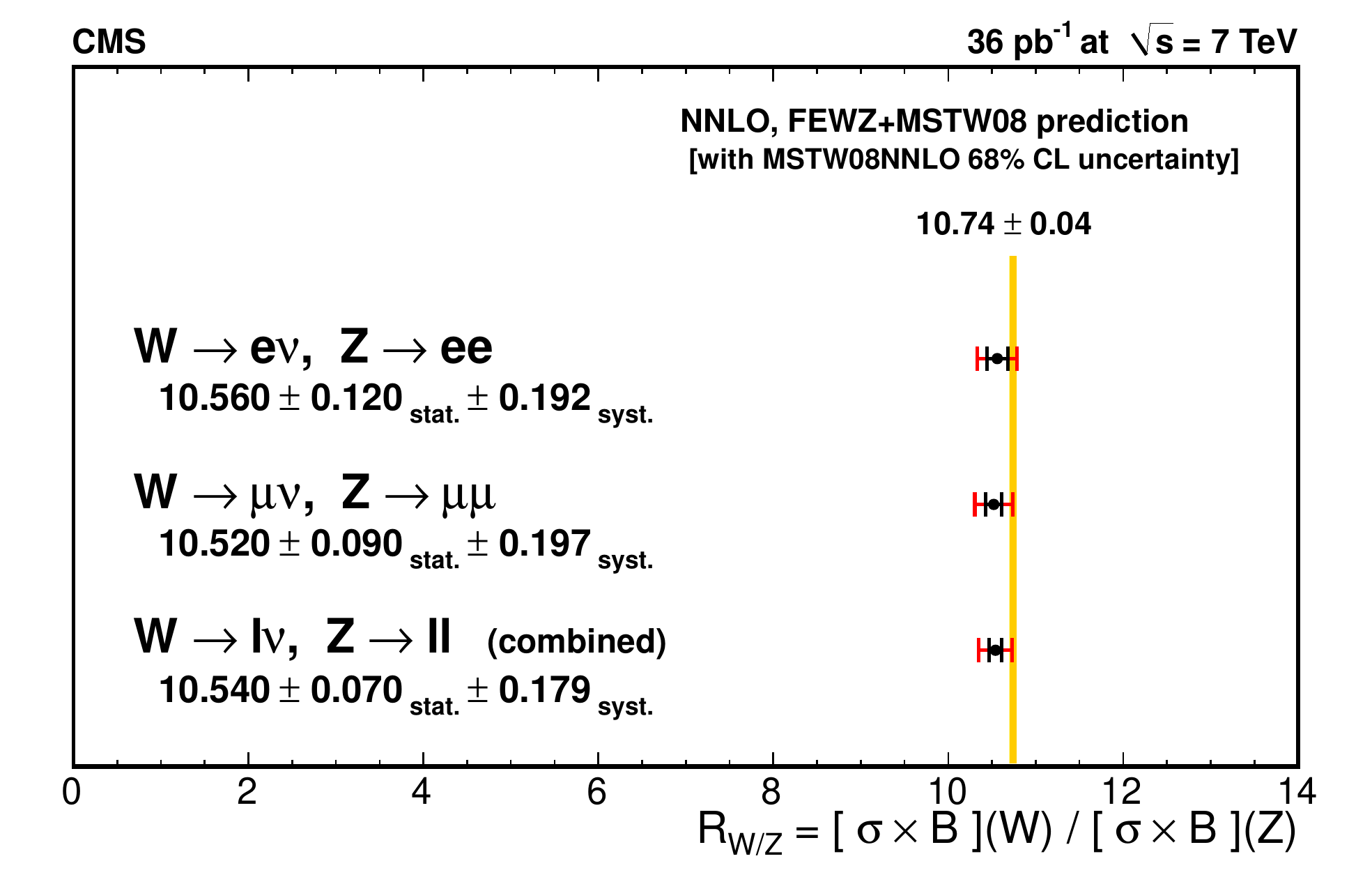}
  \includegraphics[width=0.495\textwidth]{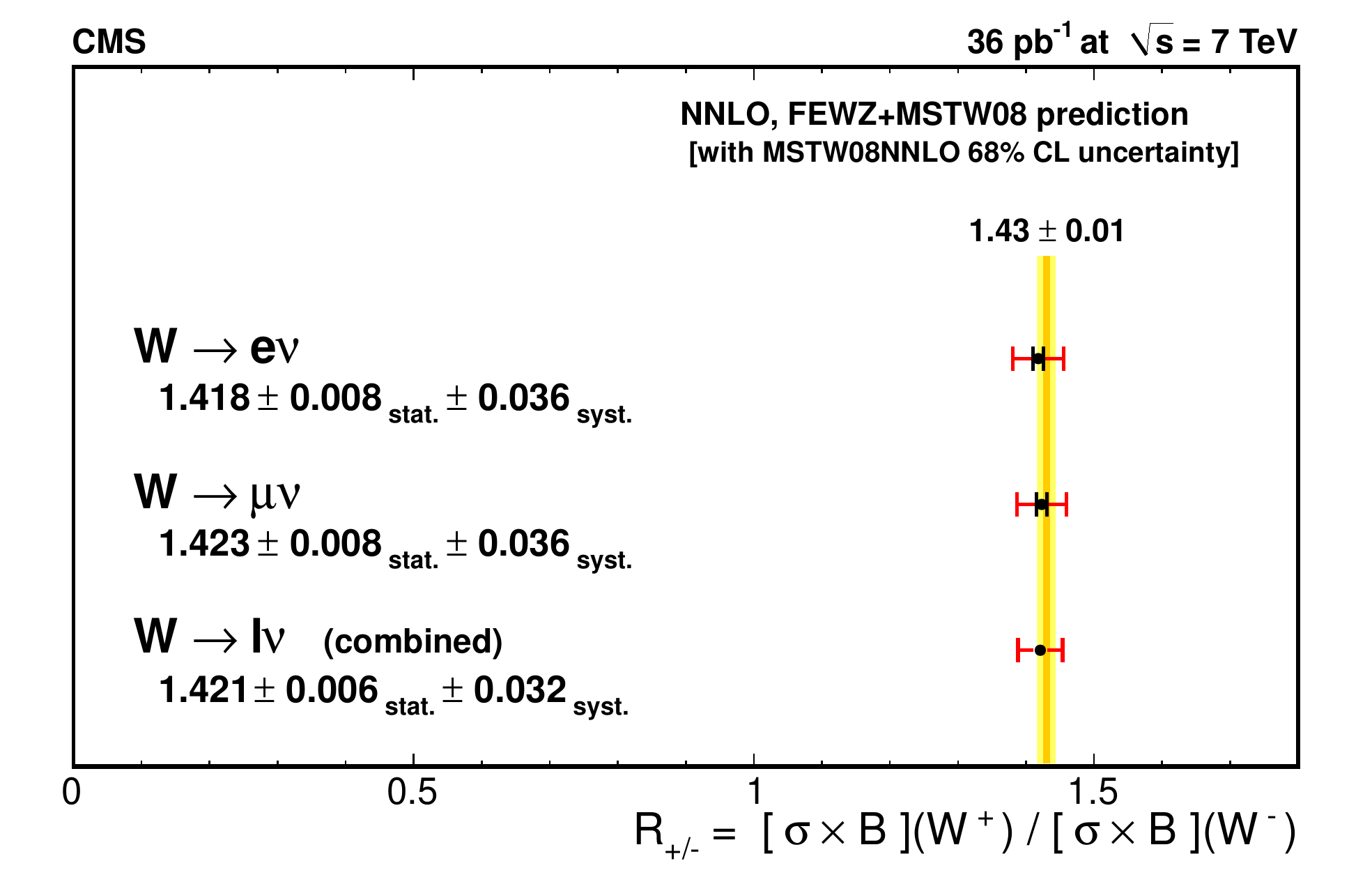}
\caption[.]{\label{fig:R_LEPstylePlots}
Summary of the measurements of the ratios of W to Z and $\Wp$ to $\Wm$ production cross sections.
Measurements in the electron and muon channels, and combined, are compared to the
theoretical predictions computed at the NNLO in QCD with recent PDF sets.
Statistical uncertainties are represented as a black error bars, while the red error bars also include systematic uncertainties.
Luminosity uncertainties cancel in the ratios.
}
\end{center}
\end{figure}

Summaries of the measurements are given in Figs.~\ref{fig:WZ_LEPstylePlots},
 \ref{fig:WPM_LEPstylePlots}, and~\ref{fig:R_LEPstylePlots},
illustrating the consistency of the measurements
in the electron and muon channels, as well as confirming the
theoretical predictions computed at the NNLO in QCD with state-of-the-art
PDF sets. For each reported measurement, the statistical error is represented in black and
the total experimental uncertainty, obtained by adding in quadrature the statistical and
systematic uncertainties, in dark blue. For the cross-section measurements, the luminosity
uncertainty is added to the experimental uncertainty, and is represented in green.
The dark-yellow vertical line represents the theoretical prediction, and the light-yellow
vertical band is the theoretical uncertainty, interpreted as a 68$\%$ confidence interval,
as described earlier.

\begin{figure}
\begin{center}
  \includegraphics[width=0.7\textwidth]{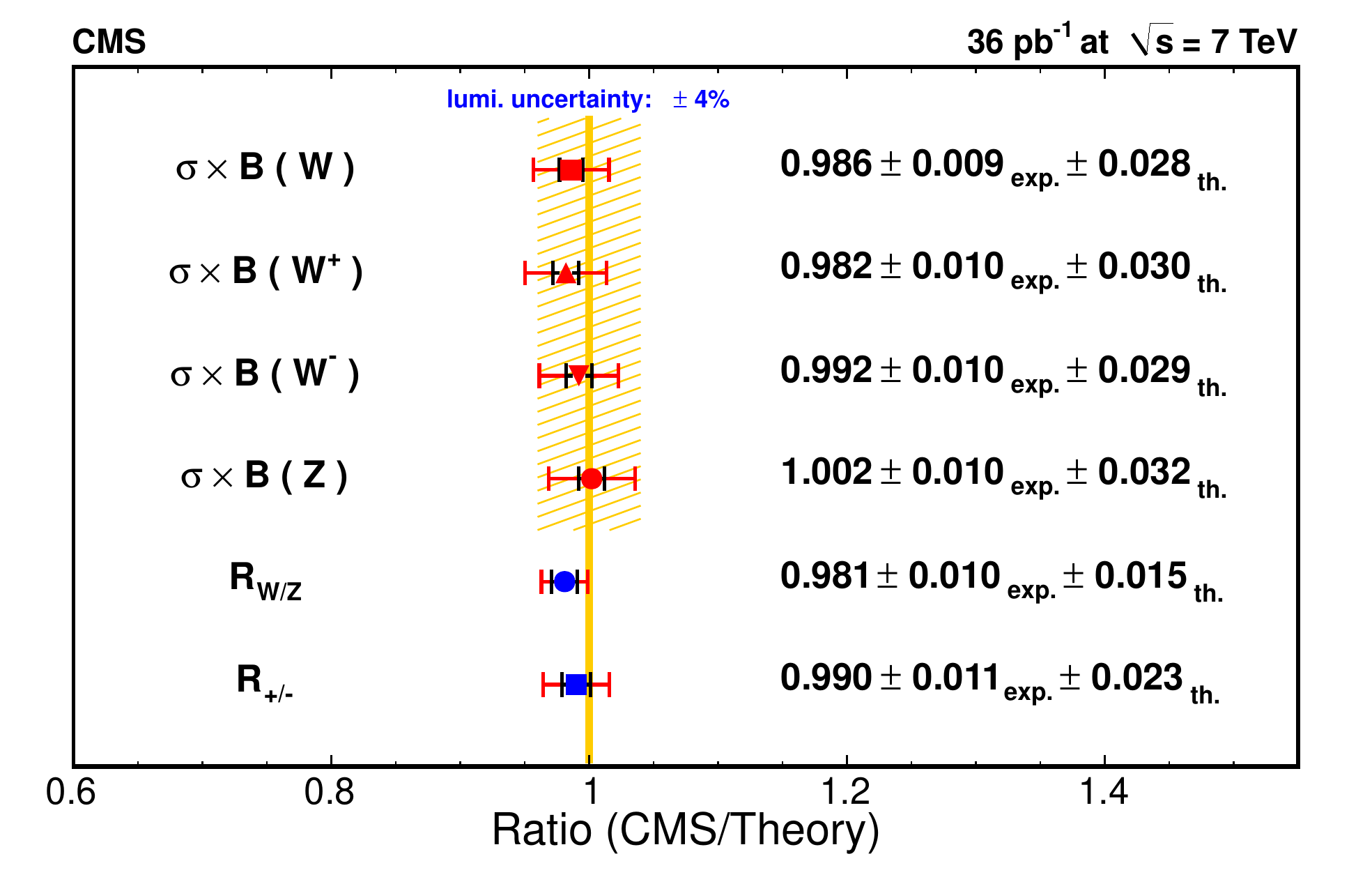}
\caption[.]{\label{fig:RatioCMSTHY}
Summary of ratios of the CMS measurements to the theoretical predictions.
The experimental uncertainties are represented as black error bars, while the
red error bars also include the combining of theoretical uncertainties
on the predictions and measured quantities. The yellow band around the vertical yellow line at one
represent the luminosity uncertainty (4\%) that affects the cross-section measurements.
}
\end{center}
\end{figure}

The ratios of the measurements to the theoretical
predictions are listed in Table~\ref{tab:RatioCMSTHY}
and displayed in Fig.~\ref{fig:RatioCMSTHY}. The experimental
uncertainty (``exp.'') is computed as the sum in quadrature of the
statistical uncertainty and the systematic uncertainties aside
from the luminosity uncertainty and the
theoretical uncertainties associated with the acceptance.
The theoretical uncertainty (``th.'') is computed by adding in
quadrature the theoretical uncertainties of the acceptance (or the
acceptance ratio) and the NNLO prediction, assuming that they are
uncorrelated.

\begin{table} %
\begin{center}
\caption[.]{ Summary of ratios of CMS measurements to the theoretical predictions. }
\begin{tabular}{|l|c|}
\hline
Quantity & Ratio (CMS/Theory) \\
\hline\hline
$\SIGBRSHORT{\Wpm}$   & $\RATCMSTHYWI$ \\
$\SIGBRSHORT{\Wp}$     & $\RATCMSTHYWP$ \\
$\SIGBRSHORT{\Wm}$     & $\RATCMSTHYWM$ \\
$\SIGBRSHORT{\Zo}$       & $\RATCMSTHYZ$  \\
$\SIGBRSHORT{\Wo}/\SIGBRSHORT{\Zo}$     & $\RATCMSTHYWZ$ \\
$\SIGBRSHORT{\Wp}/\SIGBRSHORT{\Wm}$ & $\RATCMSTHYWW$ \\
\hline
\end{tabular}
\label{tab:RatioCMSTHY}
\end{center}
\end{table}

\begin{figure}
\begin{center}
\includegraphics[width=0.8\textwidth]{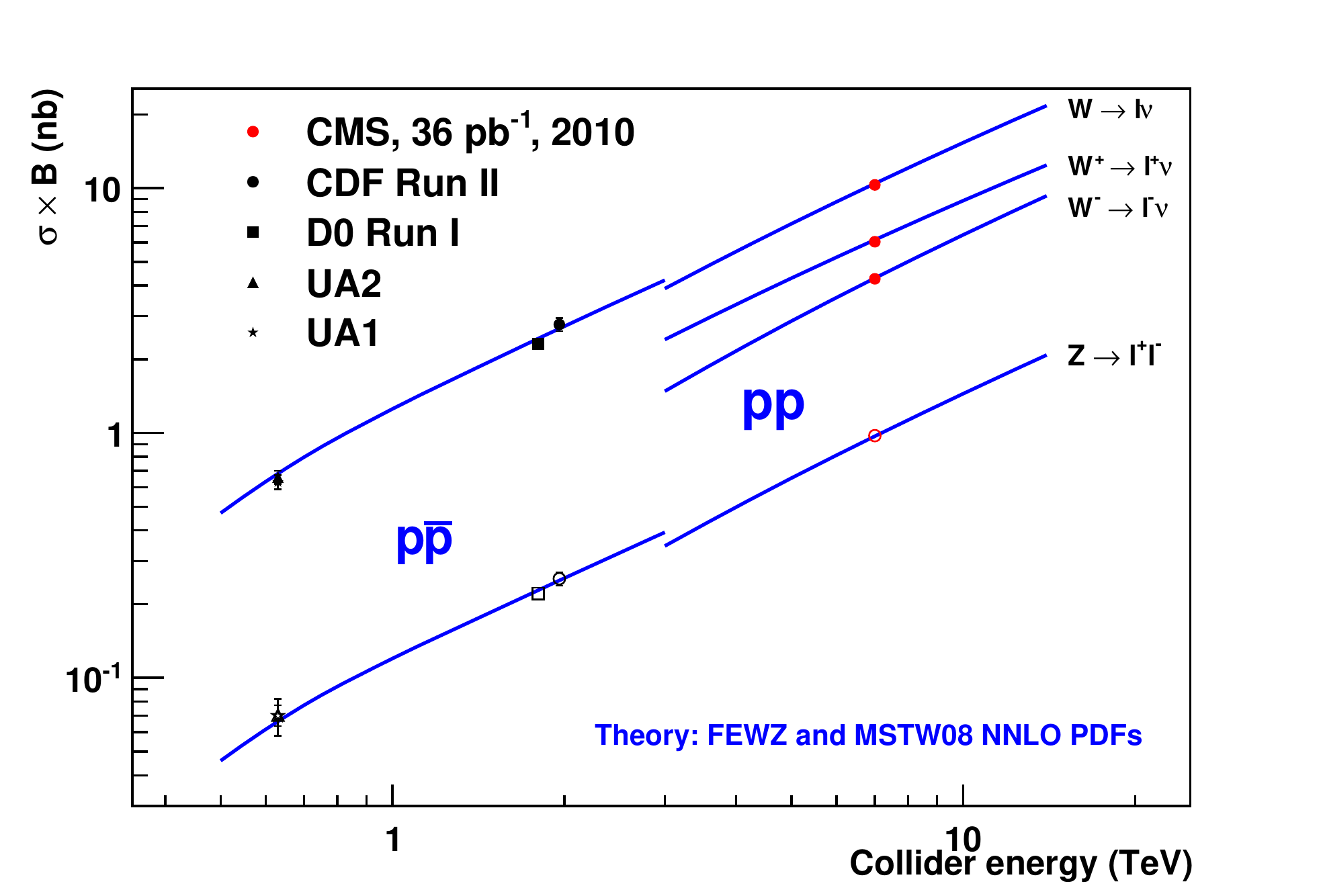}
\caption[.]{\label{fig:WZsigmas}
Measurements of inclusive W and Z production cross sections
times branching ratios as a function of center-of-mass energy for CMS and experiments
at lower-energy colliders. The lines are the NNLO theory predictions.
}
\end{center}
\end{figure}

Figure~\ref{fig:WZsigmas} shows the CMS W and Z cross section measurements together
with measurements at lower center-of-mass energy hadron colliders.
The predicted increase of the cross sections with center of mass energy is confirmed
by our measurements.

\par
Table~\ref{tab:res-restricted-xsections} reports the cross sections as
measured within the fiducial and kinematic acceptance, thereby eliminating the PDF uncertainties
from the results. In effect, these
uncertainties are transferred to the theoretical predictions,
allowing for a cleaner separation of experimental and
theoretical uncertainties.
For each channel the fiducial and kinematic acceptance
is defined as the fraction of events with lepton $\pt$ greater than $25~\GeV$
($20~\GeV$ for $\Zmm$), including no final-state QED radiation,
and with pseudorapidity in the range $|\eta|<2.5$ for electrons
and $|\eta|<2.1$ for muons.

\begin{table}[hbt] %
\begin{center}
\caption[.]{Summary of production cross section measurements
in restricted fiducial and kinematic acceptances. The $\pt$ and $|\eta|$
requirements restricting the acceptance for electrons and muons,
and the resulting acceptance values, are also given. The quoted uncertainties on
the acceptances (evaluated without FSR effect) are due to the PDF uncertainties.
\label{tab:res-restricted-xsections} }
\begin{tabular}{|l|r|c|c|}
\hline
\multicolumn{1}{|c|}{Channel} &
\multicolumn{1}{c|}{$\sigma \times {\cal B}$ in acceptance $A$ (nb)} &
\multicolumn{2}{c|}{$A$}
\\
\hline
\hline
$\Wen$  & 
{\footnotesize $5.688 \pm 0.016\,\mathrm{(stat.)} \pm 0.096\,\mathrm{(syst.)} \pm 0.228\,\mathrm{(lumi.)}$} & $0.543 \pm 0.003$ & \\
$\Wpen$ & 
{\footnotesize $3.404 \pm 0.012\,\mathrm{(stat.)} \pm 0.067\,\mathrm{(syst.)} \pm 0.136\,\mathrm{(lumi.)}$} & $0.554 \pm 0.004$ & $\pt>25\GeV$  \\
$\Wmen$ & 
{\footnotesize $2.284 \pm 0.010\,\mathrm{(stat.)} \pm 0.043\,\mathrm{(syst.)} \pm 0.091\,\mathrm{(lumi.)}$} & $0.527 \pm 0.006$ & $|\eta|<2.5$   \\
$\Zee$  & 
{\footnotesize $0.452 \pm 0.005\,\mathrm{(stat.)} \pm 0.010\,\mathrm{(syst.)} \pm 0.018\,\mathrm{(lumi.)}$} & $0.456 \pm 0.004$ & \\
\hline \hline
$\Wmn$  & {\footnotesize \WMISIGBRXGD }
& $0.465 \pm 0.004$ & {\multirow{3}{*}{\begin{tabular}{c}$\pt>25\GeV$\\$|\eta|<2.1$ \end{tabular}}} \\
$\Wpmn$ & {\footnotesize \WMPSIGBRXGD }
& $0.471 \pm 0.004$ &   \\
$\Wmmn$ & {\footnotesize \WMMSIGBRXGD }
& $0.457 \pm 0.007$ &   \\
\hline
{\multirow{2}{*}{$\Zmm$}}  & 
{\multirow{2}{*}{{\footnotesize $0.396 \pm 0.003\, \mathrm{(stat.)} \pm 0.007\, \mathrm{(syst.)} \pm 0.016\, \mathrm{(lumi.)}$}}} &
{\multirow{2}{*}{$0.409 \pm 0.005$}} & $\pt>20\GeV$\\
 & & & $|\eta|<2.1$ \\
\hline
\end{tabular}
\end{center}
\end{table}

\subsection{\texorpdfstring{Extraction of ${\cal B}(\Wln)$ and $\Gamma(\Wo)$}{Extraction of B(W->l n) and Gamma(W)}}
\label{sec:Gamma_W}

The precise value of the ratio of the W and Z cross sections
obtained from the combination of the measurements in the electron
and muon final states can be used to determine the SM
parameters ${\cal B}(\Wln)$ and $\Gamma(\Wo)$.

The ratio of W and Z cross sections can be written as
\begin{displaymath}
  R = \frac{\sigma(\pp \rightarrow \Wo X)}{\sigma(\pp \rightarrow \Zo X)}\,\,
  \frac{{\cal B}(\Wln)}{{\cal B}(\Zll)}\,.
\end{displaymath}

In order to estimate the value of ${\cal B}(\Wln)$
the predicted ratio of the W and Z production cross sections and the
measured value of the ${\cal B}(\Zll)$ are needed.
The NNLO prediction of the ratio, based on the MSTW08 PDFs, is $\sigma_\Wo$/$\sigma_\Zo$ = 3.34 $\pm$ 0.08.
The current measured value for ${\cal B}(\Zll)$ is 0.033658$\pm$0.000023
~\cite{PDG}. Those values lead to an indirect estimation of
\begin{displaymath}
{\cal B}(\Wln) = 0.106 \pm 0.003 \,,
\end{displaymath}
in agreement with the measured value, ${\cal B}(\Wln) = 0.1080 \pm 0.0009$~\cite{PDG}.

Using the SM value for the leptonic partial width,
$\Gamma(\Wln) = 226.6\pm 0.2$~MeV~\cite{GammaW-SM,GammaW-SM-New},
an indirect measurement of the total $\Gamma(\Wo)$ can be obtained through the formula
\begin{displaymath}
  {\cal B}(\Wln) = \frac{\Gamma(\Wln)}{\Gamma(\Wo)}\,.
\end{displaymath}

Based on the above values we obtain
\begin{displaymath}
\Gamma(\Wo) = 2144 \pm 62 ~~\mathrm{MeV}  \,.
\end{displaymath}

The SM prediction is 2093 $\pm$ 2 MeV~\cite{GammaW-SM-New} and the world average of experimental
results is 2085 $\pm$ 42 MeV~\cite{PDG}. The indirect measurement of $\Gamma(\Wo)$ is in good agreement
with the world average and the theoretical prediction, as well as other published measurements.

\section{Summary}
\label{sec:conclusions}

Measurements of the inclusive $\Wo$ and $\Zo$ production cross sections have been performed 
using a data sample of pp collision events at $\sqrt{s}=7\TeV$ 
collected with the CMS detector at the LHC in 2010 and 
corresponding to an integrated luminosity of 36~pb$^{-1}$. 
The inclusive production cross sections of $\Wp$ and $\Wm$ have been measured separately 
as well as the ratios of the $\Wp/\Wm$ and $\Wo/\Zo$ production cross sections.  
All measurements are dominated by systematic uncertainties,  
the main uncertainty originating from the integrated luminosity (4\%), which cancels in the ratios.
Experimental systematic uncertainties range from
0.7 to 1.8\%, and theoretical uncertainties range from 0.9 to 2.1\%.
The measurement of the $\Wo/\Zo$ cross-section ratio also leads to an
indirect determination of $\Gamma(\Wo)$, which is in agreement with the
current world average. 

The results agree with the ATLAS measurement~\cite{WZATLAS:2010} 
and with previous CMS results~\cite{WZCMS:2010}.
All measurements are consistent with the SM NNLO predictions.

\newpage
\section*{Acknowledgements}
\label{sec:acknowledgements}

\hyphenation{Bundes-ministerium Forschungs-gemeinschaft Forschungs-zentren} 
We wish to thank G.\ Watt for providing the theoretical predictions and for
our fruitful discussions.
We wish to congratulate our colleagues in the CERN accelerator departments 
for the excellent performance of the LHC machine. We thank the technical 
and administrative staff at CERN and other CMS institutes. This work was 
supported by the Austrian Federal Ministry of Science and Research; 
the Belgium Fonds de la Recherche Scientifique, 
and Fonds voor Wetenschappelijk Onderzoek; the Brazilian Funding 
Agencies (CNPq, CAPES, FAPERJ, and FAPESP); the Bulgarian Ministry 
of Education and Science; CERN; the Chinese Academy of Sciences, 
Ministry of Science and Technology, and National Natural Science 
Foundation of China; the Colombian Funding Agency (COLCIENCIAS); 
the Croatian Ministry of Science, Education and Sport; the Research 
Promotion Foundation, Cyprus; the Estonian Academy of Sciences and NICPB; 
the Academy of Finland, Finnish Ministry of Education and Culture, 
and Helsinki Institute of Physics; the Institut National de 
Physique Nucl\'eaire et de Physique des Particules~/~CNRS, 
and Commissariat \`a l'\'Energie Atomique et aux \'Energies Alternatives~/~CEA, France; 
the Bundesministerium f\"ur Bildung und Forschung, Deutsche Forschungsgemeinschaft, 
and Helmholtz-Gemeinschaft Deutscher Forschungszentren, Germany; 
the General Secretariat for Research and Technology, Greece; 
the National Scientific Research Foundation, and National Office for 
Research and Technology, Hungary; the Department of Atomic Energy and the 
Department of Science and Technology, India; the Institute for Studies in 
Theoretical Physics and Mathematics, Iran; the Science Foundation, Ireland; 
the Istituto Nazionale di Fisica Nucleare, Italy; the Korean Ministry of Education, 
Science and Technology and the World Class University program of NRF, Korea; 
the Lithuanian Academy of Sciences; the Mexican Funding Agencies (CINVESTAV, CONACYT, 
SEP, and UASLP-FAI); the Ministry of Science and Innovation, New Zealand; the Pakistan 
Atomic Energy Commission; the State Commission for Scientific Research, Poland; 
the Funda\c{c}\~ao para a Ci\^encia e a Tecnologia, Portugal; JINR (Armenia, Belarus, 
Georgia, Ukraine, Uzbekistan); the Ministry of Science and Technologies of the Russian 
Federation, the Russian Ministry of Atomic Energy and the Russian Foundation for Basic 
Research; the Ministry of Science and Technological Development of Serbia; 
the Ministerio de Ciencia e Innovaci\'on, and Programa Consolider-Ingenio 2010, Spain; 
the Swiss Funding Agencies (ETH Board, ETH Zurich, PSI, SNF, UniZH, Canton Zurich, and SER); 
the National Science Council, Taipei; the Scientific and Technical Research Council of Turkey, 
and Turkish Atomic Energy Authority; the Science and Technology Facilities Council, UK; 
the US Department of Energy, and the US National Science Foundation.

Individuals have received support from the Marie-Curie programme and the European Research 
Council (European Union); the Leventis Foundation; the A. P. Sloan Foundation; the Alexander 
von Humboldt Foundation; the Associazione per lo Sviluppo Scientifico e Tecnologico del 
Piemonte (Italy); the Belgian Federal Science Policy Office; the Fonds pour la 
Formation \`a la Recherche dans l'Industrie et dans l'Agriculture (FRIA-Belgium); 
the Agentschap voor Innovatie door Wetenschap en Technologie (IWT-Belgium); and the 
Council of Science and Industrial Research, India. 

\bibliography{auto_generated}
\cleardoublepage \appendix\section{The CMS Collaboration \label{app:collab}}\begin{sloppypar}\hyphenpenalty=5000\widowpenalty=500\clubpenalty=5000\textbf{Yerevan Physics Institute,  Yerevan,  Armenia}\\*[0pt]
S.~Chatrchyan, V.~Khachatryan, A.M.~Sirunyan, A.~Tumasyan
\vskip\cmsinstskip
\textbf{Institut f\"{u}r Hochenergiephysik der OeAW,  Wien,  Austria}\\*[0pt]
W.~Adam, T.~Bergauer, M.~Dragicevic, J.~Er\"{o}, C.~Fabjan, M.~Friedl, R.~Fr\"{u}hwirth, V.M.~Ghete, J.~Hammer\cmsAuthorMark{1}, S.~H\"{a}nsel, M.~Hoch, N.~H\"{o}rmann, J.~Hrubec, M.~Jeitler, W.~Kiesenhofer, M.~Krammer, D.~Liko, I.~Mikulec, M.~Pernicka, B.~Rahbaran, H.~Rohringer, R.~Sch\"{o}fbeck, J.~Strauss, A.~Taurok, F.~Teischinger, P.~Wagner, W.~Waltenberger, G.~Walzel, E.~Widl, C.-E.~Wulz
% ...
\vskip\cmsinstskip
% ...
1:~~Also at CERN, European Organization for Nuclear Research, Geneva, Switzerland\\
% ...

\end{sloppypar}
\end{document}